\documentclass[prd,aps,showpacs,nofootinbib,tightenlines]{revtex4}
\usepackage{mathrsfs}
\usepackage{amsmath}
\usepackage{amssymb}
\usepackage{epsfig}
\usepackage{graphicx}
\usepackage{booktabs}
\usepackage{multirow}
\usepackage{subfigure}
\usepackage[section]{placeins}
\usepackage{float}
\usepackage{slashed}
\usepackage{color}
\begin{document}
\newcommand{\psl}{ p \hspace{-1.8truemm}/ }
\newcommand{\nsl}{ n \hspace{-2.2truemm}/ }
\newcommand{\vsl}{ v \hspace{-2.2truemm}/ }
\newcommand{\epsl}{\epsilon \hspace{-1.8truemm}/\,  }

\title{ Mixing effects of $\eta-\eta'$ in $\Lambda_b\rightarrow \Lambda \eta^{(')}$  decays}
\author{Zhou Rui$^1$}\email[Corresponding  author: ]{jindui1127@126.com}
\author{Jia-Ming Li$^1$}
\author{Chao-Qi Zhang$^1$}
\affiliation{$^1$College of Sciences, North China University of Science and Technology,
Tangshan 063009,  China}
\date{\today}
\begin{abstract}
We perform a thorough analysis  of the  $\eta-\eta'$ mixing effects on the  $\Lambda_b\rightarrow \Lambda \eta^{(')}$ decays based on the perturbative QCD (PQCD) factorization approach.
Branching ratios, up-down asymmetries, and direct $CP$ asymmetries are computed by considering four popular mixing schemes,
such as  $\eta-\eta'$, $\eta-\eta'-\eta_c$, $\eta-\eta'-G$, and $\eta-\eta'-G-\eta_c$ mixing formalisms,
where $G$ represents the physical pseudoscalar glueball.
The PQCD predictions with the four mixing schemes does not change much for the $\eta$ channel but changes significantly for the $\eta'$ one.
In particular, the value of $\mathcal{B}(\Lambda_b\rightarrow \Lambda \eta^{'})$ in the $\eta-\eta'-G-\eta_c$  mixing scheme exceeds the present experimental bound by a factor of 2,
indicating the related mixing angles may be overestimated.
Because of the distinctive patterns of interference between $S$-wave and $P$-wave amplitudes,
the predicted up-down asymmetries for the two modes differ significantly.
The obvious discrepancies among different theoretical analyses should be clarified in the future.
The direct $CP$ violations are predicted to be at the level of a few percent
mainly because the tree contributions  of the strange and nonstrange amplitudes suffer from the color suppression and CKM  suppression.
Finally, as a by-product, we investigate the $\Lambda_b\rightarrow \Lambda \eta_c$ process,
which has a large branching ratio of order $10^{-4}$,
promising to be measured by the LHCb experiment.
Our findings are useful for constraining the mixing parameters,
comprehending the $\eta^{(')}$ configurations, and instructing experimental measurements.
\end{abstract}

\pacs{13.25.Hw, 12.38.Bx, 14.40.Nd }


\maketitle

\section{Introduction}
The phenomenon of mixing in the $\eta-\eta'$ system is an interesting subject in hadron physics.
In the exact $SU(3)$ flavor-symmetry limit, the pseudoscalar meson $\eta$ would be a pure flavor octet and $\eta'$ a flavor singlet.
However, it is known that the $SU(3)$ flavor symmetry breaking will lead to $\eta-\eta'$ mixing,
which can be either described in the octet-singlet basis~\cite{Leutwyler:1997yr,Kaiser:2000gs}  or in the quark-flavor basis~\cite{Feldmann:1998vh,Feldmann:1998sh}.
Occasionally, allowing for the heavier charm component in the $\eta$ and $\eta'$  mesons~\cite{Harari:1975ie,Tsai:2011dp},
one has to consider the $c\bar c$ components in the mixing basis.
In the QCD case, apart from the quark-antiquark combinations,
pure gluon configurations, such as  the two-gluon states, can also form an $SU(3)$ singlet,
which allows a possible gluonic admixture in $\eta'$ mesons~\cite{Escribano:2007cd,Ahmady:1997fa,Beneke:2002jn,Williamson:2006hb,Mathieu:2009sg,KLOE:2006guu,Ke:2010htz,Fleischer:2011ib}.
The $\eta-\eta'$ mixing phenomenon could then be generalized to include more states, such as glueballs and $\eta_c$ mesons~\cite{Feldmann:1999uf,Zhu:2018bwp}.
In this respect, a better knowledge of the quark and gluon components inside the $\eta$ and $\eta'$ states
 deepens our understanding of nonperturbative QCD dynamics in flavor physics~\cite{DiDonato:2011kr}
and the beauty hadron decays to $\eta,\eta'$  pseudoscalar mesons can be used to shed light on these phenomena.

In the $B$ meson sector, the observed hierarchy of $\mathcal{B}(B\rightarrow \eta'K)\gg \mathcal{B}(B\rightarrow \eta K)$~\cite{pdg2022} has attracted much attention
and many solutions have been proposed~\cite{Beneke:2002jn,Ahmady:1998ws,Du:1997hs,Halperin:1997as,Petrov:1997yf,Yang:2000ce,Khalil:2003bi}.
The large difference suggests a contribution to $\mathcal{B}(B\rightarrow \eta'K)$ via the $SU(3)$ singlet component of the $\eta$ and $\eta'$.
It has been known that the gluonic and charm contents of the light pseudoscalar mesons $\eta$ and $\eta'$ may have a crucial
impact on studies of many hadronic processes~\cite{DiDonato:2011kr}.
Various mixing mechanisms in $B\rightarrow \eta^{(')}$ decays have been explored in the context of perturbative QCD factorization (PQCD),
in which the transverse momenta of valence quarks are included to regulate the end-point singularities;
see, e.g., Refs.z~\cite{Kou:2001pm,Charng:2006zj,Xiao:2008sw,Tsai:2011dp,prd87094003,Akeroyd:2007fy}.
The earlier PQCD  predictions at leading order  for $B\rightarrow \eta^{(')}K$ without the flavor-singlet amplitudes are higher (lower) than the measured values~\cite{Kou:2001pm}.
Although the partial next-to-leading order (NLO) contributions are included in~\cite{Xiao:2008sw},
the gap between theory and experiment is still not completely understood.
In Ref.~\cite{Charng:2006zj}, the authors examined the gluonic contribution to the $B\rightarrow \eta,\eta'$  transition form factors
and found that it is numerically neglected.
A mixing scheme for $\eta$, $\eta'$, and the pseudoscalar glueball was proposed in~\cite{Cheng:2008ss},
in which the formalism for the $\eta-\eta'-G$ mixing was set up.
Three years later, this trimixing formalism was further extended to the $\eta-\eta'-G-\eta_c$  tetramixing by including the $\eta_c$ meson in Ref.~\cite{Tsai:2011dp}.
They discovered that the $\eta_c$-mixing effects enhance the PQCD predictions for  $\mathcal{B}(B\rightarrow \eta'K)$ by $18\%$
but not for  $\mathcal{B}(B\rightarrow \eta K)$,
and they claimed that the puzzle of the above distinctive pattern can be resolved.
In Ref.~\cite{prd87094003},  three different mixing schemes for the $\eta-\eta'$ system were taken into account,
and the PQCD calculations for the $B\rightarrow \eta^{(')}K$  decays were improved to the NLO level.
It is found that the NLO PQCD predictions in the $\eta-\eta'-G$ mixing scheme provide a nearly perfect interpretation of the measured values.

The $\eta-\eta'-G$ mixing scheme was also applied to the $B\rightarrow J/\psi \eta^{(')}$ decays~\cite{Liu:2012ib} in PQCD.
A large gluon contribution was advocated from the analysis of
relative probabilities of the $B_s\rightarrow J/\psi \eta^{'}$ and $B_s\rightarrow J/\psi \eta$ decays.
However, the subsequent measurements from LHCb~\cite{LHCb:2012cw,LHCb:2014oms}
hint at a small gluonic component in the $\eta'$ meson.
It is then worthwhile to examine whether these  mixing schemes can explain the measurements well in the baryon
reactions involving $\eta$ or $\eta'$ mesons, such as $\Lambda_b\rightarrow\Lambda \eta^{(')}$ decays.

Searches for the $\Lambda_b\rightarrow\Lambda \eta$ and $\Lambda_b\rightarrow\Lambda \eta^{'}$ decays have been performed by the LHCb~\cite{LHCb:2015kmm} Collaboration.
The branching ratio of the former was measured to be $\mathcal{B}(\Lambda_b\rightarrow\Lambda \eta)=9.3^{+7.3}_{-5.3}\times 10^{-6}$ at the level of 3$\sigma$ significance,
while an upper limit for the latter mode was set as $3.1\times 10^{-6}$ at the $90\%$ confidence level.
Some predictions exist for the $\Lambda_b\rightarrow\Lambda \eta^{(')}$  decays.
Within the framework of the light-front quark model (LFQM)~\cite{Wei:2009np},
the branching ratios were estimated to be at the order of $10^{-8}$  in the absence of penguin contributions.
In Ref~\cite{prd99054020}, based on the QCD factorization (QCDF),
the branching ratios were predicted to be in the ranges $\mathcal{B}(\Lambda_b\rightarrow\Lambda \eta) \sim (3-6)\times 10^{-7}$
and $\mathcal{B}(\Lambda_b\rightarrow\Lambda \eta') \sim (3-7)\times 10^{-6}$ with large theoretical uncertainties,
while in Ref.~\cite{epjc76399} the branching ratios were calculated to be
$\mathcal{B}(\Lambda_b\rightarrow\Lambda \eta, \Lambda \eta') = (1.47\pm0.35, 1.83\pm0.58)\times 10^{-6}$, exploiting the generalized factorization approach (GFA).
In an earlier paper~\cite{plb598203}, a wider range, $(1.8-19.0)\times 10^{-6}$, was estimated by using different models for the $\Lambda_b\rightarrow\Lambda$ form factors.

Our purpose in the present paper is to
probe the $\eta-\eta'$ mixing in the $\Lambda_b\rightarrow\Lambda \eta^{(')}$  decays by employing the PQCD approach at leading order accuracy.
Four available mixing schemes for the $\eta-\eta'$ system\textemdash namely,
$\eta-\eta'$, $\eta-\eta'-G$, $\eta-\eta'-G-\eta_c$,  and $\eta-\eta'-\eta_c$ mixing\textemdash are taken into account.
The effect of radial mixing is neglected due to the absence of $\Lambda_b\rightarrow \eta^{(')}$ form factors in the relevant processes~\cite{Datta:2002pk},
and the mixing with the pion under the isospin symmetry is not considered either.
Within these mixing schemes, we calculate the branching ratios, up-down asymmetries, and direct $CP$ violations for $\Lambda_b\rightarrow\Lambda \eta^{(')}$
and investigate the scheme dependence of the theoretical predictions.

The paper is organized as follows.
In Sec.~\ref{sec:framework},
we first discuss the four mixing schemes as well as the related mixing angles and review the hadronic light-cone distribution amplitudes (LCDAs).
Then, we briefly present the effective Hamiltonian and kinematics for the PQCD calculations.
We  show the PQCD predictions for the branching ratios,  up-down asymmetries and direct $CP$ asymmetries of the relevant decays with four different mixing schemes in Sec.~\ref{sec:results}.
A summary will be given in the last section.
The Appendix is devoted to details for the computation of the decay amplitudes  within PQCD.


\section{Theoretical framework}\label{sec:framework}

\subsection{ $\eta-\eta'$  mixing phenomenon}
This section is devoted to the phenomenological aspects of $\eta-\eta'$  mixing.
In this work we consistently use the quark-flavor mixing basis rather than the singlet-octet mixing basis
since fewer two-parton twist-3  meson distribution amplitudes need to be introduced~\cite{Charng:2006zj}.
Following the analysis of Refs.~\cite{Feldmann:1998vh,prd87094003,Cheng:2008ss,Tsai:2011dp}, we first introduce four different $\eta-\eta'$ mixing schemes.
In the conventional Feldmann-Kroll-Stech (FKS) scheme~\cite{Feldmann:1998vh,Feldmann:1998sh} for the $\eta-\eta'$  mixing,
the physical neutral pseudoscalar mesons $\eta^{(')}$ can be represented as a superposition of isosinglet states,
\begin{eqnarray}\label{eq:2mixing}
\left(
\begin{array}{c}
\eta\\
\eta'
\end{array}
\right)=
\left(
\begin{array}{cc}
c\phi & -s\phi\\
s\phi &  c\phi
\end{array}
\right)
\left(
\begin{array}{c}
\eta_q\\
\eta_s
\end{array}
\right),
\end{eqnarray}
with shorthand $(c, s) \equiv(\cos, \sin)$ and $\phi$ being the mixing angle.
Here, $\eta_q=(u\bar u+d\bar d)/\sqrt{2}$ and  $\eta_s=s\bar s$ are the so-called nonstrange and strange quark-flavor states, respectively.
The  presence of  only one mixing angle in this case is due to the Okubo-Zweig-Iizuka (OZI) suppressed contributions being neglected~\cite{Feldmann:1999uf}.
For the details of the two-angle mixing scheme for $\eta-\eta'$ system, see  Refs.~\cite{Schechter:1992iz,Escribano:2005qq}.

Alternatively, allowing for another heavy-quark charm $c \bar c$ component in the $\eta$ and $\eta'$,
the conventional FKS formalism can be generalized naturally to the trimixing of $\eta-\eta'-\eta_c$  in the $q\bar q-s\bar s-c\bar c$ basis.
The physical states are related to the flavor states via~\cite{Feldmann:1998vh}
\begin{eqnarray}\label{eq:3mixingp}
\left(
\begin{array}{c}
\eta\\
\eta'\\
\eta_c
\end{array}
\right)=
\left(
\begin{array}{ccc}
c\phi  & -s\phi  &-\theta_cs \theta_y\\
s\phi  & c\phi   &\theta_cc \theta_y \\
 -\theta_c s (\phi-\theta_y)&-\theta_c c (\phi-\theta_y)  &1
\end{array}
\right)
\left(
\begin{array}{c}
\eta_q\\
\eta_s\\
\eta_{c}
\end{array}
\right),
\end{eqnarray}
where $\theta_c$ and $\theta_y$ are two new mixing angles related to the charm decay constants of the $\eta^{(')}$ mesons.

In QCD, gluons may form a bound state, called gluonium, that can mix with neutral mesons~\cite{KLOE:2006guu}.
By including a possible  pseudoscalar glueball state $\eta_g$ in the $\eta^{(')}$ mesons~\cite{Cheng:2008ss,Fleischer:2011ib},
the FKS  mixing scheme can be extended to the $\eta-\eta'-G$ mixing formalism,
where $G$ denotes the physical pseudoscalar glueball.
Using the quark-flavor basis, we can write~\cite{Cheng:2008ss,Liu:2012ib}
\begin{eqnarray}\label{eq:3mixing}
\left(
\begin{array}{c}
\eta\\
\eta'\\
G
\end{array}
\right)=
\left(
\begin{array}{ccc}
c\phi+s\theta s\theta_i(1-c \phi_G) & -s\phi+s\theta c\theta_i(1- c \phi_G)&-s \theta s \phi_G\\
s\phi-c\theta s\theta_i(1-c \phi_G) & c\phi-c\theta c\theta_i(1-c \phi_G)&c \theta s \phi_G\\
-s \theta_i s \phi_G &-c \theta_i s \phi_G &c \phi_G
\end{array}
\right)
\left(
\begin{array}{c}
\eta_q\\
\eta_s\\
\eta_g
\end{array}
\right),
\end{eqnarray}
where $\theta_i=54.7^\circ$ is the ideal mixing angle between the octet-singlet  and  the quark-flavor states in the SU(3) flavor-symmetry  limit~\cite{DiDonato:2011kr,Bramon:1997va}.
Here, $\theta$ is related to $\phi$ by  $\theta=\phi-\theta_i$, and
$\phi_G$ is the mixing angle for the gluonium contribution.
We assume that the glueball only mixes with the flavor-singlet $\eta_1$ but not with the flavor-octet $\eta_8$,
so the two mixing angles $\phi$ and $\phi_G$ are sufficient to describe the mixing matrix in Eq.~(\ref{eq:3mixing}).
It has been verified that the contribution from the gluonic distribution amplitudes in the  $\eta^{(')}$ meson is negligible for $B$ meson transition form factors~\cite{Charng:2006zj}.
Hence, we still suppose that the $\eta$ and $\eta^{'}$ mesons are produced via the nonstrange (strange) component in the baryon decays under the $\eta-\eta'-G$ mixing.

In~\cite{Tsai:2011dp}, the authors combined
the above two trimixings by considering the tetramixing of $\eta-\eta'-G-\eta_c$,
 which is described by a $4\times4$ mixing matrix.
It was assumed that the heavy-flavor state only mixes with the pseudoscalar glueball; then the transformation reads
\begin{eqnarray}\label{eq:4mixing}
\left(
\begin{array}{c}
\eta\\
\eta'\\
G\\
\eta_c
\end{array}
\right)=
\left(
\begin{array}{cccc}
c\theta c\theta_i-s\theta s\theta_ic\phi_G & -c\theta s\theta_i-s\theta c\theta_ic \phi_G&-s \theta s \phi_Gc\phi_C&-s \theta s \phi_Gs\phi_C\\
s\theta c\theta_i+c\theta s\theta_ic\phi_G & -s\theta s\theta_i+c\theta c\theta_i c \phi_G&c \theta s \phi_G c\phi_C&c \theta s \phi_Gs\phi_C\\
s\theta_is\phi_G&-c\theta_is\phi_G&c\phi_Gc\phi_C&c\phi_Gs\phi_C\\
0&0&-s\phi_C&c\phi_C
\end{array}
\right)
\left(
\begin{array}{c}
\eta_q\\
\eta_s\\
\eta_g\\
\eta_c
\end{array}
\right),
\end{eqnarray}
where the new angle $\phi_C$ is the mixing angle between the glueball and $\eta_c$ components.
It can be easily seen   that the  $\eta-\eta'-G-\eta_c$  tetramixing formalism reduces to the $\eta-\eta'-G$ and  FKS schemes
in the $\phi_C\rightarrow 0$ and $\phi_{C,G}\rightarrow 0$ limits, respectively.

As the mixing of $\eta$ and $\eta'$ is still not completely clear at the moment,
they may be mixed with the radial excitations,
leading to more complicated mixing formalism.
In the following analysis, we ignore other possible admixtures from radial excitations.
In addition, we assume that isospin symmetry is exact ($m_u = m_d \ll m_s$);
the mixing with $\pi$\textemdash such as the $\pi-\eta$ mixing~\cite{Gross:1979ur},
the trimixing of $\pi-\eta-\eta'$~\cite{Gusbin:1980gd,Qian:2009dc}, and the tetramixing of $\pi-\eta-\eta'-\eta_c$~\cite{Peng:2011ue}\textemdash are not considered here.

\subsection{Light-cone distribution amplitudes}\label{sec:LCDAs}
The hadronic LCDAs are important in PQCD calculations,
which describe the momentum fraction distribution of valence quarks inside hadrons.
There are various models of the $\Lambda_b$ and $\Lambda$ baryon LCDAs available in the
literature~\cite{plb665197,J. High Engry Phys.112013191,epjc732302,plb738334,J. High Engry Phys.022016179,Ali:2012zza,zpc42569,Liu:2014uha,Liu:2008yg,J. High Engry Phys.020702016,prd89094511,epja55116}.
In this work, we adopt the exponential model LCDAs for the $\Lambda_b$ baryon~\cite{J. High Engry Phys.112013191} and  Chernyak-Ogloblin-Zhitnitsky (COZ) model for the $\Lambda$~\cite{zpc42569},
whose explicit expressions can be found in the previous work~\cite{Rui:2022sdc,Rui:2022jwu,Han:2022srw} and shall not be repeated here.
It has been confirmed that the models employed lead to reasonable numerical results for the  $\Lambda_b\rightarrow\Lambda$ form factor with fewer free parameters~\cite{Rui:2022sdc}.

Two-parton quark components for the $\eta_{s,c}$ mesons are defined via the nonlocal matrix elements~\cite{Charng:2006zj,Chen:2005ht,Rui:2016opu}
\begin{eqnarray}\label{eq:etas}
\langle \eta_{s}(q)|\bar s_\beta(z)s_\alpha(0)|0\rangle &=&-\frac{i}{\sqrt{2 N_c}}\int_0^1dye^{iyq\cdot z}\gamma_5[\rlap{/}{q}\phi^A_{s}(y)+m^{s}_0\phi^P_{s}(y)+m^{s}_0(\rlap{/}{v}\rlap{/}{n}-1)\phi^T_{s}(y)]_{\alpha\beta},\nonumber\\
\langle \eta_{c}(q)|\bar c_\beta(z)c_\alpha(0)|0\rangle &=&-\frac{i}{\sqrt{2 N_c}}\int_0^1dye^{iyq\cdot z}\gamma_5[\rlap{/}{q}\phi^v_{\eta_c}(y)+m_{\eta_c}\phi^s_{\eta_c}(y)],
\end{eqnarray}
where $N_c$ is the number of colors. Here, $\eta_q$ can be obtained by substituting $s$ for $d$ in Eq.~(\ref{eq:etas}) and multiplying by a factor of $1/\sqrt{2}$.
The two light-cone vectors $n=(1,0,\textbf{0}_T)$ and $v=(0,1,\textbf{0}_T)$ satisfy $n\cdot v=1$.
Note that $m_0^{q,s}$ are the chiral enhancement scales associated with the twist-3 LCDAs,
which can be expressed in terms of the decay constants $f_{q,s}$  and the mixing angles.
Their values can be fixed by solving for the mass matrix in different mixing schemes~\cite{Feldmann:1998vh},
which will be given in the next section.

The  models for the various twist distribution amplitudes have been determined in~\cite{Ball:2004ye,Sun:2008ew},
\begin{eqnarray}\label{eq:mlcdas}
\phi^A_{q,s}(y)&=&\frac{f_{q,s}}{2\sqrt{2N_c}}6y(1-y)[1+a_2C_2^{3/2}(u)+a_4C_4^{3/2}(u)],\nonumber\\
\phi^P_{q,s}(y)&=&\frac{f_{q,s}}{2\sqrt{2N_c}}[1+(30\eta_3- \frac{5}{2}\rho^2_{q,s})C_2^{1/2}(u)-3(\eta_3\omega_3+\frac{9}{20}\rho^2_{q,s}(1+6a_2))C_4^{1/2}(u)],\nonumber\\
\phi^T_{q,s}(y)&=&\frac{f_{q,s}}{2\sqrt{2N_c}}(1-2y)[1+6(5\eta_3-\frac{1}{2}\eta_3\omega_3-\frac{7}{20}\rho^2_{q,s}-\frac{3}{5}\rho^2_{q,s}a_2)(1-10y+10y^2)],\nonumber\\
\phi^v_{\eta_c}(y,b)&=&\frac{f_{\eta_c}}{2\sqrt{2N_c}}N^vy(1-y)\exp\{-\frac{m_c}{\omega}y(1-y)[(\frac{1-2y}{2y(1-y)})^2+\omega^2b^2]\},\nonumber\\
\phi^s_{\eta_c}(y,b)&=&\frac{f_{\eta_c}}{2\sqrt{2N_c}}N^s\exp\{-\frac{m_c}{\omega}y(1-y)[(\frac{1-2y}{2y(1-y)})^2+\omega^2b^2]\},
\end{eqnarray}
where we resort to SU(3) symmetry and use the same Gegenbauer moments for the $\eta_q$ and $\eta_s$.
This approximation is reasonable since the main SU(3) breaking in Gegenbauer moments is due to the nonzero odd terms and that in $a_2$ is subleading~\cite{Ball:2007hb}.
Here, we include the SU(3)-breaking effect via the decay constants, the chiral scales, and the parameters $\rho_{q,s}$ in the LCDAs.
These nonperturbative parameters are not all independent.
The two mass ratios $\rho_{q,s}$ are related to the respective chiral scales by $\rho_{q,s}=2m_{q,s}/m_0^{q,s}$, with $m_{q,s}$ being the current quark masses.
The two parameters  $\eta_3=0.015$ and $\omega_3=-3$ are the same as for the pion distribution amplitudes~\cite{Ball:2004ye}.
The Gegenbauer polynomials $C(u)$  are given as
\begin{eqnarray}
 C_2^{3/2}(u)=\frac{3}{2}(5u^2-1),\quad C_2^{1/2}(u)=\frac{1}{2}(3u^2-1), \quad C_4^{1/2}(u)=\frac{1}{8}(3-30u^2+35u^4),\quad C_4^{3/2}(u)=\frac{15}{8}(1-14u^2+21u^4),
\end{eqnarray}
with $u=2y-1$.
The shape parameter $\omega=0.6$ GeV is taken from~\cite{Sun:2008ew}.
Note that $f_{\eta_c}$ and $m_{\eta_c}$ are  the decay constant and mass of the $\eta_c$ meson, respectively.
The two normalization constants $N^{v,s}$ are determined by~\cite{Sun:2008ew}
\begin{eqnarray}
\int^1_0 dy \phi^{v,s}_{\eta_c}(y,b=0)=\frac{f_{\eta_c}}{2\sqrt{2N_c}}.
\end{eqnarray}

\subsection{ PQCD calculation}
\begin{figure}[!htbh]
	\begin{center}
		\vspace{0.01cm} \centerline{\epsfxsize=15cm \epsffile{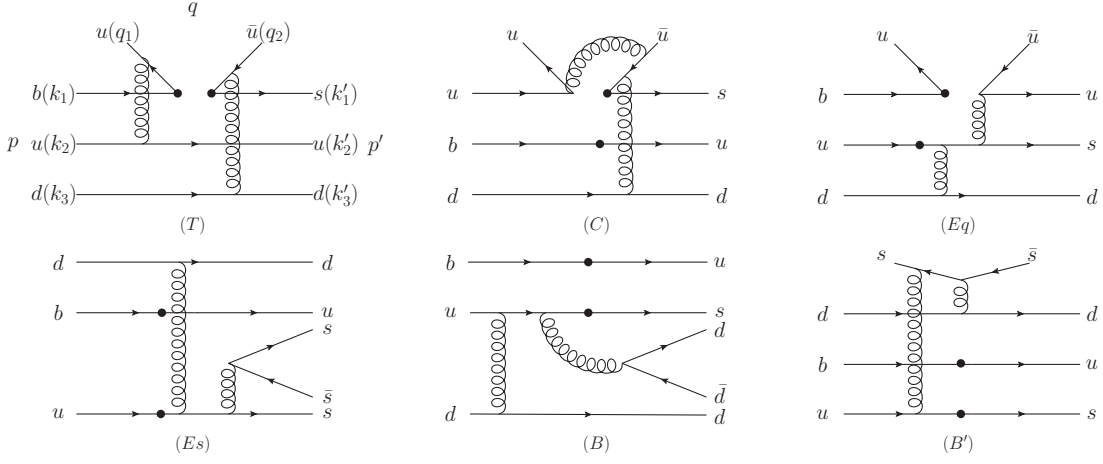}}
		\setlength{\abovecaptionskip}{1.0cm}
		\caption{Sample topological diagrams responsible for the decay $\Lambda_b\rightarrow\Lambda\eta^{(')}$.
Here, $T$ denotes the external W emission diagram, $C$
represents the internal W emission diagram, $E_{q(s)}$ labels the W exchange diagram with nonstrange (strange) components contributing to the $\eta^{(')}$ mesons,
$B$ denotes the bow-tie-type W exchange diagram, and $B'$ represents the diagram that can be obtained from the W exchange diagram by exchanging the two identical
quarks in the final states.}
		\label{fig:FeynmanT}
	\end{center}
\end{figure}
The PQCD approach has been developed and successfully applied to deal with the $\Lambda_b$ hadronic
decays~\cite{prd59094014,prd61114002,cjp39328,prd74034026,prd65074030,prd80034011,Han:2022srw,Zhang:2022iun,Rui:2022sdc,Rui:2022jwu}.
In the PQCD picture, the decay amplitudes can be calculated by the convolution of the nonperturbative, universal LCDAs and the perturbative hard scattering amplitude.
After defining the nonperturbative LCDAs in the last subsection,
we are ready to calculate the decay amplitudes of the strong coupling constant at leading order.
Various topological diagrams responsible for the considered decays  are presented in Fig.~\ref{fig:FeynmanT}.
The labels $T$, $C$, $E$, and $B$ refer to  external W emission, internal W emission, W exchange, and  bow-tie topologies, respectively.
The subscript $q(s)$ of $E$ corresponds to the contribution from the nonstrange (strange) component in the $\eta^{(')}$ mesons.
Exchanging two identical quarks in the final-state baryon and meson for the $E$- or $B$-type diagram,
we obtain a new topology denoted by $B'$ as exhibited in the last diagram of Fig.~\ref{fig:FeynmanT}.
We draw one representative Feynman diagram for each topology here;
for a more complete set of Feynman diagrams, refer to our previous work~\cite{Rui:2022sdc,Rui:2022jwu,Zhang:2022iun}.

In the $\Lambda_b$ rest frame, we choose the $\Lambda_b(\Lambda)$ baryon momentum $p(p')$ and the meson momentum $q$ in the light-cone coordinates:
\begin{eqnarray}\label{eq:pq}
p=\frac{M}{\sqrt{2}}(1,1,\textbf{0}_{T}), \quad p'=\frac{M}{\sqrt{2}}(f^+,f^-,\textbf{0}_{T}), \quad
q=\frac{M}{\sqrt{2}}\left(1-f^+,1-f^-,\textbf{0}_{T}\right),
\end{eqnarray}
with $M$ being the $\Lambda_b$ baryon mass.
The factors $f^{\pm}$ can be derived from the on-shell conditions $p'^2=m_\Lambda^2$ and $q^2=m^2$ for the final-state hadrons, which yield
\begin{eqnarray}
f^\pm=\frac{1}{2}\left(1-r^2+r_{\Lambda}^2 \pm \sqrt{(1-r^2+r_{\Lambda }^2)^2-4r_{\Lambda}^2}\right),
\end{eqnarray}
with the mass ratios $r_{(\Lambda)}=m_{(\Lambda)}/M$.
The spectator momenta inside the initial and final states  are parametrized as
\begin{eqnarray}
k_1&=&\left(\frac{M}{\sqrt{2}},\frac{M}{\sqrt{2}}x_1,\textbf{k}_{1T}\right),\quad
k_2=\left(0,\frac{M}{\sqrt{2}}x_2,\textbf{k}_{2T}\right),\quad
k_3=\left(0,\frac{M}{\sqrt{2}}x_3,\textbf{k}_{3T}\right),\nonumber\\
k_1'&=&\left(\frac{M}{\sqrt{2}}f^+x_1',0,\textbf{k}'_{1T}\right),\quad
k_2'=\left(\frac{M}{\sqrt{2}}f^+x_2',0,\textbf{k}'_{2T}\right),\quad
k_3'=\left(\frac{M}{\sqrt{2}}f^+x_3',0,\textbf{k}'_{3T}\right),\nonumber\\
q_1&=&\left(\frac{M}{\sqrt{2}}y(1-f^+),\frac{M}{\sqrt{2}}y(1-f^-),\textbf{q}_{T}\right),\nonumber\\
q_2&=&\left(\frac{M}{\sqrt{2}}(1-y)(1-f^+),\frac{M}{\sqrt{2}}(1-y)(1-f^-),-\textbf{q}_{T}\right),
\end{eqnarray}
where  $x^{(')}_{1,2,3}$ and $y$ are the parton longitudinal momentum fractions, and
$\textbf{k}^{(')}_{1T,2T,3T}$ and $\textbf{q}_T$ are the corresponding transverse momenta.
The momentum conservation implies the relations
\begin{eqnarray}
\sum_{l=1}^3x^{(')}_l=1,\quad \sum_{l=1}^3\textbf{k}^{(')}_{lT}=0.
\end{eqnarray}
Here, all the kinematical variables are labeled in the first diagram of Fig.~\ref{fig:FeynmanT}.

Based on the operator product expansion, the effective weak-interaction Hamiltonian for the $b\rightarrow s$ transition reads~\cite{Buchalla:1995vs}
\begin{eqnarray}
\mathcal{H}_{eff}&=&\frac{G_F}{\sqrt{2}} \{V_{Qb}V^*_{Qs}[C_1(\mu)O^Q_1(\mu)+C_2(\mu)O^Q_2(\mu)]-\sum_{k}V_{tb}V^*_{ts}C_k(\mu)O_k(\mu)\}+\mathrm{H.c.},
\end{eqnarray}
where $V_{Qb,Qs,tb,ts}$ are the Cabibbo-Kobayashi-Maskawa (CKM) matrix elements with $Q=u$ for $\eta_{q,s}$ and $Q=c$ for $\eta_{c}$.
Here, $G_F$ is the Fermi constant, and $\mu$ is the factorization scale.
The sum over $k$ comprises the QCD penguin operators $O_{3-6}$ and the electroweak penguin operators $O_{7-10}$.
The $C_i$'s are the Wilson coefficients which encode the short-distance physics.
The four-quark operators $O_i$ describing the hard electroweak process in $b$ quark decays read
\begin{eqnarray}
O^Q_1&=& \bar{Q}_\alpha \gamma_\mu(1-\gamma_5) b_\beta  \otimes \bar{s}_\beta  \gamma^\mu(1-\gamma_5) Q_\alpha, \nonumber\\
O^Q_2&=& \bar{Q}_\alpha \gamma_\mu(1-\gamma_5) b_\alpha \otimes \bar{s}_\beta  \gamma^\mu(1-\gamma_5) Q_\beta,  \nonumber\\
O_3&=& \bar{s}_\beta  \gamma_\mu(1-\gamma_5) b_\beta  \otimes \sum_{q'} \bar{q}'_\alpha \gamma^\mu(1-\gamma_5) q'_\alpha, \nonumber\\
O_4&=& \bar{s}_\beta  \gamma_\mu(1-\gamma_5) b_\alpha \otimes \sum_{q'} \bar{q}'_\alpha \gamma^\mu(1-\gamma_5) q'_\beta,  \nonumber\\
O_5&=& \bar{s}_\beta  \gamma_\mu(1-\gamma_5) b_\beta  \otimes \sum_{q'} \bar{q}'_\alpha \gamma^\mu(1+\gamma_5) q'_\alpha, \nonumber\\
O_6&=& \bar{s}_\beta  \gamma_\mu(1-\gamma_5) b_\alpha \otimes \sum_{q'} \bar{q}'_\alpha \gamma^\mu(1+\gamma_5) q'_\beta,  \nonumber\\
O_7&=&   \frac{3}{2}\bar{s}_\beta \gamma_\mu(1-\gamma_5) b_\beta  \otimes \sum_{q'} e_{q'}\bar{q}'_\alpha \gamma^\mu(1+\gamma_5) q'_\alpha, \nonumber\\
O_8&=&   \frac{3}{2}\bar{s}_\beta \gamma_\mu(1-\gamma_5) b_\alpha \otimes \sum_{q'} e_{q'}\bar{q}'_\alpha \gamma^\mu(1+\gamma_5) q'_\beta,  \nonumber\\
O_9&=&   \frac{3}{2}\bar{s}_\beta \gamma_\mu(1-\gamma_5) b_\beta  \otimes \sum_{q'} e_{q'}\bar{q}'_\alpha \gamma^\mu(1-\gamma_5) q'_\alpha, \nonumber\\
O_{10}&=&\frac{3}{2}\bar{s}_\beta \gamma_\mu(1-\gamma_5) b_\alpha \otimes \sum_{q'} e_{q'}\bar{q}'_\alpha \gamma^\mu(1-\gamma_5) q'_\beta.
\end{eqnarray}
where the sum over $q'$ runs over the quark fields that are active at the scale $\mu=\mathcal{O}(m_b)$.

The decay amplitudes of $\Lambda_b\rightarrow\Lambda \eta_q$, $\Lambda_b\rightarrow\Lambda \eta_s$, and $\Lambda_b\rightarrow\Lambda \eta_c$
(namely, nonstrange $\mathcal{M}_{q}$, strange $\mathcal{M}_{s}$, and charm $\mathcal{M}_{c}$)
are given by sandwiching $\mathcal{H}_{eff}$ with the initial and final states,
\begin{eqnarray}
\mathcal{M}_{q,s,c}=\langle \Lambda \eta_{q,s,c}|\mathcal{H}_{eff}|\Lambda_b\rangle,
\end{eqnarray}
which can be further expanded with the Dirac spinors as
\begin{eqnarray}\label{eq:kq}
\mathcal{M}=\bar {\Lambda} (p')[M_S+M_P \gamma_5]\Lambda_b(p),
\end{eqnarray}
where $M_S$ and $M_P$  correspond to the parity-violating $S$-wave and parity-conserving  $P$-wave amplitudes, respectively.
Their generic factorization formula can be written as
\begin{eqnarray}\label{eq:amp}
M_{S/P}=\frac{f_{\Lambda_b}\pi^2 G_F}{18\sqrt{3}}\sum_{R_{ij}}
\int[d^3x][d^3x']dy[\mathcal{D}b]_{R_{ij}}
\alpha_s^2(t_{R_{ij}})\Omega_{R_{ij}}(b,b',b_q)e^{-S_{R_{ij}}}\sum_{\sigma=LL,LR,SP}a_{R_{ij}}^{\sigma}H^{\sigma}_{R_{ij}}(x,x',y),
\end{eqnarray}
where the summation extends over all possible diagrams $R_{ij}$.
Here, $a_{R_{ij}}^{\sigma}$ denotes the product of the CKM matrix elements and the Wilson coefficients,
where the superscripts $\sigma=LL$,  $LR$, and $SP$ refer to the contributions from $(V-A)(V-A)$, $(V-A)(V+A)$, and $(S-P)(S+P)$ operators, respectively.
Notice that an overall factor 8 from  $a_{R_{ij}}^{\sigma}$ has been absorbed into the coefficient in Eq.~(\ref{eq:amp}) for convenience.
The explicit forms of the Sudakov factors $S_{R_{ij}}$ can be found in~\cite{Rui:2022jwu}.
Now that $H^{\sigma}_{R_{ij}}$ is the numerator of the hard amplitude depending on the spin structure of the final state, and
$\Omega_{R_{ij}}$ is the Fourier transformation of the denominator of the hard amplitude from the $k_T$ space to its conjugate $b$ space.
The impact parameters $b$, $b'$, and $b_q$ are conjugate to the parton transverse momenta $k_T$, $k'_T$, and $q_T$, respectively.
The integration measure of the momentum fractions is defined as
\begin{eqnarray}
[d^3x^{(')}]=dx^{(')}_1dx^{(')}_2dx^{(')}_3\delta(1-x^{(')}_1-x^{(')}_2-x^{(')}_3),
\end{eqnarray}
where the $\delta$ functions enforce momentum conservation.
The quantities associated with a specific diagram\textemdash such as $H_{R_{ij}}$, $a_{R_{ij}}$, $[\mathcal{D}b]_{R_{ij}}$, and $t_{R_{ij}}$\textemdash are collected in Appendix.
The decay branching ratios, up-down asymmetries, and direct $CP$ asymmetries of the relevant decays are given as~\cite{Cheng:1996cs,Zhang:2022iun}
\begin{eqnarray}\label{eq:alp}
\mathcal{B}&=&\frac{|P|\tau_{\Lambda_b}}{8\pi}[(1+r)^2|M_S|^2+(1-r)^2|M_P|^2], \nonumber\\
\alpha &=&-\frac{2(1-r^2) Re[M_S^*M_P]}{(1+r)^2|M_S|^2+(1-r)^2|M_P|^2},\nonumber\\
A_{CP}&=&\frac{\Gamma(\Lambda_b \rightarrow \Lambda \eta^{(')})-\Gamma(\bar{\Lambda}_b \rightarrow \bar{\Lambda} \eta^{(')})}
{\Gamma(\Lambda_b \rightarrow \Lambda \eta^{(')})+\Gamma(\bar{\Lambda}_b \rightarrow \bar{\Lambda} \eta^{(')})},
\end{eqnarray}
where $|P|$ is the magnitude of the three-momentum of the $\Lambda$ baryon in the rest frame of the $\Lambda_b$ baryon.

\section{NUMERICAL ANALYSIS}\label{sec:results}
In this section, we perform a numerical analysis for the
branching ratios, up-down asymmetries, and direct $CP$ asymmetries within various mixing schemes.
As we have discussed before, there are four available mixing schemes for the $\eta-\eta'$ system,
denoted as S1, S2, S3, and S4, respectively.
We first collect the scheme dependent input parameters as follows:\\
\begin{itemize}
  \item $\eta-\eta'$ mixing (S1)~\cite{Feldmann:1998vh},
\begin{eqnarray}\label{eq:s1}
f_q=1.07 f_\pi, \quad f_s=1.34f_\pi, \quad \phi=39.3^\circ\pm 1.0^\circ,
\quad  m_0^{q}=1.07 \text{GeV},\quad  m_0^{s}=1.92 \text{GeV}.
\end{eqnarray}
  \item $\eta-\eta'-G$ mixing (S2)~\cite{Cheng:2008ss},
\begin{eqnarray}\label{eq:s2}
f_q=f_\pi, \quad f_s=1.3f_\pi, \quad \phi=43.7^\circ, \quad \phi_G=12^\circ\pm 13^\circ,
\quad  m_0^{q}=2.08 \text{GeV},\quad  m_0^{s}=1.76 \text{GeV}.
\end{eqnarray}
  \item $\eta-\eta'-G-\eta_c$ mixing (S3)~\cite{Tsai:2011dp},
\begin{eqnarray}\label{eq:s3}
f_q&=&f_\pi, \quad f_s=1.3f_\pi, \quad \phi=43.7^\circ, \quad \phi_G=12^\circ\pm 13^\circ,
\quad  m_0^{q}=2.08 \text{GeV},\quad  m_0^{s}=1.76 \text{GeV}, \nonumber\\
 \phi_C&=&11^\circ, \quad  f_{\eta_c}=0.42 \text{GeV}, \quad m_{\eta_c}=2.98 \text{GeV}.
\end{eqnarray}
  \item  $\eta-\eta'-\eta_c$ mixing (S4)~\cite{Feldmann:1998vh},
  \begin{eqnarray}\label{eq:s4}
f_q&=&1.07 f_\pi, \quad f_s=1.34f_\pi, \quad \phi=39.3^\circ\pm 1.0^\circ,\quad  m_0^{q}=1.79 \text{GeV},\quad  m_0^{s}=1.91 \text{GeV},\nonumber\\
 m_{\eta_c}&=&2.95 \text{GeV},\quad \theta_c=-1^\circ, \quad \theta_y=-21.2^\circ, \quad  f_{\eta_c}=0.405 \text{GeV}.
\end{eqnarray}
\end{itemize}
To obtain the chiral enhancement scales $m_0^{q,s}$,
we  also need the light quark mass as input.
Because meson distribution amplitudes are defined at 1 GeV,
we take $m_q(1 \text{GeV})=5.6$ MeV~\cite{Ball:2006wn} and $m_s(1 \text{GeV})=0.13$ GeV~\cite{prd87094003}.
The relevant masses (GeV) and the CKM parameters in Wolfenstein parametrization are taken from the Particle Data Group~\cite{pdg2022}.
Their current values are
\begin{eqnarray}
M=5.6196, \quad  m_{\Lambda}=1.116, \quad  m_b=4.8, \quad  m_c=1.275,\quad  m_\eta=0.548 ,\quad  m_{\eta'}=0.958,
\end{eqnarray}
and
\begin{eqnarray}
 \lambda =0.22650, \quad  A=0.790,  \quad \bar{\rho}=0.141, \quad \bar{\eta}=0.357.
\end{eqnarray}
The lifetime of the $\Lambda_b$ baryon is taken as 1.464ps.
For the pion decay constant, we use $f_\pi=0.131$ GeV~\cite{Ball:2004ye}, and
for the Gegenbauer moments, we choose $a_2=0.44\pm 0.22$ and $a_4=0.25$~\cite{Kou:2001pm}.
We neglect the scale dependence of the chiral scales and the Gegenbauer moments in the default calculation.

\begin{table}
\footnotesize
\caption{Numerical results of various topological amplitudes of $\Lambda_b\rightarrow \Lambda \eta_{q,s,c}$  in units of $10^{-10}$.
The last column is their sum. Only central values are presented here.}
\label{tab:amp}
\begin{tabular}[t]{lcccccc}
\hline\hline
Scheme    & $T$           & $C$           & $E$           & $B$          & $B'$           & Total          \\ \hline
$\Lambda_b\rightarrow \Lambda \eta_q$ &&&&&&\\
$M_S$(S1)  & $-3.8+i1.2$  & $3.2-i15.3$  & $-1.4+i2.3$  & $-2.4+i3.1$  & $0.7-i0.3$ & $-3.7-i9.0$  \\
$M_P$(S1)  & $-4.0-i16.0$ & $-3.3+i12.5$  & $-0.3+i2.6$ & $-1.8-i6.8$  & $-0.6+i0.3$ & $-10.0-i7.4$  \\
$M_S$(S2,S3)   &$-3.4+i0.4$ &$3.1-i20.9$ &$-1.4+i3.3$  &$-3.4+i5.5$  &$0.6-i0.3$  &$-4.5-i12.0$  \\
$M_P$(S2,S3)   &$-4.8-i14.7$ &$-2.6+i17.1$ &$-1.7+i2.2$ &$-5.1-i7.6$ &$-0.6+i0.3$ &$-14.8-i2.7$ \\
$M_S$(S4)    &$-3.8+i1.2$ &$4.4-i20.9$ &$-1.1+i3.0$ &$-3.6+i4.3$ &$0.7-i0.3$ &$-3.4-i12.7$ \\
$M_P$(S4)    &$-4.1-i16.0$ &$-2.4+i16.7$ &$-0.2+i2.6$ &$-4.7-i7.1$ &$-0.6+i0.3$ &$-12.0-i3.5$ \\
$\Lambda_b\rightarrow \Lambda \eta_s$ &&&&&&\\
$M_S$(S1)  & $-6.1+i15.5$  & $\cdots$  & $4.0-i8.5$  & $\cdots$ & $0.6-i0.2$ & $-1.5+i6.8$  \\
$M_P$(S1)  & $8.9-i64.1$   & $\cdots$  & $-6.6+i4.1$  & $\cdots$ & $-0.6+i0.3$ & $1.7-i59.7$  \\
$M_S$(S2,S3)    &$-4.1+i9.5$ &$\cdots$  &$2.7-i5.6$ &$\cdots$  &$0.4-i0.2$ &$-1.0+i3.7$\\
$M_P$(S2,S3)    &$6.6-i43.2$ &$\cdots$  &$-4.5+i2.1$ &$\cdots$  &$-0.4+i0.2$ &$1.7-i40.9$\\
$M_S$(S4)    &$-5.8+i14.8$ &$\cdots$  &$4.0-i8.5$ &$\cdots$  &$0.6-i0.2$ &$-1.2+i6.1$\\
$M_P$(S4)    &$10.5-i63.8$ &$\cdots$  &$-6.0+i4.9$ &$\cdots$  &$-0.6+i0.3$ &$3.9-i58.6$\\
$\Lambda_b\rightarrow \Lambda \eta_c$ &&&&&&\\
$M_S$(S3)   &$42.8-i369.9$ &$\cdots$  &$\cdots$ &$\cdots$  &$0.05+i0.02$ &$42.9-i369.9$\\
$M_P$(S3)    &$80.9-i583.9$ &$\cdots$  &$\cdots$ &$\cdots$  &$-0.05+i0.01$ &$80.8-i583.9$\\
$M_S$(S4)    &$35.4-i356.0$ &$\cdots$  &$\cdots$ &$\cdots$  &$0.04+i0.01$ &$35.4-i356.0$\\
$M_P$(S4)    &$74.3-i576.9$ &$\cdots$  &$\cdots$ &$\cdots$  &$-0.05+i0.01$ &$74.3-i576.9$\\
\hline\hline
\end{tabular}
\end{table}

As stressed before, the PQCD calculations are performed in the quark-flavor basis.
The contribution of various topological diagrams to a specific process is determined by the quark-flavor composition of the particles involved in the decay.
For example, the nonstrange amplitude $\mathcal{M}_{q}$ receives contributions from all five topological diagrams,
while the strange one $\mathcal{M}_{s}$ has no contributions from the $C$- and $B$-type diagrams.
Note that the W exchange $bu\rightarrow su$ transition contributes to
$\mathcal{M}_{q}$ and  $\mathcal{M}_{s}$  through the $E_q$ and $E_s$ diagrams, respectively.
As for the $\Lambda_b \rightarrow \Lambda \eta_c$ decay, besides the dominant $T$-type diagrams contributing to $\mathcal{M}_{c}$,
it can proceed via $B'$ diagrams if one replaces the $s\bar s$ pair with $c \bar c$ in the last diagram of Fig.~\ref{fig:FeynmanT}.

According to Eq.~(\ref{eq:amp}), we first give the numerical results of various topology contributions to the
$S$-wave and $P$-wave amplitudes for the $\Lambda_b \rightarrow \Lambda \eta_{q,s,c}$ processes within four mixing schemes in Table~\ref{tab:amp}.
The differences among these solutions can be ascribed to the chiral enhancement scales related to the decay constants and the mixing angles.
The drastic sensitivity of the  chiral enhancement scales to the choice of mixing schemes will be reflected in the spread of our predictions for the decay amplitudes.
Referring to Eqs.~(\ref{eq:s1}) and (\ref{eq:s2}),
we can see that the chiral scale $m_0^q$ in S2 is almost twice as large as that in S1,
causing distinct nonstrange amplitudes for the two schemes.
 Analogously, the $m_0^s$ in S1 and S4 are almost equal,
resulting in a  tiny variation  in the strange amplitudes.
Likewise,  the same mixing parameters are used in S2 and S3 as shown in Eqs.~(\ref{eq:s2}) and (\ref{eq:s3});
the calculated amplitudes are exactly the same.
Numerical differences between S3 and S4 for the charm amplitude arise from a different choice of mass and decay constants of the $\eta_c$ meson as shown in Eqs.~(\ref{eq:s3}) and (\ref{eq:s4}).
One may wonder why the $T$-type nonstrange amplitudes yield the same results under S1 and S4 despite the fact that the parameter $m_0^q$ differs between the two schemes.
The interpretation  is not trivial.
We know that the chiral scales are proportional to the twist-3 meson LCDAs,
which do not contribute to the nonstrange amplitude via the external W emission diagram at the current theoretical accuracy
(see the expression of $H^{\sigma}_{T_{c7}}$ in Table~\ref{tab:amppc}, for example)
because only the external W emission from $(V-A)(V-A)$ and $(S-P)(S+P)$ operators survives for the $b\rightarrow s u \bar u$ and $b\rightarrow s d \bar d$ transitions.
Nevertheless, the strange amplitude receives additional twist-3 contributions arising from the W emission diagrams
through the $b\rightarrow s s \bar s$ transition  with the $(V-A)(V+A)$ operators inserted.
As a result,  the $m_0^s$ term appears in the strange amplitude, resulting in the different $T$-type strange amplitudes between S1 and S4.

We now proceed to discuss the relative sizes of various topological amplitudes.
From Table~\ref{tab:amp}, we observe that the $\Lambda_b \rightarrow \Lambda \eta_q$ process is dominated by $T$ and $C$.
As the interference between $T$ and $C$ is destructive, the contributions from the exchange diagrams, such as $B$ and $E$,  are in fact important and non-negligible.
Similar to the case of $\Lambda_b\rightarrow \Lambda \phi$~\cite{Rui:2022jwu},
the $\Lambda_b \rightarrow \Lambda \eta_s$ decay is governed by $T$ and $E$, which are of similar sizes.
The $\Lambda_b \rightarrow \Lambda \eta_c$ process is dominated by the $T$-type diagrams, and its amplitudes are larger than the (non)strange ones by 1 order of magnitude.
The contributions from $B'$-type exchange diagrams are predicted to be negligibly small in all the three processes.

It is worth noting that the penguin operators could be inserted into the diagrams in Fig.~\ref{fig:FeynmanT} via the Fierz transformation.
We do not distinguish between the tree and penguin contributions in  Table~\ref{tab:amp}.
The tree contributions of the strange and nonstrange amplitudes are expected to be small due to the CKM suppressed compared to the penguin ones.
If we turn off the penguin contributions,  their total amplitudes will decrease by 1 or 2 orders of magnitude.
This feature is different for the charm one,
which is triggered by the quark decay  $b\rightarrow sc\bar c$.
The large CKM matrix element $V_{cb}V_{cs}^*$ enhances the tree contributions, which dominate over the penguin ones.

Utilizing the values of Table~\ref{tab:amp} in conjunction with various mixing formalisms,
one can calculate the $S$- and $P$-wave amplitudes of $\Lambda_b\rightarrow \Lambda \eta^{(')}$,
whose numerical results are displayed in Table~\ref{tab:br}.
Branching ratios,  up-down asymmetries,  and the direct $CP$ asymmetries are shown in the last three columns.
There are four uncertainties.
The first quoted uncertainty is due to the shape parameters $\omega_0$ in the  $\Lambda_b$ LCDAs with $10\%$ variation.
The second uncertainty is caused by the variation of the Gegenbauer moment $a_2=0.44\pm 0.22$ in the leading-twist LCDAs of $\eta_{q,s}$.
Since the Gegenbauer moment $a_2$ is not yet well determined, the possible $50\%$ variation leads to large changes of our predictions.
The third one refers to the uncertainty of the mixing angles $\phi(\phi_G)$ as shown in Eqs.~(\ref{eq:s1})\textendash (\ref{eq:s4}).
Note that the chiral enhancement scale $m_0^q$ changes rapidly with the mixing angles,
so this uncertainty can be classified as the hadronic parameter uncertainty.
The last one is from the hard scale $t$ varying from $0.8t$ to $1.2t$.
The scale-dependent uncertainty
can be reduced only if the next-to-leading order contributions in the PQCD approach are included.
One can see that these large hadronic parameter uncertainties have a crucial influence on the PQCD calculations.
The up-down asymmetries from large theoretical errors, especially for the values in S4, due to complex interference effects, which will be detailed later.
It is interesting that $\mathcal{B}(\Lambda_b\rightarrow\Lambda \eta')$ in S3  is more sensitive to $\phi_G$.
The phenomena could be ascribed to the sizable $\eta_c$ mixing effect in S3.
According to Eq.~(\ref{eq:4mixing}), the charm amplitude for the $\eta'$ mode
suffers from the suppression from the mixing factor $\cos \theta \sin \phi_G\sin\phi_C=0.039^{+0.040}_{-0.042}$,
where the large uncertainty is due to the angle $\phi_G$, which varies in a conservative range $\phi_G=12^\circ\pm 13^\circ$.
Moreover, the large $\Lambda_b\rightarrow \Lambda \eta_c$ amplitude, as indicated in Table~\ref{tab:amp}, can compensate for this  suppression
and give a sizable impact on the $\Lambda_b\rightarrow \Lambda \eta'$  decay.
For the $\eta$ mode, the corresponding  mixing factor is $-\sin \theta \sin \phi_G\sin\phi_C=-0.0076^{+0.0078}_{-0.0082}$,
which is smaller by a factor of 5.
It follows that it does not have much impact on the decay rates involving $\eta$.

\begin{table}
\footnotesize
\caption{Magnitudes of the $S$- and $P$-wave amplitudes (in units of $10^{-10}$), branching ratios (in units of $10^{-6}$),
 up-down asymmetries, and  direct $CP$ violations of $\Lambda_b\rightarrow \Lambda \eta^{(')}$  decays with different mixing schemes.
The errors for these entries correspond to the shape parameter $\omega_0$, Gegenbauer moment $a_2$, mixing angles $\phi(\phi_G)$, and the hard scale $t$, respectively.}
\label{tab:br}
\begin{tabular}[t]{lcccccc}
\hline\hline
Scheme& $M_S$     & $M_P$  &$\mathcal{B}(10^{-6})$  & $\alpha$     & $A_{CP}(\%)$                         \\ \hline
$\Lambda_b\rightarrow \Lambda \eta$  &&&& \\
S1  & $-1.9 -i11.2 $ & $-8.9 +i32.1$
& $2.13^{+0.35+0.01+0.46+1.03}_{-0.41-0.01-0.26-0.60}$ &$0.734^{+0.087+0.102+0.000+0.004}_{-0.041-0.266-0.090-0.097}$ &$1.9^{+0.2+1.8+4.0+3.3}_{-2.5-4.8-0.0-0.0}$\\
S2  & $-2.5-i11.2$ & $-11.8 +i26.4$
& $1.73^{+0.39+0.00+0.15+0.41}_{-0.54-0.19-0.68-0.42}$ &$0.702^{+0.012+0.182+0.000+0.091}_{-0.018-0.088-0.074-0.041}$ &$8.8^{+0.0+0.0+0.0+3.2}_{-6.7-9.7-7.7-6.9}$\\
S3  & $-2.2-i14.0$ & $-11.2 +i22.0$
& $1.61^{+0.35+0.00+0.27+0.42}_{-0.52-0.18-0.70-0.41}$ &$0.800^{+0.039+0.119+0.028+0.064}_{-0.000-0.028-0.130-0.013}$ &$8.1^{+0.0+0.0+0.0+3.0}_{-8.6-7.0-2.9-5.7}$ \\
S4 & $-2.1-i11.5$ & $-12.2 +i37.7$
& $2.86^{+1.02+0.05+0.83+1.34}_{-0.64-0.24-0.53-0.81}$ &$0.650^{+0.000+0.047+0.032+0.025}_{-0.073-0.267-0.130-0.027}$ &$5.7^{+0.0+0.0+0.0+0.0}_{-1.5-7.8-3.5-4.0}$\\
$\Lambda_b\rightarrow \Lambda \eta'$  &&&& \\
S1 & $-3.5-i0.5 $ & $-5.0 -i50.9$
 & $3.94^{+0.92+1.20+0.00+1.21}_{-0.65-0.94-0.30-1.14}$ &$-0.046^{+0.127+0.344+0.154+0.098}_{-0.148-0.261-0.094-0.000}$ &$5.5^{+0.0+1.4+0.0+0.0}_{-3.4-2.8-4.5-6.6}$\\
S2  & $-3.7-i5.4 $ & $-8.7 -i30.9$
 & $1.68^{+0.25+0.50+0.00+0.49}_{-0.55-0.55-0.15-0.53}$ &$-0.531^{+0.247+0.226+0.354+0.030}_{-0.154-0.173-0.173-0.014}$ &$3.7^{+0.0+1.1+2.0+3.7}_{-9.8-5.4-2.4-0.1}$\\
S3 & $-2.1-i19.8$ & $-5.6 -i53.6$
& $5.67^{+1.34+1.19+6.48+1.37}_{-1.65-1.22-4.20-1.25}$ &$-0.847^{+0.068+0.042+0.180+0.002}_{-0.055-0.047-0.030-0.007}$ &$0.3^{+0.0+1.0+6.2+3.0}_{-5.5-0.6-2.7-0.4}$\\
S4  & $-3.6+i2.3 $ & $-5.7 -i38.1$
& $2.28^{+0.46+0.84+0.52+0.93}_{-0.25-0.18-0.00-0.36}$ &$0.134^{+0.193+0.490+0.278+0.154}_{-0.193-0.249-0.260-0.006}$ &$-0.4^{+5.6+8.7+9.2+11.2}_{-0.0-0.0-0.0-0.0}$\\
\hline\hline
\end{tabular}
\end{table}

We now discuss the sensitivity of the branching ratios of the $\Lambda_b\rightarrow \Lambda \eta^{(')}$ to different mixing schemes.
From Table~\ref{tab:br}, one can see clearly that  the results of the $\eta$ mode are less sensitive to the mixing schemes,
which suggests small gluonic and $\eta_c$ components in the $\eta$ meson.
The marginal differences among various schemes can be more or less traced to the different chiral enhancement scales as already emphasized.
However, various schemes lead to very different branching ratios for the $\eta'$ channel.
For example, the central values vary from $1.68\times 10^{-6}$ for S2 to $5.67\times 10^{-6}$ for S3.
The biggest branching ratio from S3 is ascribed to the fact that in the $\eta-\eta'-G-\eta_c$ mixing scheme,
the  tree dominated $\Lambda_b\rightarrow \Lambda \eta_c$ amplitude, induced from the $b\rightarrow sc\bar c$ transition,
can contribute to the $\Lambda_b\rightarrow \Lambda \eta^{(')}$ decays
through the mixing matrix as indicated in Eq.~(\ref{eq:4mixing}).
As stated above, the large  charm amplitude can compensate for the tiny mixing factor,
which implies that the component of $\eta_c$ in the $\eta'$ meson is important.
Our results indicate that the obtained branching ratios for $\Lambda_b\rightarrow \Lambda \eta $ and $\Lambda_b\rightarrow \Lambda \eta'$
are comparable in magnitude.
This observation is different from the pattern of $\mathcal{B}(B\rightarrow K \eta)$ and $\mathcal{B}(B\rightarrow K \eta')$,
where the former is about an order of magnitude smaller than the latter.

In Ref.~\cite{Hsu:2007qc}, the authors point out that few-percent OZI violating effects, neglected in the FKS scheme,
 could enhance the chiral scale $m_0^q$ sufficiently,
which accommodates the dramatically different data of the  $B\rightarrow K\eta^{(')}$   branching ratios in the PQCD approach.
It is therefore interesting to see whether this effect can modify the pattern of $\Lambda_b\rightarrow \Lambda \eta^{(')}$ branching ratios
 and improve the agreement with the current data.
It should be noted that the inclusion of the OZI violating effects implies two additional twist-2 meson distribution amplitudes associated with the OZI violating decay constants that
need to be considered,
but their contributions turn out to be insignificant~\cite{Hsu:2007qc}.
Hence, we can simply concentrate on the effect of the modified parameter set.
Using the central values of $f_q=1.10f_\pi, f_s=1.46f_\pi, \phi=36.84^\circ, m_0^q=4.32\text{GeV}, m_0^s=1.94\text{GeV}$ from~\cite{Hsu:2007qc} as inputs,
we derive
\begin{eqnarray}
\mathcal{B}(\Lambda_b\rightarrow \Lambda \eta)&=& 5.45\times 10^{-6}, \nonumber\\
\mathcal{B}(\Lambda_b\rightarrow \Lambda \eta')&=& 3.91\times 10^{-6}.
\end{eqnarray}
We will see later that by including the OZI violating effects in S1,
$\mathcal{B}(\Lambda_b\rightarrow \Lambda \eta)$ tend to be large,
while $\mathcal{B}(\Lambda_b\rightarrow \Lambda \eta')$ tend to be small, as favored by the experiments.

For the up-down asymmetries of the $\eta$ mode, all four solutions basically exhibit a similar pattern in size and sign.
However,  the observation is different for the $\eta'$ channel:
From Table~\ref{tab:br}  we see that S2 and S3 give  large and negative asymmetries,
while the central values in S1 and S4 are small in magnitude but with opposite signs.
These features can be understood by the following observation.
We learn that $\alpha$ describes the interference between the $S$-wave and $P$-wave amplitudes from Eq.~(\ref{eq:alp}).
According to the mixing matrices described in the previous section,
both the $S$-wave and $P$-wave amplitudes in $\Lambda_b\rightarrow \Lambda \eta^{(')}$ decays
can be written as the linear superposition of strange and nonstrange amplitudes through the mixing angle.
It should be noted that the nonstrange contents of the $\eta$ and $\eta'$ mesons have the same sign in S1,
while the strange ones are opposite in sign.
This means the interference between  the strange and nonstrange amplitudes is always destructive in $\Lambda_b\rightarrow \Lambda \eta$
but constructive in  $\Lambda_b\rightarrow \Lambda \eta^{'}$.
In addition, compared to the strange amplitude,
the nonstrange amplitude acquires additional sizable contributions from the internal W emission diagrams as shown in Fig.~\ref{fig:FeynmanT},
which leads to different patterns of the $S$-wave and $P$-wave contributions in the strange and nonstrange amplitudes.
Numerically, one can see from Table~\ref{tab:br} that the $P$-wave component dominates over the $S$-wave one in the $\eta'$ mode,
whereas they are comparable in the $\eta$ one.
The above combined effects cause the imaginary parts of the $S$-wave amplitudes of $\Lambda_b\rightarrow \Lambda \eta$ and $\Lambda_b\rightarrow \Lambda \eta'$ to
have the same sign, while the $P$-wave ones have the opposite sign as exhibited in Table~\ref{tab:br}.
Consequently, the up-down asymmetries of  $\Lambda_b\rightarrow \Lambda \eta$ and $\Lambda_b\rightarrow \Lambda \eta'$ are of opposite sign.
The feature in the S2 scheme can be explained in a similar way.
The interference pattern is more complicated with the inclusion of the charm content in the $\eta^{(')}$ mesons within S3 and S4.
We have learned from Eqs.~(\ref{eq:4mixing}) and (\ref{eq:3mixingp}) that the charm contents of the $\eta^{(')}$ meson for the two mixing schemes are opposite (same) in sign.
This difference has very little effect on the $\eta$ mode because of the strong suppression from the mixing factors;
however, it has an important influence on the $\eta'$ one as discussed before.
We predict a large and negative  $\alpha(\Lambda_b\rightarrow \Lambda \eta')$ in S3 but a small and positive one in S4.

Since the decays under consideration are dominated by the penguin contribution and the tree amplitudes are color and CKM factor suppressed,
their direct $CP$ asymmetries are not large, less than $10\%$.
Although the additional tree amplitudes stemming from the $b\rightarrow s c\bar c$ transition are included in S3 and S4,
the weak phase of $V_{cb}V_{cs}^*$ is zero at the order of $\lambda^2$,
which is the same as the penguin one, $V_{tb}V_{ts}^*$.
The enhancement arising from the charm content in fact leads to a smaller tree-over-penguin ratio,
and thus the direct $CP$ asymmetries of the $\eta'$ process in S3 and S4 are further reduced to less than one percent.

\begin{table}
\caption{Various predictions in the literature of the $\Lambda_b\rightarrow \Lambda \eta^{(')}$ decays.
Experimental measurements are taken from Ref~\cite{LHCb:2015kmm}.}
\label{tab:com}
\begin{tabular}[t]{lccccccc}
\hline\hline
&LFQM~\cite{Wei:2009np} &QCDF~\cite{prd99054020}     & GFA~\cite{epjc76399} & GFA~\cite{prd95093001} &QCDSR~\cite{plb598203} &PM~\cite{plb598203} & LHCb~\cite{LHCb:2015kmm}                         \\ \hline
$\mathcal{B}(\Lambda_b\rightarrow \Lambda \eta)$ & $5.46\times 10^{-8}$  & $4.39\times 10^{-7}$ &$(1.47\pm 0.35)\times 10^{-6}$ &$(1.59^{+0.45}_{-0.29})\times 10^{-6}$
&$11.47\times 10^{-6}$&$2.95\times 10^{-6}$ &$9.3^{+7.3}_{-5.3}\times 10^{-6}$\\
$\mathcal{B}(\Lambda_b\rightarrow \Lambda \eta')$ & $2.29\times 10^{-8}$ & $4.03\times 10^{-6}$ &$(1.83\pm 0.58)\times 10^{-6}$  &$(1.90^{+0.74}_{-0.36})\times 10^{-6}$
&$11.33\times 10^{-6}$&$3.24\times 10^{-6}$ &$<3.1\times 10^{-6}$\\
$\alpha(\Lambda_b\rightarrow \Lambda \eta)$ &-1 &$0.24^{+0.19}_{-0.12}$ &$\cdots$&$\cdots$&$\cdots$&$\cdots$&$\cdots$\\
$\alpha(\Lambda_b\rightarrow \Lambda \eta')$ &-1 &$0.99^{+0.00}_{-0.03}$ &$\cdots$&$\cdots$&$\cdots$&$\cdots$&$\cdots$\\
$A_{CP}(\Lambda_b\rightarrow \Lambda \eta)$ &$\cdots$ &$(-3.4^{+0.6}_{-0.4})\%$ &$\cdots$&$(0.4\pm0.2)\%$ &$\cdots$&$\cdots$&$\cdots$\\
$A_{CP}(\Lambda_b\rightarrow \Lambda \eta')$ &$\cdots$ &$(1.0^{+0.1}_{-0.2})\%$ &$\cdots$&$(1.6\pm0.1)\%$ &$\cdots$&$\cdots$&$\cdots$\\
\hline\hline
\end{tabular}
\end{table}


The comparisons with different theoretical models and the experimental data are presented in Table~\ref{tab:com}.
There is a wide spread in the branching ratios predicted by the various model calculations, ranging from $10^{-8}$ to $10^{-5}$.
The LFQM calculations~\cite{Wei:2009np} give the lowest predictions for the branching ratios because only the contributions from the tree operators were considered in their calculations,
which implies the penguin contributions play leading roles in the relevant processes.
In the absence of the penguin contributions, our central values of the branching ratios for the $\eta$ and $\eta'$ modes in S1 will be reduced to $8.92\times10^{-8}$ and $4.14\times10^{-8}$, respectively, which seem to be comparable to the results of LFQM~\cite{Wei:2009np}.
The two solutions of Ref.~\cite{plb598203} are evaluated by using two different $\Lambda_b\rightarrow \Lambda$ form factors.
The branching ratios for the form factors calculated in the pole model (PM) agree with our PQCD predictions within the S1 and S4 mixing schemes.
The two results from GFA~\cite{epjc76399,prd95093001} are basically consistent with each other and close to our values in S2.
It is also observed in Table~\ref{tab:com} that
most of the approaches give predictions of the same order of magnitude for the decay rates of the two modes,
except for the predictions of Ref.~\cite{prd99054020},
in which the branching ratio for the $\eta'$ mode is much larger than that of $\eta$ by 1 order of magnitude due to the additional enhanced factor for the $\eta'$ mode.

It should be noted that the previous theoretical calculations are based on the S1 scheme and do not take into account the contributions from the exchange amplitudes.
As seen in Table~\ref{tab:amp}, the emission amplitudes generally dominate over the exchange ones in PQCD calculations,
so we can drop all the exchange amplitudes and focus on the S1 mixing scenario to obtain a simple picture for the relevant decays.
The  resulting central values of the branching ratios are $\mathcal{B}(\Lambda_b\rightarrow \Lambda \eta)=3.89\times 10^{-6}$
 and $\mathcal{B}(\Lambda_b\rightarrow \Lambda \eta')=4.13\times 10^{-6}$,
 indicating that our predictions in Table~\ref{tab:amp} can be roughly reproduced under this simple picture.
This implies other processes  governed by the $T$ and $C$ topologies  have the branching ratios of $10^{-6}$ as well.
For the charmless decays $\Lambda_b\rightarrow\Lambda M$, there are six more processes with $M=\pi,\rho,K,K^*,\phi,\omega$, which are dominated by the emission diagrams.
The $\rho$ and $\pi$ modes belong to the isospin-violating decays; thus, they have no QCD penguin contributions.
The $K$ and $K^*$ modes are induced by the $b\rightarrow d$ transitions, which are CKM suppressed.
As such, the four processes should have substantially lower rates than $10^{-6}$.
The $\phi$ mode is studied in PQCD~\cite{Rui:2022jwu}, and its branching ratio is predicted to be of the order $10^{-6}$, which is comparable to the data.
There is currently no PQCD prediction or experimental data for the $\omega$ mode.
Because the $\omega$ meson has the same quark content as the $\eta_q$ meson, we estimate its branching ratio in PQCD to be $10^{-6}$.
This estimation is likely to be beneficial in experimental searches for these modes.

Unlike the branching ratios, up-down asymmetries, and $CP$ asymmetries in the relevant decays have received little attention in theoretical and experimental studies.
The estimates based on QCDF give $\alpha(\Lambda_b\rightarrow \Lambda \eta)=0.24^{+0.19}_{-0.12}$
and $\alpha(\Lambda_b\rightarrow \Lambda \eta')=0.99^{+0.00}_{-0.03}$~\cite{prd99054020},
while the LFQM calculations yield $\alpha(\Lambda_b\rightarrow \Lambda \eta)=\alpha(\Lambda_b\rightarrow \Lambda \eta^{'})=-1$~\cite{Wei:2009np}.
It can be observed that the available theoretical predictions on the up-down asymmetries vary and differ even in sign.
Hence, an accurate measurement of the up-down asymmetry  will enable us to discern different models.
For the direct $CP$ violation, nearly all of the current theoretical predictions are small, less than $10\%$ in magnitude.

From the experimental data shown in Table~\ref{tab:com},
it is clear that LHCb's measurement of $\mathcal{B}(\Lambda_b\rightarrow \Lambda \eta)$~\cite{LHCb:2015kmm} is generally larger than the theoretical expectations.
Although the prediction of Ref.~\cite{plb598203} is in accordance with the  central  value  of the data,
its value of $\mathcal{B}(\Lambda_b\rightarrow \Lambda \eta')$ exceeds the present experimental bound by a factor of 3.
The PQCD results of $\mathcal{B}(\Lambda_b\rightarrow \Lambda \eta')$  based on
the S1 and S3 mixing schemes are also large compared to the experimental upper limit.
In particular, the latter is larger by a factor of 2,
which indicates the mixing angles $\phi_G$ and/or $\phi_C$ may be overestimated.
Of course, the measurement was performed in 2015, and the experimental error was also quite large.
Moreover, there are still no available data on the up-down and direct $CP$ asymmetries at the moment.
We look forward to more experimental efforts to improve the accuracy of the relevant measurements.

As a by-product, we have predicted the decay branching ratio and up-down asymmetry of the  $\Lambda_b\rightarrow \Lambda \eta_c$ mode
by useing the values of the charm amplitudes in Table~\ref{tab:amp}.
Explicitly, we obtain
\begin{align}
\begin{split}
\mathcal{B}(\Lambda_b\rightarrow \Lambda \eta_c)= \left \{
\begin{array}{ll}
(6.83^{+2.10+1.37}_{-0.99-0.24})\times10^{-4} &  \text{for} \quad \text{S3} \\
(6.55^{+1.73+0.91}_{-1.32-0.59})\times10^{-4} &  \text{for} \quad \text{S4}
\end{array}
\right.
\end{split}
\end{align}
and 
\begin{align}
\begin{split}
\alpha(\Lambda_b\rightarrow \Lambda \eta_c)= \left \{
\begin{array}{ll}
-0.998^{+0.002+0.001}_{-0.000-0.000} &  \text{for} \quad \text{S3} \\
-0.996^{+0.000+0.001}_{-0.000-0.001} &  \text{for} \quad \text{S4},
\end{array}
\right.
\end{split}
\end{align}
where the first and second sets of error bars are due to
the shape parameter $\omega_0=0.40\pm0.04$ GeV in the $\Lambda_b$ baryon LCDAs
and the hard scale $t$ varying from  $0.8t$  to $1.2t$, respectively.
Our branching ratio is much larger than the values of $(2.47^{+0.33+0.42+0.67}_{-0.19-0.47-0.23})\times10^{-4}$ in QCDF~\cite{prd99054020}
and $(1.5\pm0.9)\times10^{-4}$ in GFA~\cite{Hsiao:2015cda}.
This is not surprising because the PQCD prediction on the branching ratio of the $J/\psi$ mode presented in~\cite{Rui:2022sdc}
is also generally larger than the corresponding values from~\cite{prd99054020,Hsiao:2015cda}
due to the significant nonfactorizable contributions.
All of these theoretical predictions are at the order of $10^{-4}$,
which can be accessible to the experiments at the LHCb.
The predicted up-down asymmetry is  nearly $100\%$ and negative, which is consistent with the value of $-0.99\pm 0.00$ obtained in~\cite{prd99054020}.
Since both the tree and penguin amplitudes have no weak phase to order $\lambda^2$,
the direct $CP$ violation for the $\Lambda_b\rightarrow \Lambda \eta_c$ process is predicted to be zero.

\section{conclusion}\label{sec:sum}
Decays of $b$ hadrons to two-body final states containing an  $\eta$ or $\eta'$  meson are of great phenomenological importance.
These processes could provide useful information about the $\eta-\eta'$ mixing and the structure of the  $\eta$ and $\eta'$ mesons,
which  is still a long-standing question in the literature.
In this work, we have carried out a systematic study of the penguin-dominant $\Lambda_b\rightarrow \Lambda \eta^{(')}$ decays in the PQCD approach.
The calculations are performed in the quark-flavor mixing basis,
in which we first give the PQCD predictions on the nonstrange, strange, and charm amplitudes including various topological contributions.
It is observed that the nonstrange amplitude  is dominated by the $T$- and $C$-type diagrams.
As the interference between $T$ and $C$ is destructive, the contributions from the exchange diagrams, such as $B$ and $E$,  are in fact important and non-negligible.
The strange amplitude is governed by $T$ and $E$, which are of similar sizes.
The charm one is dominated by the $T$-type diagrams, and its amplitudes are larger than the (non)strange ones by 1 order of magnitude.
Furthermore, the contributions from $B'$-type exchange diagrams are predicted to be negligibly small for all three amplitudes.

Utilizing the four available mixing schemes for the $\eta-\eta'$ system\textemdash namely  $\eta-\eta'$, $\eta-\eta'-G$, $\eta-\eta'-G-\eta_c$,  and $\eta-\eta'-\eta_c$ mixing,
denoted, respectively, as S1, S2, S3, and S4\textemdash
we evaluate the branching ratios, up-down asymmetries, and direct $CP$ asymmetries for $\Lambda_b\rightarrow\Lambda \eta^{(')}$ decays
and investigate  the scheme dependence of our theoretical predictions.
We find that  the results of the $\eta$ mode are less sensitive to the mixing schemes,
which implies small gluonic and $\eta_c$ components in the $\eta$ meson.
However, various schemes lead to quite different predictions on both the branching ratio and up-down asymmetry for the channel involving $\eta'$.
For instance, $\mathcal{B}(\Lambda_b\rightarrow \Lambda \eta')$ increases by a factor of 3 from S2 to S3,
while $ \alpha(\Lambda_b\rightarrow \Lambda \eta')$ varies from $-0.046$ for S1 to $-0.847$ for S3  and even flips signs in S4.
The large discrepancy among these solutions  suggests the $\eta'$ mode is very useful in discriminating various mixing schemes.

We consider theoretical uncertainties arising from the shape parameter $\omega_0$, Gegenbauer moment $a_2$, mixing angles $\phi(\phi_G)$, and the hard scale $t$.
We show that the nontrivial dependence of the PQCD calculations on the nonperturbative hadronic parameters,
which are poorly determined at present.
In particular, $\mathcal{B}(\Lambda_b\rightarrow \Lambda \eta')$  is extremely sensitive to the variation of $\phi_G$ and thus a good candidate for constraining the mixing parameters,
once it is measured with sufficient accuracy.
The scale-dependent uncertainty also gives large uncertainties to the branching ratios,
which can be reduced only if the next-to-leading order contributions in the PQCD approach are known.

We also compare our results with predictions of the other theoretical approaches as well as existing experimental data.
Various model estimations  on the  branching ratios span fairly wide ranges from $10^{-8}$ to $10^{-5}$.
Our branching ratios for the S2 scheme are consistent with the GFA calculations,
while the S4 ones are close to the results of the PM.
The predicted central values of $\mathcal{B}(\Lambda_b\rightarrow \Lambda \eta)$ in various schemes are generally lower than the measurement from LHCb.
The inclusion of OZI violating effects can enhance $\mathcal{B}(\Lambda_b\rightarrow \Lambda \eta)$ by a factor of 2.5 and improve the agreement with the current data.
The PQCD results of $\mathcal{B}(\Lambda_b\rightarrow \Lambda \eta')$  based on S1 and S3 mixing schemes exceed the present experimental bound.
Note that the measured branching ratios also have large uncertainties.
In general, the values in S4 appear to be more preferred by the current data among these solutions.
For the up-down asymmetries, there are considerable deviations among PQCD, QCDF, and LFQM estimates, which should be clarified in the future.
On the other hand, since the tree contributions suffer from the color suppression and CKM  suppression,
the obtained direct $CP$ asymmetries are less than $10\%$, in comparison with the numbers from QCDF and GFA.
At the moment, there is neither experimental information on  the up-down asymmetries nor on direct $CP$ asymmetries.
It will be interesting to see the updated measurements on the two decay modes.

Finally, we explore the decay of $\Lambda_b\rightarrow \Lambda \eta_c$.
The estimated branching ratio is  at the $10^{-4}$ level with an up-down asymmetry close to $-1$,
which may shed light on future measurements.

\begin{acknowledgments}
We wish to acknowledge discussions with Hsiang-nan Li.
This work is supported by the National Natural Science Foundation
of China under Grants No. 12075086 and  No. 11605060 and the  Hebei Natural Science Foundation
under Grants No.A2021209002 and  No.A2019209449.
\end{acknowledgments}

\begin{appendix}
\section{FACTORIZATION FORMULAS}\label{sec:for}
Following the conventions in Ref.~\cite{Zhang:2022iun},
we provide some details about the factorization formulas in Eq.~(\ref{eq:amp}) for the nonstrange amplitude, which were not given before.
As the strange and charm processes have the same decay topologies as $\Lambda_b\rightarrow \Lambda \phi$ and $\Lambda_b\rightarrow \Lambda J/\psi$ modes, respectively,
one can find the relevant formulas in our previous work~\cite{Rui:2022sdc,Rui:2022jwu}.
The combinations of the Wilson coefficients $a_{R_{ij}}^{\sigma}$ and the $b$ dependent quantities $[\mathcal{D}b]_{R_{ij}}$ and $\Omega_{R_{ij}}$
are given in Tables~\ref{tab:wilson} and~\ref{tab:bb}, respectively,
and the auxiliary functions $h_{1,2,3}$ and the Bessel function $K_0$ can be found in~\cite{Zhang:2022iun}.

The hard scale $t$ for each diagram is chosen as the maximal virtuality of internal particles including the factorization scales in a hard amplitude:
\begin{eqnarray}
t_{R_{ij}}=\max(\sqrt{|t_A|},\sqrt{|t_B|},\sqrt{|t_C|},\sqrt{|t_D|},w,w',w_q),
\end{eqnarray}
where the expressions of $t_{A,B,C,D}$ are listed in Table~\ref{tab:ttt}.
The factorization scales $w$, $w'$, and $w_q$ are defined by
 \begin{eqnarray}
w^{(')}=\min(\frac{1}{b^{(')}_1},\frac{1}{b^{(')}_2},\frac{1}{b^{(')}_3}),\quad w_q=\frac{1}{b_q},
\end{eqnarray}
 with the variables
 \begin{eqnarray}
b^{(')}_1=|b^{(')}_2-b^{(')}_3|,
\end{eqnarray}
and the other $b^{(')}_l$ defined by permutation.
Here, we only present the results of $[\mathcal{D}b]_{R_{ij}}$, $\Omega_{R_{ij}}$, and $t_{A,B,C,D}$ for the $C$, $B$, and $E_q$ diagrams.
The remaining results are the same as those for $\Lambda_b\rightarrow \Lambda \phi$ and can be found in~\cite{Rui:2022jwu}.

In Table~\ref{tab:amppc}, we give the expressions of $H^{\sigma}_{R_{ij}}$ for a representative set of diagrams for each type, as shown in Fig.~\ref{fig:FeynmanT},
while those for others can be derived in an analogous way.

\begin{table} [H]
	\footnotesize
	\caption{The expressions of $a^{LL}$ and $a^{SP}$ in the  $\Lambda_b\rightarrow \Lambda\eta_q$ decay.
For convenience, we have extracted an overall coefficient 8, which is absorbed into the prefactor in Eq.~(\ref{eq:amp}).}
	\newcommand{\tabincell}[2]{\begin{tabular}{@{}#1@{}}#2\end{tabular}}
	\label{tab:wilson}
	\begin{tabular}[t]{lccc}
		\hline\hline
		$R_{ij}$        &$a^{LL}$          &$a^{SP}$             \\ \hline
		$T_{a1,a2,a3,a5,b1,b2,b4}$       &$\frac{1}{3}V_{ub}V_{us}^*[C_1+\frac{1}{3}C_2]-\frac{1}{3}V_{tb}V_{ts}^*[2C_3+\frac{2}{3}C_4+\frac{1}{2}C_9+\frac{1}{6}C_{10}]$
		&$-\frac{1}{3}V_{tb}V_{ts}^*[2C_5+\frac{2}{3}C_6+\frac{1}{2}C_7+\frac{1}{6}C_8]$\\		
		$T_{a6,a7,b6,b7,c1,c2,d1,d2}$
		&$\frac{1}{3}V_{ub}V_{us}^*C_2-\frac{1}{3}V_{tb}V_{ts}^*[2C_4+\frac{1}{2}C_{10}]$
		&$-\frac{1}{3}V_{tb}V_{ts}^*[2C_6+\frac{1}{2}C_8]$                 \\
		$T_{c5,c7,d6}$
		&$V_{ub}V_{us}^*[-\frac{1}{4}C_1+\frac{1}{3}C_2]+V_{tb}V_{ts}^*[\frac{1}{2}C_3-\frac{2}{3}C_4+\frac{1}{8}C_9-\frac{1}{6}C_{10}]$
		&$V_{tb}V_{ts}^*[\frac{1}{2}C_5-\frac{2}{3}C_6+\frac{1}{8}C_7-\frac{1}{6}C_8]$     \\
		
		\tabincell{l}{$C_{a1-a4,b1,b3,c1,c3,d1-d3,e1,f1}$\\$Eq_{a1-a7,e1-e4,f2}$\\$B_{a1-a4,b1,b3,c1,c4,d1,d2,d4,d5}$}
		&$\frac{1}{3}V_{ub}V_{us}^*[C_1-C_2]-\frac{1}{6}V_{tb}V_{ts}^*[4C_3-4C_4+C_9-C_{10}]$
		&$-\frac{1}{6}V_{tb}V_{ts}^*[4C_5-4C_6+C_7-C_8]$   \\
		
		\tabincell{l}{$C_{a5,b4,c4,e2},Eq_{b1,b3,b4,b7}$}
		&$-\frac{1}{3}V_{ub}V_{us}^*[2C_1+C_2]+\frac{1}{3}V_{tb}V_{ts}^*[4C_3+2C_4+C_9+\frac{1}{2}C_{10}]$
		&$\frac{1}{3}V_{tb}V_{ts}^*[4C_5+2C_6+C_7+\frac{1}{2}C_8]$  \\
		
		\tabincell{l}{$C_{a6,b6,c5,e3},Eq_{c1,c2,c3,c7}$}
		&$\frac{1}{3}V_{ub}V_{us}^*[C_1-2C_2]-\frac{1}{3}V_{tb}V_{ts}^*[2C_3+4C_4+\frac{1}{2}C_9+C_{10}]$
		&$-\frac{1}{3}V_{tb}V_{ts}^*[2C_5+4C_6+\frac{1}{2}C_7+C_8]$ \\
		
		\tabincell{l}{$C_{a7,b7,c7,d7,e4},Eq_{d1-d4,d7}$\\$B_{a7,b7,c7,d7,e1-e4}$}
		&$\frac{1}{3}V_{ub}V_{us}^*[-2C_1+2C_2]+\frac{1}{3}V_{tb}V_{ts}^*[4C_3-4C_4+C_9-C_{10}]$
		&$\frac{1}{3}V_{tb}V_{ts}^*[4C_5-4C_6+C_7-C_8]$   \\
		
		\tabincell{l}{$C_{b2},Eq_{f1},B_{b2,c5}$}
		&$\frac{1}{3}V_{ub}V_{us}^*[-\frac{5}{4}C_1-C_2]+V_{tb}V_{ts}^*[\frac{5}{6}C_3+\frac{2}{3}C_4+\frac{5}{24}C_9+\frac{1}{6}C_{10}]$
		&$V_{tb}V_{ts}^*[\frac{5}{6}C_5+\frac{2}{3}C_6+\frac{5}{24}C_7+\frac{1}{6}C_8]$   \\
		
		\tabincell{l}{$C_{b5,d4,f2},Eq_{b2,b5,b6},B_{d1}$}
		&$\frac{1}{3}V_{ub}V_{us}^*[\frac{1}{4}C_1-C_2]-V_{tb}V_{ts}^*[\frac{1}{6}C_3-\frac{2}{3}C_4+\frac{1}{24}C_9-\frac{1}{6}C_{10}]$
		&$-V_{tb}V_{ts}^*[\frac{1}{6}C_5-\frac{2}{3}C_6+\frac{1}{24}C_7-\frac{1}{6}C_8]$   \\
		
		\tabincell{l}{$C_{c2},Eq_{f3},B_{b5,c3}$}
		&$\frac{1}{3}V_{ub}V_{us}^*[C_1+\frac{5}{4}C_2]-V_{tb}V_{ts}^*[\frac{2}{3}C_3+\frac{5}{6}C_4+\frac{1}{6}C_9+\frac{5}{24}C_{10}]$
		&$-V_{tb}V_{ts}^*[\frac{2}{3}C_5+\frac{5}{6}C_6+\frac{1}{6}C_7+\frac{5}{24}C_8]$   \\
		
		\tabincell{l}{$C_{c6,f3},Eq_{c4,c5,c6}$}
		&$\frac{1}{3}V_{ub}V_{us}^*[C_1-\frac{1}{4}C_2]-V_{tb}V_{ts}^*[\frac{2}{3}C_3-\frac{1}{6}C_4+\frac{1}{6}C_9-\frac{1}{24}C_{10}]$
		&$-V_{tb}V_{ts}^*[\frac{2}{3}C_5-\frac{1}{6}C_6+\frac{1}{6}C_7-\frac{1}{24}C_8]$   \\
		
		\tabincell{l}{$C_{d5,d6,f4},Eq_{d5,d6},B_{a6,c6,d6,f1-f4}$}
		&$\frac{1}{12}V_{ub}V_{us}^*[C_1-C_2]-V_{tb}V_{ts}^*[\frac{1}{6}C_3-\frac{1}{6}C_4+\frac{1}{24}C_9-\frac{1}{24}C_{10}]$
		&$-V_{tb}V_{ts}^*[\frac{1}{6}C_5-\frac{1}{6}C_6+\frac{1}{24}C_7-\frac{1}{24}C_8]$   \\
		
		\tabincell{l}{$Eq_{f4},B_{a5,b4,c2,d3}$}
		&$-\frac{5}{12}V_{ub}V_{us}^*[C_1-C_2]+V_{tb}V_{ts}^*[\frac{5}{6}C_3-\frac{5}{6}C_4+\frac{5}{24}C_9-\frac{5}{24}C_{10}]$
		&$V_{tb}V_{ts}^*[\frac{5}{6}C_5-\frac{5}{6}C_6+\frac{5}{24}C_7-\frac{5}{24}C_8]$   \\
		
		\tabincell{l}{$C_{g1},Eq_{g2}$}
		&$0$
		&$0$   \\
		
		$C_{g2}$
		&$-\frac{3}{4}V_{ub}V_{us}^*C_1+\frac{3}{4}V_{tb}V_{ts}^*[2C_3+\frac{1}{2}C_9]$
		&$\frac{3}{4}V_{tb}V_{ts}^*[2C_5+\frac{1}{2}C_7]$   \\
		$C_{g3}$
		&$-\frac{3}{4}V_{ub}V_{us}^*[C_1-C_2]+\frac{3}{8}V_{tb}V_{ts}^*[4C_3-4C_4+C_9-C_{10}]$
		&$\frac{3}{8}V_{tb}V_{ts}^*[4C_5-4C_6+C_7-C_8]$  \\
		$C_{g4}$
		&$\frac{3}{4}V_{ub}V_{us}^*C_2-\frac{3}{8}V_{tb}V_{ts}^*[4C_4+C_{10}]$
		&$-\frac{3}{8}V_{tb}V_{ts}^*[4C_6+C_8]$   \\
		
		$Eq_{g1}$
		&$\frac{3}{4}V_{ub}V_{us}^*C_1-\frac{3}{8}V_{tb}V_{ts}^*[4C_3+C_9]$
		&$-\frac{3}{8}V_{tb}V_{ts}^*[4C_5+C_7]$   \\		
	
		$Eq_{g3}$
		&$-\frac{3}{4}V_{ub}V_{us}^*C_2+\frac{3}{8}V_{tb}V_{ts}^*[4C_4+C_{10}]$
		&$\frac{3}{8}V_{tb}V_{ts}^*[4C_6+C_8]$   \\
		
		\tabincell{l}{$Eq_{g4},B_{g1-g4}$}
		&$\frac{3}{4}V_{ub}V_{us}^*[C_1-C_2]-\frac{3}{8}V_{tb}V_{ts}^*[4C_3-4C_4+C_9-C_{10}]$
		&$-\frac{3}{8}V_{tb}V_{ts}^*[4C_5-4C_6+C_7-C_8]$   \\
						
		${B'}_{a1-a4,b1-b4}$
		&$\frac{1}{4}V_{ub}V_{us}^*[-C_1+C_2]+\frac{1}{8}V_{tb}V_{ts}^*[4C_3-4C_4+C_9-C_{10}]$
		&$\frac{1}{8}V_{tb}V_{ts}^*[4C_5-4C_6+C_7-C_8]$   \\

		\hline\hline
	\end{tabular}
\end{table}

\begin{table}[H]
	\scriptsize
	\centering
	\caption{The expressions of $[\mathcal{D}b]$ and $\Omega_{R_{ij}}$ for the $C$, $B$, and $E_q$ diagrams.}
	\label{tab:bb}
	\begin{tabular}[t]{lccc}
		\hline\hline
		$R_{ij}$ & $[\mathcal{D}b]$ &$\Omega_{R_{ij}}$\\ \hline
		$C_{a1}$ & $  d^2 \textbf{b}_2d^2\textbf{b}_3d^2\textbf{b}'_3d^2\textbf{b}_q$
		&$\frac{1}{(2\pi)^4}K_0(\sqrt{t_A}|\textbf{b}_2-\textbf{b}_3|)K_0(\sqrt{t_B}|\textbf{b}_2-\textbf{b}_3-\textbf{b}'_3+\textbf{b}_q|)K_0(\sqrt{t_C}|\textbf{b}_2-\textbf{b}'_3+\textbf{b}_q|)K_0(\sqrt{t_D}|\textbf{b}_2+\textbf{b}_q|)$ \\
		
		$C_{a2}$ &$  d^2 \textbf{b}_2d^2\textbf{b}_3d^2\textbf{b}'_3d^2\textbf{b}_q$ &$\frac{1}{(2\pi)^4}
		K_0(\sqrt{t_A}|\textbf{b}_q-\textbf{b}'_3|)
		K_0(\sqrt{t_B}|\textbf{b}_3-\textbf{b}_2+\textbf{b}'_3-\textbf{b}_q|)
		K_0(\sqrt{t_C}|\textbf{b}_3|)
		K_0(\sqrt{t_D}|\textbf{b}_2+\textbf{b}_q|)$ \\
		
		$C_{a3}$ &$  d^2\textbf{b}_2 d^2\textbf{b}_3 d^2\textbf{b}_q$
		&$K_0(\sqrt{t_A}|\textbf{b}_2|)h_2(\textbf{b}_3-\textbf{b}_2-\textbf{b}_q,\textbf{b}_2+\textbf{b}_q,t_B,t_C,t_D)$ \\
		
		$C_{a4}$ &$  d^2\textbf{b}_2 d^2\textbf{b}_3 d^2\textbf{b}_q$
		&$K_0(\sqrt{t_B}|\textbf{b}_3|)h_2(\textbf{b}_q,\textbf{b}_2+\textbf{b}_q,t_A,t_C,t_D)$ \\
		
		$C_{a5}$ &$  d^2 \textbf{b}_2d^2\textbf{b}_3d^2\textbf{b}'_2d^2\textbf{b}_q$ &$\frac{1}{(2\pi)^4}
		K_0(\sqrt{t_A}|\textbf{b}_q-\textbf{b}'_2|)
		K_0(\sqrt{t_B}|\textbf{b}_3|)
		K_0(\sqrt{t_C}|\textbf{b}'_2|)
		K_0(\sqrt{t_D}|\textbf{b}_q+\textbf{b}_2|)$ \\
		
		$C_{a6}$ &$  d^2 \textbf{b}_2d^2\textbf{b}_3d^2\textbf{b}'_2d^2\textbf{b}_q$ &$\frac{1}{(2\pi)^4}
		K_0(\sqrt{t_A}|\textbf{b}_q+\textbf{b}'_2|)
		K_0(\sqrt{t_B}|\textbf{b}_3|)
		K_0(\sqrt{t_C}|\textbf{b}'_2|)
		K_0(\sqrt{t_D}|\textbf{b}_q+\textbf{b}_2|)$ \\
		
		$C_{a7}$ &$  d^2 \textbf{b}_2d^2\textbf{b}_3d^2\textbf{b}'_3d^2\textbf{b}_q$ &$\frac{1}{(2\pi)^4}
		K_0(\sqrt{t_A}|\textbf{b}_q|)
		K_0(\sqrt{t_B}|\textbf{b}_3|)
		K_0(\sqrt{t_C}|\textbf{b}_3-\textbf{b}_2+\textbf{b}'_3-\textbf{b}_q|)
		K_0(\sqrt{t_D}|\textbf{b}_3+\textbf{b}'_3|)$ \\
		
		$C_{b1}$ &$  d^2 \textbf{b}_2d^2\textbf{b}_3d^2\textbf{b}'_2d^2\textbf{b}'_3$ &$\frac{1}{(2\pi)^4}
		K_0(\sqrt{t_A}|\textbf{b}_2-\textbf{b}_3|)
		K_0(\sqrt{t_B}|\textbf{b}'_2-\textbf{b}'_3|)
		K_0(\sqrt{t_C}|\textbf{b}_3+\textbf{b}'_3|)
		K_0(\sqrt{t_D}|\textbf{b}'_2|)$ \\
		
		$C_{b2}$ &$  d^2 \textbf{b}_2d^2\textbf{b}_3d^2\textbf{b}'_2d^2\textbf{b}'_3$ &$\frac{1}{(2\pi)^4}
		K_0(\sqrt{t_A}|\textbf{b}_2+\textbf{b}'_3|)
		K_0(\sqrt{t_B}|\textbf{b}_3+\textbf{b}'_2|)
		K_0(\sqrt{t_C}|\textbf{b}_3+\textbf{b}'_3|)
		K_0(\sqrt{t_D}|\textbf{b}'_2|)$ \\
		
		$C_{b3}$ &$  d^2 \textbf{b}_2d^2\textbf{b}_3d^2\textbf{b}'_2d^2\textbf{b}'_3$ &$\frac{1}{(2\pi)^4}
		K_0(\sqrt{t_A}|\textbf{b}_2|)
		K_0(\sqrt{t_B}|\textbf{b}'_2-\textbf{b}'_3|)
		K_0(\sqrt{t_C}|\textbf{b}_3+\textbf{b}'_3|)
		K_0(\sqrt{t_D}|\textbf{b}'_2|)$ \\
		
		$C_{b4}$ &$  d^2\textbf{b}_2 d^2\textbf{b}_3 d^2\textbf{b}'_2$
		&$K_0(\sqrt{t_B}|\textbf{b}_3+\textbf{b}'_2|)h_2(\textbf{b}_2,\textbf{b}'_2,t_A,t_C,t_D)$ \\
		
		$C_{b5}$ &$  d^2\textbf{b}_2 d^2\textbf{b}_3 d^2\textbf{b}'_2$
		&$K_0(\sqrt{t_A}|\textbf{b}_2+\textbf{b}'_2|)h_2(\textbf{b}_3+\textbf{b}'_2,\textbf{b}'_2,t_B,t_C,t_D)$ \\
		
		$C_{b6}$ &$  d^2 \textbf{b}_2d^2\textbf{b}_3d^2\textbf{b}'_2d^2\textbf{b}'_3$ &$\frac{1}{(2\pi)^4}
		K_0(\sqrt{t_A}|\textbf{b}_3-\textbf{b}_2+\textbf{b}'_3|)
		K_0(\sqrt{t_B}|\textbf{b}'_2-\textbf{b}'_3|)
		K_0(\sqrt{t_C}|\textbf{b}_3+\textbf{b}'_3|)
		K_0(\sqrt{t_D}|\textbf{b}'_2-\textbf{b}'_3-\textbf{b}_3|)$ \\
		
		$C_{b7}$ &$  d^2 \textbf{b}_2d^2\textbf{b}_3d^2\textbf{b}'_2d^2\textbf{b}_q$ &$\frac{1}{(2\pi)^4}
		K_0(\sqrt{t_A}|\textbf{b}_q|)
		K_0(\sqrt{t_B}|\textbf{b}_3+\textbf{b}'_2|)
		K_0(\sqrt{t_C}|\textbf{b}_2+\textbf{b}_q|)
		K_0(\sqrt{t_D}|\textbf{b}'_2|)$ \\
		
		$C_{c1}$ &$  d^2 \textbf{b}_2d^2\textbf{b}_3d^2\textbf{b}'_2d^2\textbf{b}'_3$ &$\frac{1}{(2\pi)^4}
		K_0(\sqrt{t_A}|\textbf{b}_2-\textbf{b}_3|)
		K_0(\sqrt{t_B}|\textbf{b}'_3|)
		K_0(\sqrt{t_C}|\textbf{b}_3-\textbf{b}'_2+\textbf{b}'_3|)
		K_0(\sqrt{t_D}|\textbf{b}'_2|)$ \\
		
		$C_{c2}$ &$  d^2 \textbf{b}_2d^2\textbf{b}_3d^2\textbf{b}'_2d^2\textbf{b}'_3$ &$\frac{1}{(2\pi)^4}
		K_0(\sqrt{t_A}|\textbf{b}'_2-\textbf{b}'_3-\textbf{b}_2|)
		K_0(\sqrt{t_B}|\textbf{b}_3-\textbf{b}'_2|)
		K_0(\sqrt{t_C}|\textbf{b}_3-\textbf{b}'_2-\textbf{b}'_3|)
		K_0(\sqrt{t_D}|\textbf{b}'_2|)$ \\
		
		$C_{c3}$ &$  d^2 \textbf{b}_2d^2\textbf{b}_3d^2\textbf{b}'_2d^2\textbf{b}'_3$ &$\frac{1}{(2\pi)^4}
		K_0(\sqrt{t_A}|\textbf{b}_2|)
		K_0(\sqrt{t_B}|\textbf{b}'_3|)
		K_0(\sqrt{t_C}|\textbf{b}'_2-\textbf{b}'_3-\textbf{b}_3|)
		K_0(\sqrt{t_D}|\textbf{b}'_2|)$ \\
		
		$C_{c4}$ &$  d^2 \textbf{b}_2d^2\textbf{b}_3d^2\textbf{b}'_2d^2\textbf{b}'_3$ &$\frac{1}{(2\pi)^4}
		K_0(\sqrt{t_A}|\textbf{b}_2-\textbf{b}_3+\textbf{b}'_2-\textbf{b}'_3|)
		K_0(\sqrt{t_B}|\textbf{b}'_3|)
		K_0(\sqrt{t_C}|\textbf{b}'_2-\textbf{b}'_3-\textbf{b}_3|)
		K_0(\sqrt{t_D}|\textbf{b}_3+\textbf{b}'_3|)$ \\
		
		$C_{c5}$ &$  d^2\textbf{b}_2 d^2\textbf{b}_3 d^2\textbf{b}'_2$
		&$K_0(\sqrt{t_B}|\textbf{b}_3-\textbf{b}'_2|)h_2(\textbf{b}_2,\textbf{b}'_2,t_A,t_C,t_D)$ \\
		
		$C_{c6}$ &$  d^2\textbf{b}_2 d^2\textbf{b}_3 d^2\textbf{b}'_2$
		&$K_0(\sqrt{t_A}|\textbf{b}_2-\textbf{b}'_2|)h_2(\textbf{b}_3-\textbf{b}'_2,\textbf{b}'_2,t_B,t_C,t_D)$ \\
		
		$C_{c7}$ &$  d^2 \textbf{b}_2d^2\textbf{b}_3d^2\textbf{b}'_2d^2\textbf{b}_q$ &$\frac{1}{(2\pi)^4}
		K_0(\sqrt{t_A}|\textbf{b}_q|)
		K_0(\sqrt{t_B}|\textbf{b}_3-\textbf{b}'_2|)
		K_0(\sqrt{t_C}|\textbf{b}_2+\textbf{b}_q|)
		K_0(\sqrt{t_D}|\textbf{b}'_2|)$ \\
		
		$C_{d1}$ &$  d^2 \textbf{b}_2d^2\textbf{b}_3d^2\textbf{b}'_3d^2\textbf{b}_q$ &$\frac{1}{(2\pi)^4}
		K_0(\sqrt{t_A}|\textbf{b}_2-\textbf{b}_3|)
		K_0(\sqrt{t_B}|\textbf{b}_2+\textbf{b}'_3+\textbf{b}_q|)
		K_0(\sqrt{t_C}|\textbf{b}_3+\textbf{b}'_3|)
		K_0(\sqrt{t_D}|\textbf{b}_2+\textbf{b}_q|)$ \\
		
		$C_{d2}$ &$  d^2 \textbf{b}_2d^2\textbf{b}_3d^2\textbf{b}'_3d^2\textbf{b}_q$ &$\frac{1}{(2\pi)^4}
		K_0(\sqrt{t_A}|\textbf{b}_2+\textbf{b}'_3|)
		K_0(\sqrt{t_B}|\textbf{b}_2-\textbf{b}_3+\textbf{b}_q|)
		K_0(\sqrt{t_C}|\textbf{b}_3+\textbf{b}'_3|)
		K_0(\sqrt{t_D}|\textbf{b}_2+\textbf{b}_q|)$ \\
		
		$C_{d3}$ &$  d^2 \textbf{b}_2d^2\textbf{b}_3d^2\textbf{b}'_3d^2\textbf{b}_q$ &$\frac{1}{(2\pi)^4}
		K_0(\sqrt{t_A}|\textbf{b}_2|)
		K_0(\sqrt{t_B}|\textbf{b}_2-\textbf{b}_3+\textbf{b}_q|)
		K_0(\sqrt{t_C}|\textbf{b}_3+\textbf{b}'_3|)
		K_0(\sqrt{t_D}|\textbf{b}_3-\textbf{b}_2+\textbf{b}'_3-\textbf{b}_q|)$ \\
		
		$C_{d4}$ &$  d^2 \textbf{b}_2d^2\textbf{b}_3d^2\textbf{b}'_2d^2\textbf{b}_q$ &$\frac{1}{(2\pi)^4}
		K_0(\sqrt{t_A}|\textbf{b}_2+\textbf{b}'_2|)
		K_0(\sqrt{t_B}|\textbf{b}_2-\textbf{b}_3+\textbf{b}_q|)
		K_0(\sqrt{t_C}|\textbf{b}'_2|)
		K_0(\sqrt{t_D}|\textbf{b}_2+\textbf{b}_q|)$ \\
		
		$C_{d5}$ &$  d^2 \textbf{b}_2d^2\textbf{b}_3d^2\textbf{b}'_2d^2\textbf{b}_q$ &$\frac{1}{(2\pi)^4}
		K_0(\sqrt{t_A}|\textbf{b}_2-\textbf{b}'_2|)
		K_0(\sqrt{t_B}|\textbf{b}_2-\textbf{b}_3+\textbf{b}_q|)
		K_0(\sqrt{t_C}|\textbf{b}'_2|)
		K_0(\sqrt{t_D}|\textbf{b}_2+\textbf{b}_q|)$ \\
		
		$C_{d6}$ &$  d^2\textbf{b}_2 d^2\textbf{b}_3 d^2\textbf{b}_q$
		&$K_0(\sqrt{t_B}|\textbf{b}_2-\textbf{b}_3+\textbf{b}_q|)h_2(\textbf{b}_2,\textbf{b}_2+\textbf{b}_q,t_A,t_C,t_D)$ \\
		
		$C_{d7}$ &$  d^2\textbf{b}_2 d^2\textbf{b}_3 d^2\textbf{b}_q$
		&$K_0(\sqrt{t_A}|\textbf{b}_q|)h_2(\textbf{b}_3-\textbf{b}_2-\textbf{b}_q,\textbf{b}_2+\textbf{b}_q,t_B,t_C,t_D)$ \\
		
		$C_{e1}$ &$  d^2 \textbf{b}_2d^2\textbf{b}_3d^2\textbf{b}'_3d^2\textbf{b}_q$ &$\frac{1}{(2\pi)^4}
		K_0(\sqrt{t_A}|\textbf{b}_2-\textbf{b}_3|)
		K_0(\sqrt{t_B}|\textbf{b}_q-\textbf{b}_3-\textbf{b}'_3|)
		K_0(\sqrt{t_C}|\textbf{b}_2-\textbf{b}_3-\textbf{b}'_3+\textbf{b}_q|)
		K_0(\sqrt{t_D}|\textbf{b}_3+\textbf{b}'_3|)$ \\
		
		$C_{e2}$ &$  d^2 \textbf{b}_2d^2\textbf{b}_3d^2\textbf{b}'_2d^2\textbf{b}_q$ &$\frac{1}{(2\pi)^4}
		K_0(\sqrt{t_A}|\textbf{b}_2-\textbf{b}_3|)
		K_0(\sqrt{t_B}|\textbf{b}_q-\textbf{b}'_2|)
		K_0(\sqrt{t_C}|\textbf{b}_2+\textbf{b}_q|)
		K_0(\sqrt{t_D}|\textbf{b}'_2|)$ \\
		
		$C_{e3}$ &$  d^2 \textbf{b}_2d^2\textbf{b}_3d^2\textbf{b}'_2d^2\textbf{b}_q$ &$\frac{1}{(2\pi)^4}
		K_0(\sqrt{t_A}|\textbf{b}_2-\textbf{b}_3|)
		K_0(\sqrt{t_B}|\textbf{b}'_2+\textbf{b}_q|)
		K_0(\sqrt{t_C}|\textbf{b}'_2|)
		K_0(\sqrt{t_D}|\textbf{b}_3-\textbf{b}'_2+\textbf{b}'_3|)$ \\
		
		$C_{e4}$ &$  d^2 \textbf{b}'_2d^2\textbf{b}'_3d^2\textbf{b}_q$ &$K_0(\sqrt{t_A}|\textbf{b}_2+\textbf{b}'_3|)K_0(\sqrt{t_B}|\textbf{b}_q|)h_1(\textbf{b}_2+\textbf{b}_q,t_C,t_D)$ \\
				
		$C_{f1}$ &$  d^2 \textbf{b}_2d^2\textbf{b}_3d^2\textbf{b}'_3d^2\textbf{b}_q$ &$\frac{1}{(2\pi)^4}
		K_0(\sqrt{t_A}|\textbf{b}_q-\textbf{b}'_3|)
		K_0(\sqrt{t_B}|\textbf{b}_2|)
		K_0(\sqrt{t_C}|\textbf{b}_2-\textbf{b}_3-\textbf{b}'_3+\textbf{b}_q|)
		K_0(\sqrt{t_D}|\textbf{b}_3+\textbf{b}'_3|)$ \\
		
		$C_{f2}$ &$  d^2 \textbf{b}_2d^2\textbf{b}_3d^2\textbf{b}'_2d^2\textbf{b}_q$ &$\frac{1}{(2\pi)^4}
		K_0(\sqrt{t_A}|\textbf{b}_3+\textbf{b}_q|)
		K_0(\sqrt{t_B}|\textbf{b}_2+\textbf{b}'_2|)
		K_0(\sqrt{t_C}|\textbf{b}_2+\textbf{b}_q|)
		K_0(\sqrt{t_D}|\textbf{b}'_2|)$ \\
		
		$C_{f3}$ &$  d^2 \textbf{b}_2d^2\textbf{b}_3d^2\textbf{b}'_2d^2\textbf{b}_q$ &$\frac{1}{(2\pi)^4}
		K_0(\sqrt{t_A}|\textbf{b}_3+\textbf{b}_q|)
		K_0(\sqrt{t_B}|\textbf{b}_2-\textbf{b}'_2|)
		K_0(\sqrt{t_C}|\textbf{b}_2+\textbf{b}_q|)
		K_0(\sqrt{t_D}|\textbf{b}'_2|)$ \\
		
		$C_{f4}$ &$  d^2 \textbf{b}_2d^2\textbf{b}_3d^2\textbf{b}_q$ &$h_3(\textbf{b}_q-\textbf{b}_2+\textbf{b}_3,\textbf{b}_q,\textbf{b}_2,t_A,t_B,t_C,t_D)$ \\
		
		$C_{g1}$ &$  d^2\textbf{b}_2 d^2\textbf{b}_3 d^2\textbf{b}_q$
		&$K_0(\sqrt{t_D}|\textbf{b}_2+\textbf{b}_q|)h_2(\textbf{b}_2-\textbf{b}_3,\textbf{b}_2-\textbf{b}_3-\textbf{b}'_3+\textbf{b}_q,t_A,t_B,t_C)$ \\
		
		$C_{g2}$ &$  d^2\textbf{b}_2 d^2\textbf{b}_3 d^2\textbf{b}'_3$
		&$K_0(\sqrt{t_D}|\textbf{b}_3+\textbf{b}'_3|)h_2(\textbf{b}_2,\textbf{b}_3,t_A,t_B,t_C)$ \\
		
		$C_{g3}$ &$  d^2\textbf{b}_2 d^2\textbf{b}'_3 d^2\textbf{b}_q$
		&$K_0(\sqrt{t_D}|\textbf{b}_2+\textbf{b}_q|)h_2(\textbf{b}_q,\textbf{b}_2+\textbf{b}'_3+\textbf{b}_q,t_A,t_B,t_C)$ \\
		
		$C_{g4}$ &$  d^2\textbf{b}'_2 d^2\textbf{b}'_3 d^2\textbf{b}_q$
		&$K_0(\sqrt{t_D}|\textbf{b}'_2|)h_2(\textbf{b}'_2+\textbf{b}_q,\textbf{b}'_3,t_A,t_B,t_C)$ \\
		
		$Eq_{a1}$ &$  d^2 \textbf{b}_2d^2\textbf{b}_3d^2\textbf{b}'_3d^2\textbf{b}_q$ &$\frac{1}{(2\pi)^4}
		K_0(\sqrt{t_A}|\textbf{b}_2-\textbf{b}_3|)
		K_0(\sqrt{t_B}|\textbf{b}_3+\textbf{b}'_3-\textbf{b}_q|)
		K_0(\sqrt{t_C}|\textbf{b}_2|)
		K_0(\sqrt{t_D}|\textbf{b}_3+\textbf{b}'_3|)$ \\
		
		$Eq_{a2}$ &$  d^2\textbf{b}_3 d^2\textbf{b}'_3 d^2\textbf{b}_q$
		&$K_0(\sqrt{t_B}|\textbf{b}_3+\textbf{b}'_3-\textbf{b}_q|)h_2(\textbf{b}'_3,\textbf{b}_3+\textbf{b}'_3,t_A,t_C,t_D)$ \\
		
		$Eq_{a3}$ &$  d^2\textbf{b}_3 d^2\textbf{b}'_3 d^2\textbf{b}_q$
		&$K_0(\sqrt{t_A}|\textbf{b}'_3|)h_2(\textbf{b}_3+\textbf{b}'_3,\textbf{b}_q,t_B,t_C,t_D)$ \\
		
		$Eq_{a4}$ &$  d^2 \textbf{b}_2d^2\textbf{b}_3d^2\textbf{b}'_3d^2\textbf{b}_q$ &$\frac{1}{(2\pi)^4}
		K_0(\sqrt{t_A}|\textbf{b}_3|)
		K_0(\sqrt{t_B}|\textbf{b}_2-\textbf{b}_3-\textbf{b}'_3+\textbf{b}_q|)
		K_0(\sqrt{t_C}|\textbf{b}_2|)
		K_0(\sqrt{t_D}|\textbf{b}_2-\textbf{b}_3-\textbf{b}'_3|)$ \\
		
		$Eq_{a5}$ &$  d^2 \textbf{b}_2d^2\textbf{b}_3d^2\textbf{b}'_3d^2\textbf{b}_q$ &$\frac{1}{(2\pi)^4}
		K_0(\sqrt{t_A}|\textbf{b}'_2-\textbf{b}'_3+\textbf{b}_q|)
		K_0(\sqrt{t_B}|\textbf{b}'_2|)
		K_0(\sqrt{t_C}|\textbf{b}_3-\textbf{b}'_2+\textbf{b}'_3-\textbf{b}_q|)
		K_0(\sqrt{t_D}|\textbf{b}_3+\textbf{b}'_3|)$ \\
		
		$Eq_{a6}$ &$  d^2 \textbf{b}_3d^2\textbf{b}'_2d^2\textbf{b}'_3d^2\textbf{b}_q$ &$\frac{1}{(2\pi)^4}
		K_0(\sqrt{t_A}|\textbf{b}_3-\textbf{b}_q|)
		K_0(\sqrt{t_B}|\textbf{b}'_2|)
		K_0(\sqrt{t_C}|\textbf{b}_3-\textbf{b}'_2-\textbf{b}'_3-\textbf{b}_q|)
		K_0(\sqrt{t_D}|\textbf{b}_3+\textbf{b}'_3|)$ \\
		
		$Eq_{a7}$ &$  d^2 \textbf{b}_3d^2\textbf{b}'_2d^2\textbf{b}'_3d^2\textbf{b}_q$ &$\frac{1}{(2\pi)^4}
		K_0(\sqrt{t_A}|\textbf{b}'_2-\textbf{b}'_3|)
		K_0(\sqrt{t_B}|\textbf{b}_3+\textbf{b}'_3-\textbf{b}_q|)
		K_0(\sqrt{t_C}|\textbf{b}_3-\textbf{b}'_2+\textbf{b}'_3-\textbf{b}_q|)
		K_0(\sqrt{t_D}|\textbf{b}_3+\textbf{b}'_3|)$ \\
		
		$Eq_{b1}$ &$  d^2\textbf{b}_2 d^2\textbf{b}_3 d^2\textbf{b}_q$
		&$K_0(\sqrt{t_A}|\textbf{b}_2-\textbf{b}_3|)h_2(\textbf{b}_q,\textbf{b}_2-\textbf{b}_3,t_B,t_C,t_D)$ \\
		
		$Eq_{b2}$ &$  d^2\textbf{b}_2 d^2\textbf{b}_3 d^2\textbf{b}_q$
		&$K_0(\sqrt{t_B}|\textbf{b}_2-\textbf{b}_q|)h_2(\textbf{b}_3,\textbf{b}_2,t_A,t_C,t_D)$ \\
		
		$Eq_{b3}$ &$  d^2 \textbf{b}_2d^2\textbf{b}_3d^2\textbf{b}'_2d^2\textbf{b}_q$ &$\frac{1}{(2\pi)^4}
		K_0(\sqrt{t_A}|\textbf{b}_3-\textbf{b}'_2-\textbf{b}_q|)
		K_0(\sqrt{t_B}|\textbf{b}_2-\textbf{b}_q|)
		K_0(\sqrt{t_C}|\textbf{b}'_2+\textbf{b}_q|)
		K_0(\sqrt{t_D}|\textbf{b}_2|)$ \\
		
		$Eq_{b4}$ &$  d^2 \textbf{b}_2d^2\textbf{b}_3d^2\textbf{b}'_3d^2\textbf{b}_q$ &$\frac{1}{(2\pi)^4}
		K_0(\sqrt{t_A}|\textbf{b}_3|)
		K_0(\sqrt{t_B}|\textbf{b}_2-\textbf{b}_3-\textbf{b}'_3-\textbf{b}_q|)
		K_0(\sqrt{t_C}|\textbf{b}_3+\textbf{b}'_3|)
		K_0(\sqrt{t_D}|\textbf{b}_2-\textbf{b}_3-\textbf{b}'_3|)$ \\
		
		$Eq_{b5}$ &$  d^2 \textbf{b}_2d^2\textbf{b}_3d^2\textbf{b}'_2d^2\textbf{b}_q$ &$\frac{1}{(2\pi)^4}
		K_0(\sqrt{t_A}|\textbf{b}_3+\textbf{b}'_2+\textbf{b}_q|)
		K_0(\sqrt{t_B}|\textbf{b}_2+\textbf{b}'_2|)
		K_0(\sqrt{t_C}|\textbf{b}'_2+\textbf{b}_q|)
		K_0(\sqrt{t_D}|\textbf{b}_2|)$ \\
		
		$Eq_{b6}$ &$  d^2 \textbf{b}_2d^2\textbf{b}_3d^2\textbf{b}'_2d^2\textbf{b}_q$ &$\frac{1}{(2\pi)^4}
		K_0(\sqrt{t_A}|\textbf{b}_3-\textbf{b}_q|)
		K_0(\sqrt{t_B}|\textbf{b}_2+\textbf{b}'_2|)
		K_0(\sqrt{t_C}|\textbf{b}'_2+\textbf{b}_q|)
		K_0(\sqrt{t_D}|\textbf{b}_2|)$ \\
		
		$Eq_{b7}$ &$  d^2 \textbf{b}_2d^2\textbf{b}_3d^2\textbf{b}'_2d^2\textbf{b}_q$ &$\frac{1}{(2\pi)^4}
		K_0(\sqrt{t_A}|\textbf{b}_3+\textbf{b}'_2|)
		K_0(\sqrt{t_B}|\textbf{b}_2-\textbf{b}_q|)
		K_0(\sqrt{t_C}|\textbf{b}'_2+\textbf{b}_q|)
		K_0(\sqrt{t_D}|\textbf{b}_2|)$ \\
		
		$Eq_{c1}$ &$  d^2 \textbf{b}_2d^2\textbf{b}_3d^2\textbf{b}'_3d^2\textbf{b}_q$ &$\frac{1}{(2\pi)^4}
		K_0(\sqrt{t_A}|\textbf{b}_2-\textbf{b}_3|)
		K_0(\sqrt{t_B}|\textbf{b}_3+\textbf{b}'_3-\textbf{b}_q|)
		K_0(\sqrt{t_C}|\textbf{b}_2-\textbf{b}_3-\textbf{b}'_3|)
		K_0(\sqrt{t_D}|\textbf{b}_3+\textbf{b}'_3|)$ \\
		
		$Eq_{c2}$ &$  d^2 \textbf{b}_2d^2\textbf{b}_3d^2\textbf{b}'_3d^2\textbf{b}_q$ &$\frac{1}{(2\pi)^4}
		K_0(\sqrt{t_A}|\textbf{b}'_3|)
		K_0(\sqrt{t_B}|\textbf{b}_2-\textbf{b}_q|)
		K_0(\sqrt{t_C}|\textbf{b}_2-\textbf{b}_3-\textbf{b}'_3|)
		K_0(\sqrt{t_D}|\textbf{b}_2|)$ \\
		
		$Eq_{c3}$ &$  d^2\textbf{b}_3 d^2\textbf{b}'_2 d^2\textbf{b}'_3$
		&$K_0(\sqrt{t_A}|\textbf{b}_3|)h_2(\textbf{b}'_2,\textbf{b}_3+\textbf{b}'_3,t_B,t_C,t_D)$ \\
		
		$Eq_{c4}$ &$  d^2\textbf{b}_3 d^2\textbf{b}'_2 d^2\textbf{b}'_3$
		&$K_0(\sqrt{t_B}|\textbf{b}_3-\textbf{b}'_2+\textbf{b}'_3|)h_2(\textbf{b}'_3,\textbf{b}_3+\textbf{b}'_3,t_A,t_C,t_D)$ \\
		
		$Eq_{c5}$ &$  d^2 \textbf{b}_3d^2\textbf{b}'_2d^2\textbf{b}'_3d^2\textbf{b}_q$ &$\frac{1}{(2\pi)^4}
		K_0(\sqrt{t_A}|\textbf{b}'_2-\textbf{b}'_3+\textbf{b}_q|)
		K_0(\sqrt{t_B}|\textbf{b}'_2-\textbf{b}_3-\textbf{b}'_3|)
		K_0(\sqrt{t_C}|\textbf{b}'_2+\textbf{b}_q|)
		K_0(\sqrt{t_D}|\textbf{b}_3+\textbf{b}'_3|)$ \\
		
		$Eq_{c6}$ &$  d^2 \textbf{b}_3d^2\textbf{b}'_2d^2\textbf{b}'_3d^2\textbf{b}_q$ &$\frac{1}{(2\pi)^4}
		K_0(\sqrt{t_A}|\textbf{b}'_3+\textbf{b}_q|)
		K_0(\sqrt{t_B}|\textbf{b}_3-\textbf{b}'_2+\textbf{b}'_3|)
		K_0(\sqrt{t_C}|\textbf{b}'_2+\textbf{b}_q|)
		K_0(\sqrt{t_D}|\textbf{b}_3+\textbf{b}'_3|)$ \\
		
		$Eq_{c7}$ &$  d^2 \textbf{b}_3d^2\textbf{b}'_2d^2\textbf{b}'_3d^2\textbf{b}_q$ &$\frac{1}{(2\pi)^4}
		K_0(\sqrt{t_A}|\textbf{b}'_2-\textbf{b}'_3|)
		K_0(\sqrt{t_B}|\textbf{b}_3-\textbf{b}'_3+\textbf{b}_q|)
		K_0(\sqrt{t_C}|\textbf{b}'_2+\textbf{b}_q|)
		K_0(\sqrt{t_D}|\textbf{b}_3+\textbf{b}'_3|)$ \\

		\hline\hline
	\end{tabular}
\end{table}

\begin{table}[H]
    \scriptsize
	\centering
	TABLE~\ref{tab:bb} (continued) \\
\begin{tabular}[t]{lccc}
\hline\hline	
		$Eq_{d1}$ &$  d^2 \textbf{b}_2d^2\textbf{b}_3d^2\textbf{b}'_2d^2\textbf{b}_q$ &$\frac{1}{(2\pi)^4}
		K_0(\sqrt{t_A}|\textbf{b}_2-\textbf{b}_3|)
		K_0(\sqrt{t_B}|\textbf{b}_q|)
		K_0(\sqrt{t_C}|\textbf{b}_2)
		K_0(\sqrt{t_D}|\textbf{b}'_2+\textbf{b}_q|)$ \\
		
		$Eq_{d2}$ &$  d^2 \textbf{b}_3d^2\textbf{b}'_2d^2\textbf{b}'_3d^2\textbf{b}_q$ &$\frac{1}{(2\pi)^4}
		K_0(\sqrt{t_A}|\textbf{b}'_3|)
		K_0(\sqrt{t_B}|\textbf{b}_q|)
		K_0(\sqrt{t_C}|\textbf{b}_3+\textbf{b}'_3|)
		K_0(\sqrt{t_D}|\textbf{b}_3-\textbf{b}'_2+\textbf{b}'_3-\textbf{b}_q|)$ \\
		
		$Eq_{d3}$ &$  d^2 \textbf{b}_3d^2\textbf{b}'_2d^2\textbf{b}'_3d^2\textbf{b}_q$ &$\frac{1}{(2\pi)^4}
		K_0(\sqrt{t_A}|\textbf{b}_3|)
		K_0(\sqrt{t_B}|\textbf{b}_q|)
		K_0(\sqrt{t_C}|\textbf{b}_3+\textbf{b}'_3|)
		K_0(\sqrt{t_D}|\textbf{b}'_2+\textbf{b}_q|)$ \\
		
		$Eq_{d4}$ &$  d^2\textbf{b}'_2 d^2\textbf{b}'_3 d^2\textbf{b}_q$
		&$K_0(\sqrt{t_B}|\textbf{b}_q|)h_2(\textbf{b}'_2-\textbf{b}'_3+\textbf{b}_q,\textbf{b}'_2-\textbf{b}_q,t_A,t_C,t_D)$ \\
		
		$Eq_{d5}$ &$  d^2\textbf{b}'_2 d^2\textbf{b}'_3 d^2\textbf{b}_q$
		&$K_0(\sqrt{t_A}|\textbf{b}'_2-\textbf{b}'_3+\textbf{b}_q|)h_2(\textbf{b}'_2,\textbf{b}'_2+\textbf{b}_q,t_B,t_C,t_D)$ \\
		
		$Eq_{d6}$ &$  d^2\textbf{b}'_2 d^2\textbf{b}'_3 d^2\textbf{b}_q$
		&$h_3(\textbf{b}'_3-\textbf{b}_q,\textbf{b}'_2,\textbf{b}'_2-\textbf{b}_q,t_A,t_B,t_C,t_D)$ \\
		
		$Eq_{d7}$ &$  d^2 \textbf{b}'_2d^2\textbf{b}'_3d^2\textbf{b}_q$ &$K_0(\sqrt{t_A}|\textbf{b}_2-\textbf{b}'_3|)K_0(\sqrt{t_B}|\textbf{b}_q|)h_1(\textbf{b}'_2+\textbf{b}_q,t_C,t_D)$ \\
		
		$Eq_{e1}$ &$  d^2 \textbf{b}_2d^2\textbf{b}_3d^2\textbf{b}'_3d^2\textbf{b}_q$ &$\frac{1}{(2\pi)^4}
		K_0(\sqrt{t_A}|\textbf{b}_3+\textbf{b}_q|)
		K_0(\sqrt{t_B}|\textbf{b}_2-\textbf{b}_3|)
		K_0(\sqrt{t_C}|\textbf{b}_3+\textbf{b}'_3|)
		K_0(\sqrt{t_D}|\textbf{b}_2|)$ \\
		
		$Eq_{e2}$ &$  d^2 \textbf{b}_3d^2\textbf{b}'_2d^2\textbf{b}'_3d^2\textbf{b}_q$ &$\frac{1}{(2\pi)^4}
		K_0(\sqrt{t_A}|\textbf{b}_3-\textbf{b}_q|)
		K_0(\sqrt{t_B}|\textbf{b}'_3|)
		K_0(\sqrt{t_C}|\textbf{b}_3-\textbf{b}'_2+\textbf{b}'_3-\textbf{b}_q|)
		K_0(\sqrt{t_D}|\textbf{b}'_2+\textbf{b}_q|)$ \\
		
		$Eq_{e3}$ &$  d^2 \textbf{b}_2d^2\textbf{b}_3d^2\textbf{b}'_3d^2\textbf{b}_q$ &$\frac{1}{(2\pi)^4}
		K_0(\sqrt{t_A}|\textbf{b}'_3+\textbf{b}_q|)
		K_0(\sqrt{t_B}|\textbf{b}_3|)
		K_0(\sqrt{t_C}|\textbf{b}_2-\textbf{b}_3-\textbf{b}'_3|)
		K_0(\sqrt{t_D}|\textbf{b}_2|)$ \\
		
		$Eq_{e4}$ &$  d^2 \textbf{b}_3d^2\textbf{b}'_2d^2\textbf{b}'_3d^2\textbf{b}_q$ &$\frac{1}{(2\pi)^4}
		K_0(\sqrt{t_A}|\textbf{b}'_2-\textbf{b}'_3|)
		K_0(\sqrt{t_B}|\textbf{b}_3+\textbf{b}'_2+\textbf{b}_q|)
		K_0(\sqrt{t_C}|\textbf{b}_3+\textbf{b}'_3|)
		K_0(\sqrt{t_D}|\textbf{b}'_2+\textbf{b}_q|)$ \\
		
		$Eq_{f1}$ &$  d^2 \textbf{b}_2d^2\textbf{b}_3d^2\textbf{b}'_3d^2\textbf{b}_q$ &$\frac{1}{(2\pi)^4}
		K_0(\sqrt{t_A}|\textbf{b}_3-\textbf{b}_q|)
		K_0(\sqrt{t_B}|\textbf{b}_2+\textbf{b}'_3|)
		K_0(\sqrt{t_C}|\textbf{b}_3+\textbf{b}'_3|)
		K_0(\sqrt{t_D}|\textbf{b}_2|)$ \\
		
		$Eq_{f2}$ &$  d^2 \textbf{b}_3d^2\textbf{b}'_2d^2\textbf{b}'_3d^2\textbf{b}_q$ &$\frac{1}{(2\pi)^4}
		K_0(\sqrt{t_A}|\textbf{b}_3-\textbf{b}_q|)
		K_0(\sqrt{t_B}|\textbf{b}'_3|)
		K_0(\sqrt{t_C}|\textbf{b}_3-\textbf{b}'_2+\textbf{b}'_3-\textbf{b}_q|)
		K_0(\sqrt{t_D}|\textbf{b}'_2+\textbf{b}_q|)$ \\
		
		$Eq_{f3}$ &$  d^2 \textbf{b}_2d^2\textbf{b}_3d^2\textbf{b}'_3d^2\textbf{b}_q$ &$\frac{1}{(2\pi)^4}
		K_0(\sqrt{t_A}|\textbf{b}_2-\textbf{b}_3+\textbf{b}_q|)
		K_0(\sqrt{t_B}|\textbf{b}_2-\textbf{b}'_3|)
		K_0(\sqrt{t_C}|\textbf{b}_2-\textbf{b}_3-\textbf{b}'_3|)
		K_0(\sqrt{t_D}|\textbf{b}_2|)$ \\
		
		$Eq_{f4}$ &$  d^2 \textbf{b}_3d^2\textbf{b}'_2d^2\textbf{b}'_3d^2\textbf{b}_q$ &$\frac{1}{(2\pi)^4}
		K_0(\sqrt{t_A}|\textbf{b}_3+\textbf{b}'_2|)
		K_0(\sqrt{t_B}|\textbf{b}'_2-\textbf{b}'_3+\textbf{b}_q|)
		K_0(\sqrt{t_C}|\textbf{b}_3+\textbf{b}'_3|)
		K_0(\sqrt{t_D}|\textbf{b}'_2+\textbf{b}_q|)$ \\
		
		$Eq_{g1}$ &$  d^2\textbf{b}_2 d^2\textbf{b}'_3 d^2\textbf{b}_q$
		&$K_0(\sqrt{t_D}|\textbf{b}_2|)h_2(\textbf{b}_2-\textbf{b}_q,\textbf{b}_2+\textbf{b}_q,t_A,t_B,t_C)$ \\
		
		$Eq_{g2}$ &$  d^2\textbf{b}_3 d^2\textbf{b}'_2 d^2\textbf{b}_q$
		&$K_0(\sqrt{t_D}|\textbf{b}'_2+\textbf{b}_q|)h_2(\textbf{b}'_2,\textbf{b}_3-\textbf{b}'_2-\textbf{b}_q,t_A,t_B,t_C)$ \\
		
		$Eq_{g3}$ &$  d^2\textbf{b}_2 d^2\textbf{b}'_3 d^2\textbf{b}_q$
		&$K_0(\sqrt{t_D}|\textbf{b}_2|)h_2(\textbf{b}_2+\textbf{b}_q,\textbf{b}_2-\textbf{b}'_3,t_A,t_B,t_C)$ \\
		
		$Eq_{g4}$ &$  d^2\textbf{b}'_2 d^2\textbf{b}'_3 d^2\textbf{b}_q$
		&$K_0(\sqrt{t_D}|\textbf{b}'_2+\textbf{b}_q|)h_2(\textbf{b}_q,\textbf{b}'_2-\textbf{b}'_3+\textbf{b}_q,t_A,t_B,t_C)$ \\
		
		$B_{a1}$ &$  d^2 \textbf{b}_2d^2\textbf{b}_3d^2\textbf{b}'_3d^2\textbf{b}_q$ &$\frac{1}{(2\pi)^4}
		K_0(\sqrt{t_A}|\textbf{b}_3|)
		K_0(\sqrt{t_B}|\textbf{b}'_3|)
		K_0(\sqrt{t_C}|\textbf{b}_2|)
		K_0(\sqrt{t_D}|\textbf{b}_2-\textbf{b}_3-\textbf{b}'_3-\textbf{b}_q|)$ \\
		
		$B_{a2}$ &$  d^2 \textbf{b}_3d^2\textbf{b}'_2d^2\textbf{b}'_3d^2\textbf{b}_q$ &$\frac{1}{(2\pi)^4}
		K_0(\sqrt{t_A}|\textbf{b}'_2-\textbf{b}'_3-\textbf{b}_q|)
		K_0(\sqrt{t_B}|\textbf{b}'_3|)
		K_0(\sqrt{t_C}|\textbf{b}_3-\textbf{b}'_2+\textbf{b}'_3+\textbf{b}_q|)
		K_0(\sqrt{t_D}|\textbf{b}_3+\textbf{b}'_3+\textbf{b}_q|)$ \\
		
		$B_{a3}$ &$  d^2 \textbf{b}_3d^2\textbf{b}'_2d^2\textbf{b}'_3d^2\textbf{b}_q$ &$\frac{1}{(2\pi)^4}
		K_0(\sqrt{t_A}|\textbf{b}_2-\textbf{b}_3|)
		K_0(\sqrt{t_B}|\textbf{b}'_3|)
		K_0(\sqrt{t_C}|\textbf{b}_2|)
		K_0(\sqrt{t_D}|\textbf{b}'_2|)$ \\
		
		$B_{a4}$ &$  d^2\textbf{b}_3 d^2\textbf{b}'_2 d^2\textbf{b}'_3$
		&$K_0(\sqrt{t_B}|\textbf{b}'_3|)h_2(\textbf{b}'_2-\textbf{b}_3,\textbf{b}'_2,t_A,t_C,t_D)$ \\
		
		$B_{a5}$ &$  d^2\textbf{b}_3 d^2\textbf{b}'_2 d^2\textbf{b}'_3$
		&$K_0(\sqrt{t_A}|\textbf{b}_3-\textbf{b}'_2|)h_2(\textbf{b}'_2-\textbf{b}'_3,\textbf{b}'_2,t_B,t_C,t_D)$ \\
		
		$B_{a6}$ &$  d^2 \textbf{b}_3d^2\textbf{b}'_2d^2\textbf{b}'_3d^2\textbf{b}_q$ &$\frac{1}{(2\pi)^4}
		K_0(\sqrt{t_A}|\textbf{b}_3-\textbf{b}'_2+\textbf{b}'_3|)
		K_0(\sqrt{t_B}|\textbf{b}_3-\textbf{b}'_2+\textbf{b}_q|)
		K_0(\sqrt{t_C}|\textbf{b}_3-\textbf{b}'_2+\textbf{b}'_3+\textbf{b}_q|)
		K_0(\sqrt{t_D}|\textbf{b}'_2|)$ \\
		
		$B_{a7}$ &$  d^2 \textbf{b}_3d^2\textbf{b}'_2d^2\textbf{b}'_3d^2\textbf{b}_q$ &$\frac{1}{(2\pi)^4}
		K_0(\sqrt{t_A}|\textbf{b}_q|)
		K_0(\sqrt{t_B}|\textbf{b}'_3|)
		K_0(\sqrt{t_C}|\textbf{b}_3-\textbf{b}'_2+\textbf{b}'_3-\textbf{b}_q|)
		K_0(\sqrt{t_D}|\textbf{b}'_2|)$ \\
		
		$B_{b1}$ &$  d^2 \textbf{b}_2d^2\textbf{b}_3d^2\textbf{b}'_3d^2\textbf{b}_q$ &$\frac{1}{(2\pi)^4}
		K_0(\sqrt{t_A}|\textbf{b}_3|)
		K_0(\sqrt{t_B}|\textbf{b}_2-\textbf{b}_3-\textbf{b}_q|)
		K_0(\sqrt{t_C}|\textbf{b}_3+\textbf{b}'_3+\textbf{b}_q|)
		K_0(\sqrt{t_D}|\textbf{b}_2-\textbf{b}_3-\textbf{b}'_3-\textbf{b}_q|)$ \\
		
		$B_{b2}$ &$  d^2 \textbf{b}_3d^2\textbf{b}'_2d^2\textbf{b}'_3d^2\textbf{b}_q$ &$\frac{1}{(2\pi)^4}
		K_0(\sqrt{t_A}|\textbf{b}_3+\textbf{b}'_2|)
		K_0(\sqrt{t_B}|\textbf{b}_2+\textbf{b}'_3|)
		K_0(\sqrt{t_C}|\textbf{b}'_2|)
		K_0(\sqrt{t_D}|\textbf{b}_2|)$ \\
		
		$B_{b3}$ &$  d^2 \textbf{b}_2d^2\textbf{b}'_3d^2\textbf{b}_q$ &$K_0(\sqrt{t_A}|\textbf{b}_2+\textbf{b}'_3+\textbf{b}_q|)h_2(\textbf{b}'_3,\textbf{b}_2,t_B,t_C,t_D)$ \\
		
		$B_{b4}$ &$  d^2 \textbf{b}_2d^2\textbf{b}'_3d^2\textbf{b}_q$ &$K_0(\sqrt{t_B}|\textbf{b}_2+\textbf{b}'_3|)h_2(\textbf{b}'_3+\textbf{b}_q,\textbf{b}_2,t_A,t_C,t_D)$ \\
		
		$B_{b5}$ &$  d^2 \textbf{b}_2d^2\textbf{b}_3d^2\textbf{b}'_3d^2\textbf{b}_q$ &$\frac{1}{(2\pi)^4}
		K_0(\sqrt{t_A}|\textbf{b}'_3+\textbf{b}_q|)
		K_0(\sqrt{t_B}|\textbf{b}_2-\textbf{b}_3-\textbf{b}_q|)
		K_0(\sqrt{t_C}|\textbf{b}_3-\textbf{b}'_3+\textbf{b}_q|)
		K_0(\sqrt{t_D}|\textbf{b}_2|)$ \\
		
		$B_{b6}$ &$  d^2 \textbf{b}_2d^2\textbf{b}_3d^2\textbf{b}'_3d^2\textbf{b}_q$ &$\frac{1}{(2\pi)^4}
		K_0(\sqrt{t_A}|\textbf{b}_3+\textbf{b}'_3|)
		K_0(\sqrt{t_B}|\textbf{b}_2-\textbf{b}_3-\textbf{b}_q|)
		K_0(\sqrt{t_C}|\textbf{b}_3+\textbf{b}'_3+\textbf{b}_q|)
		K_0(\sqrt{t_D}|\textbf{b}_2|)$ \\
		
		$B_{b7}$ &$  d^2 \textbf{b}_2d^2\textbf{b}_3d^2\textbf{b}'_3d^2\textbf{b}_q$ &$\frac{1}{(2\pi)^4}
		K_0(\sqrt{t_A}|\textbf{b}_q|)
		K_0(\sqrt{t_B}|\textbf{b}_2+\textbf{b}'_3|)
		K_0(\sqrt{t_C}|\textbf{b}_3+\textbf{b}'_3-\textbf{b}_q|)
		K_0(\sqrt{t_D}|\textbf{b}_2|)$ \\
		
		$B_{c1}$ &$  d^2 \textbf{b}_2d^2\textbf{b}_3d^2\textbf{b}'_3$ &$K_0(\sqrt{t_A}|\textbf{b}_3|)h_2(\textbf{b}_2-\textbf{b}'_3,\textbf{b}_2,t_B,t_C,t_D)$ \\
		
		$B_{c2}$ &$  d^2 \textbf{b}_2d^2\textbf{b}_3d^2\textbf{b}'_3$ &$K_0(\sqrt{t_B}|\textbf{b}_2-\textbf{b}'_3|)h_2(\textbf{b}_2-\textbf{b}_3, \textbf{b}_2,t_A,t_C,t_D)$ \\
		
		$B_{c3}$ &$  d^2 \textbf{b}_2d^2\textbf{b}_3d^2\textbf{b}'_2d^2\textbf{b}_q$ &$\frac{1}{(2\pi)^4}
		K_0(\sqrt{t_A}|\textbf{b}_2-\textbf{b}_3-\textbf{b}'_2|)
		K_0(\sqrt{t_B}|\textbf{b}_3+\textbf{b}_q|)
		K_0(\sqrt{t_C}|\textbf{b}'_2|)
		K_0(\sqrt{t_D}|\textbf{b}_2|)$ \\
		
		$B_{c4}$ &$  d^2 \textbf{b}_2d^2\textbf{b}_3d^2\textbf{b}'_3d^2\textbf{b}_q$ &$\frac{1}{(2\pi)^4}
		K_0(\sqrt{t_A}|\textbf{b}_2-\textbf{b}_3|)
		K_0(\sqrt{t_B}|\textbf{b}_3+\textbf{b}_q|)
		K_0(\sqrt{t_C}|\textbf{b}_2-\textbf{b}_3-\textbf{b}'_3-\textbf{b}_q|)
		K_0(\sqrt{t_D}|\textbf{b}_3+\textbf{b}'_3+\textbf{b}_q|)$ \\
		
		$B_{c5}$ &$  d^2 \textbf{b}_2d^2\textbf{b}_3d^2\textbf{b}'_3d^2\textbf{b}_q$ &$\frac{1}{(2\pi)^4}
		K_0(\sqrt{t_A}|\textbf{b}'_3+\textbf{b}_q|)
		K_0(\sqrt{t_B}|\textbf{b}_3+\textbf{b}_q|)
		K_0(\sqrt{t_C}|\textbf{b}_2-\textbf{b}_3-\textbf{b}'_3-\textbf{b}_q|)
		K_0(\sqrt{t_D}|\textbf{b}_2|)$ \\
		
		$B_{c6}$ &$  d^2 \textbf{b}_2d^2\textbf{b}_3d^2\textbf{b}'_3d^2\textbf{b}_q$ &$\frac{1}{(2\pi)^4}
		K_0(\sqrt{t_A}|\textbf{b}_2-\textbf{b}_3-\textbf{b}'_3|)
		K_0(\sqrt{t_B}|\textbf{b}_3+\textbf{b}_q|)
		K_0(\sqrt{t_C}|\textbf{b}_2-\textbf{b}_3+\textbf{b}'_3+\textbf{b}_q|)
		K_0(\sqrt{t_D}|\textbf{b}_2|)$ \\
		
		$B_{c7}$ &$  d^2 \textbf{b}_2d^2\textbf{b}_3d^2\textbf{b}'_3d^2\textbf{b}_q$ &$\frac{1}{(2\pi)^4}
		K_0(\sqrt{t_A}|\textbf{b}_2+\textbf{b}_q|)
		K_0(\sqrt{t_B}|\textbf{b}_2+\textbf{b}'_3|)
		K_0(\sqrt{t_C}|\textbf{b}_2-\textbf{b}_3-\textbf{b}'_3-\textbf{b}_q|)
		K_0(\sqrt{t_D}|\textbf{b}_2|)$ \\
		
		$B_{d1}$ &$  d^2 \textbf{b}_3d^2\textbf{b}'_2d^2\textbf{b}'_3d^2\textbf{b}_q$ &$\frac{1}{(2\pi)^4}
		K_0(\sqrt{t_A}|\textbf{b}_3|)
		K_0(\sqrt{t_B}|\textbf{b}'_2-\textbf{b}'_3|)
		K_0(\sqrt{t_C}|\textbf{b}_2|)
		K_0(\sqrt{t_D}|\textbf{b}_2|)$ \\
		
		$B_{d2}$ &$  d^2 \textbf{b}_3d^2\textbf{b}'_2d^2\textbf{b}'_3$ &$K_0(\sqrt{t_B}|\textbf{b}'_2-\textbf{b}'_3|)h_2(\textbf{b}_3+\textbf{b}'_2,\textbf{b}'_2,t_A,t_C,t_D)$ \\
		
		$B_{d3}$ &$  d^2 \textbf{b}_3d^2\textbf{b}'_2d^2\textbf{b}'_3$ &$K_0(\sqrt{t_A}|\textbf{b}_3+\textbf{b}'_2|)h_2(\textbf{b}'_3,\textbf{b}'_2,t_B,t_C,t_D)$ \\
		
		$B_{d4}$ &$  d^2 \textbf{b}_2d^2\textbf{b}_3d^2\textbf{b}'_2d^2\textbf{b}'_3$ &$\frac{1}{(2\pi)^4}
		K_0(\sqrt{t_A}|\textbf{b}_2-\textbf{b}_3|)
		K_0(\sqrt{t_B}|\textbf{b}'_2-\textbf{b}'_3|)
		K_0(\sqrt{t_C}|\textbf{b}_2|)
		K_0(\sqrt{t_D}|\textbf{b}'_2|)$ \\
		
		$B_{d5}$ &$  d^2 \textbf{b}_3d^2\textbf{b}'_2d^2\textbf{b}'_3d^2\textbf{b}_q$ &$\frac{1}{(2\pi)^4}
		K_0(\sqrt{t_A}|\textbf{b}'_3+\textbf{b}_q|)
		K_0(\sqrt{t_B}|\textbf{b}'_2-\textbf{b}'_3|)
		K_0(\sqrt{t_C}|\textbf{b}_3+\textbf{b}'_3+\textbf{b}_q|)
		K_0(\sqrt{t_D}|\textbf{b}_3-\textbf{b}'_2+\textbf{b}'_3+\textbf{b}_q|)$ \\
		
		$B_{d6}$ &$  d^2 \textbf{b}_3d^2\textbf{b}'_2d^2\textbf{b}'_3d^2\textbf{b}_q$ &$\frac{1}{(2\pi)^4}
		K_0(\sqrt{t_A}|\textbf{b}_3+\textbf{b}'_3|)
		K_0(\sqrt{t_B}|\textbf{b}_3+\textbf{b}'_2+\textbf{b}_q|)
		K_0(\sqrt{t_C}|\textbf{b}_3+\textbf{b}'_3+\textbf{b}_q|)
		K_0(\sqrt{t_D}|\textbf{b}'_2|)$ \\
		
		$B_{d7}$ &$  d^2 \textbf{b}_3d^2\textbf{b}'_2d^2\textbf{b}'_3d^2\textbf{b}_q$ &$\frac{1}{(2\pi)^4}
		K_0(\sqrt{t_A}|\textbf{b}_q|)
		K_0(\sqrt{t_B}|\textbf{b}'_2-\textbf{b}'_3|)
		K_0(\sqrt{t_C}|\textbf{b}_3+\textbf{b}'_3+\textbf{b}_q|)
		K_0(\sqrt{t_D}|\textbf{b}'_2|)$ \\
		
		$B_{e1}$ &$  d^2 \textbf{b}_2d^2\textbf{b}_3d^2\textbf{b}'_3d^2\textbf{b}_q$ &$\frac{1}{(2\pi)^4}
		K_0(\sqrt{t_A}|\textbf{b}_q|)
		K_0(\sqrt{t_B}|\textbf{b}_3|)
		K_0(\sqrt{t_C}|\textbf{b}_2-\textbf{b}_3-\textbf{b}'_3-\textbf{b}_q|)
		K_0(\sqrt{t_D}|\textbf{b}_2|)$ \\
		
		$B_{e2}$ &$  d^2 \textbf{b}_3d^2\textbf{b}'_2d^2\textbf{b}'_3d^2\textbf{b}_q$ &$\frac{1}{(2\pi)^4}
		K_0(\sqrt{t_A}|\textbf{b}_q|)
		K_0(\sqrt{t_B}|\textbf{b}_3+\textbf{b}'_2|)
		K_0(\sqrt{t_C}|\textbf{b}_3+\textbf{b}'_3+\textbf{b}_q|)
		K_0(\sqrt{t_D}|\textbf{b}'_2|)$ \\
		
		$B_{e3}$ &$  d^2 \textbf{b}_2d^2\textbf{b}_3d^2\textbf{b}'_3d^2\textbf{b}_q$ &$\frac{1}{(2\pi)^4}
		K_0(\sqrt{t_A}|\textbf{b}_q|)
		K_0(\sqrt{t_B}|\textbf{b}_2-\textbf{b}_3|)
		K_0(\sqrt{t_C}|\textbf{b}_3+\textbf{b}'_3+\textbf{b}_q|)
		K_0(\sqrt{t_D}|\textbf{b}_2|)$ \\
		
		$B_{e4}$ &$  d^2 \textbf{b}_3d^2\textbf{b}'_2d^2\textbf{b}'_3d^2\textbf{b}_q$ &$\frac{1}{(2\pi)^4}
		K_0(\sqrt{t_A}|\textbf{b}_q|)
		K_0(\sqrt{t_B}|\textbf{b}_3-\textbf{b}'_2|)
		K_0(\sqrt{t_C}|\textbf{b}_3-\textbf{b}'_2+\textbf{b}'_3+\textbf{b}_q|)
		K_0(\sqrt{t_D}|\textbf{b}'_2|)$ \\
		
		$B_{f1}$ &$  d^2 \textbf{b}_2d^2\textbf{b}_3d^2\textbf{b}'_3d^2\textbf{b}_q$ &$\frac{1}{(2\pi)^4}
		K_0(\sqrt{t_A}|\textbf{b}_2-\textbf{b}_3-\textbf{b}'_3|)
		K_0(\sqrt{t_B}|\textbf{b}_2-\textbf{b}'_3-\textbf{b}_q|)
		K_0(\sqrt{t_C}|\textbf{b}_2-\textbf{b}_3-\textbf{b}'_3-\textbf{b}_q|)
		K_0(\sqrt{t_D}|\textbf{b}_2|)$ \\
		
		$B_{f2}$ &$  d^2 \textbf{b}_3d^2\textbf{b}'_2d^2\textbf{b}'_3d^2\textbf{b}_q$ &$\frac{1}{(2\pi)^4}
		K_0(\sqrt{t_A}|\textbf{b}_3+\textbf{b}'_3|)
		K_0(\sqrt{t_B}|\textbf{b}'_2-\textbf{b}'_3-\textbf{b}_q|)
		K_0(\sqrt{t_C}|\textbf{b}_3+\textbf{b}'_3+\textbf{b}_q|)
		K_0(\sqrt{t_D}|\textbf{b}'_2|)$ \\
		
		$B_{f3}$ &$  d^2 \textbf{b}_2d^2\textbf{b}_3d^2\textbf{b}'_3d^2\textbf{b}_q$ &$\frac{1}{(2\pi)^4}
		K_0(\sqrt{t_A}|\textbf{b}_3+\textbf{b}'_3|)
		K_0(\sqrt{t_B}|\textbf{b}_2+\textbf{b}'_3+\textbf{b}_q|)
		K_0(\sqrt{t_C}|\textbf{b}_3+\textbf{b}'_3+\textbf{b}_q|)
		K_0(\sqrt{t_D}|\textbf{b}_2|)$ \\
		
		$B_{f4}$ &$  d^2 \textbf{b}_3d^2\textbf{b}'_2d^2\textbf{b}'_3d^2\textbf{b}_q$ &$\frac{1}{(2\pi)^4}
		K_0(\sqrt{t_A}|\textbf{b}'_2-\textbf{b}_3-\textbf{b}'_3|)
		K_0(\sqrt{t_B}|\textbf{b}'_3+\textbf{b}_q|)
		K_0(\sqrt{t_C}|\textbf{b}_3-\textbf{b}'_2+\textbf{b}'_3+\textbf{b}_q|)
		K_0(\sqrt{t_D}|\textbf{b}'_2|)$ \\
		
		$B_{g1}$ &$  d^2 \textbf{b}_2d^2\textbf{b}'_3d^2\textbf{b}_q$ &$K_0(\sqrt{t_D}|\textbf{b}_2|)h_2(\textbf{b}_2+\textbf{b}'_3+\textbf{b}_q, \textbf{b}_2-\textbf{b}'_3,t_A,t_B,t_C)$ \\
		
		$B_{g2}$ &$  d^2 \textbf{b}_2d^2\textbf{b}'_3d^2\textbf{b}_q$ &$K_0(\sqrt{t_D}|\textbf{b}'_2|)h_2(\textbf{b}'_3+\textbf{b}_q,\textbf{b}'_3,t_A,t_B,t_C)$ \\
		
		$B_{g3}$ &$  d^2 \textbf{b}_2d^2\textbf{b}_3d^2\textbf{b}_q$ &$K_0(\sqrt{t_D}|\textbf{b}_2|)h_2(\textbf{b}_3,\textbf{b}_3+\textbf{b}_q,t_A,t_B,t_C)$ \\
		
		$B_{g4}$ &$  d^2 \textbf{b}'_2d^2\textbf{b}'_3d^2\textbf{b}_q$ &$K_0(\sqrt{t_D}|\textbf{b}'_2|)h_2(\textbf{b}'_2-\textbf{b}'_3-\textbf{b}_q, \textbf{b}'_2-\textbf{b}'_3,t_A,t_B,t_C)$ \\

\hline\hline		
\end{tabular}
\end{table}

\begin{table}[H]
\tiny	
	\centering
	\caption{The virtualities of the internal gluon $t_{A,B}$ and quark $t_{C,D}$ for the $C$, $B$, and $E_q$ diagrams.}
	\newcommand{\tabincell}[2]{\begin{tabular}{@{}#1@{}}#2\end{tabular}}
	\label{tab:ttt}
	\begin{tabular}[t]{lcccc|ccccc}
		\hline\hline
 $R_{ij}$ &  $\frac{t_A}{M^2}$   &$\frac{t_B}{M^2}$  &$\frac{t_C}{M^2}$ &$\frac{t_D}{M^2}$
 & $R_{ij}$ &  $\frac{t_A}{M^2}$   &$\frac{t_B}{M^2}$  &$\frac{t_C}{M^2}$ &$\frac{t_D}{M^2}$ \\ \hline
 $C_{a1}$  &  $x_2y$      &  $y(1-x_1)$        &$y(1-x_1-x_3')$  &$y+x_3'(1-y)$
 & $Eq_{a1}$ &  $x_3x_3'$   &  $x_2'(y-1)$        &$x_3'(1-x_1)$  &$(y-1)(1-x_3')$  \\

 $C_{a2}$  &  $x_2y$      &  $y(x_2-x_3')$        &$y(1-x_1-x_3')$  &$y+x_3'(1-y)$
 &$Eq_{a2}$ &  $x_3x_3'$   &  $x_2'(y-1)$        &$x_3+y-1$  &$(y-1)(1-x_3')$ \\

 $C_{a3}$  &  $x_2y$      &  $x_3x_3'$        &$x_3+y$  &$y+x_3'(1-y)$
 &$Eq_{a3}$ &  $x_3x_3'$   &  $x_2'(y-1)$        &$x_3(1-x_2')$  &$x_3+y-1$ \\

 $C_{a4}$  &  $x_2y$      &  $x_3x_3'$        &$x_2+x_3'$  &$y+x_3'(1-y)$
 &$Eq_{a4}$ &  $x_3x_3'$   &  $x_2'(y-1)$        &$x_2+x_3'(1-x_2)$  &$(y-1)(1-x_3')$ \\

 $C_{a5}$  &  $x_2y$      &  $x_3x_3'$        &$y(x_2-x_2')$  &$x_2+x_3'$
 &$Eq_{a5}$ &  $x_3x_3'$   &  $x_2'(y-1)$        &$x_3'(x_3-y)$  &$(y-1)(1-x_3')$ \\

 $C_{a6}$  &  $x_2y$      &  $x_3x_3'$        &$y(x_2-x_1')$  &$x_2+x_3'$
 &$Eq_{a6}$ &  $x_3x_3'$   &  $(1-x_1')(x_3+y-1)$        &$x_3'(x_3+y+1)$  &$x_3+y-1$ \\

 $C_{a7}$  &  $x_2y$      &  $x_3x_3'$        &$x_2$  &$x_2+x_3'$
 &$Eq_{a7}$ &  $x_3x_3'$   &  $(1-x_1')(x_3+y-1)$        &$x_3(1-x_1')$  &$x_3+y-1$ \\

 $C_{b1}$  &  $x_2y$      &  $y(1-x_1-x_3')$        &$y(1-x_1)$  &$y(x_1+x_1')$
 &$Eq_{b1}$ &  $x_3x_3'$   &  $x_2'(y-1)$        &$x_3'(1-x_1)$  &$(1-x_1')(y-x_1)$ \\

 $C_{b2}$  &  $x_2y$      &  $y(1-x_1-x_3')$        &$y(x_2-x_3')$  &$y(x_1+x_1')$
 &$Eq_{b2}$ &  $x_3x_3'$   &  $x_2'(y-1)$        &$(y-x_1)(1-x_1')$  &$x_2'(x_2+y-1)$ \\

 $C_{b3}$  &  $x_2y$      &  $x_3x_3'y$        &$x_3+y$  &$x_3(1-x_1')$
 &$Eq_{b3}$ &  $x_3x_3'$   &  $x_2'(y-1)$        &$x_3(1-x_2')$  &$x_2'(x_2+y-1)$ \\

 $C_{b4}$  &  $x_2y$      &  $x_3x_3'$        &$(y-x_3)(x_2-x_2'-x_3')$  &$x_3(1-x_1')$
 &$Eq_{b4}$ &  $x_3x_3'$   &  $x_2'(y-1)$        &$x_2+x_3'(1-x_2)$  &$x_2'(x_2+y-1)$ \\

 $C_{b5}$  &  $x_2y$      &  $x_3x_3'y$        &$y(x_2-x_2')$  &$x_2-x_2'-x_3'(y-x_3)$
 &$Eq_{b5}$ &  $x_3x_3'$   &  $x_2'(y-1)$        &$x_3'(x_3-y)$  &$x_2'(x_2+y-1)$ \\

 $C_{b6}$  &  $x_2y$      &  $x_3x_3'$        &$y(x_2-x_1')$  &$x_3(1-x_1')$
 &$Eq_{b6}$ &  $x_3x_3'$   &  $(1-x_1')(x_3+y-1)$        &$x_3'(x_3+y+1)$  &$(1-x_1')(y-x_1)$ \\

 $C_{b7}$  &  $x_2y$      &  $x_3x_3'$        &$x_2$  &$x_3(1-x_1')$
 &$Eq_{b7}$ &  $x_3x_3'$   &  $(1-x_1')(x_3+y-1)$        &$x_3(1-x_1')$  &$(1-x_1')(y-x_1)$ \\

 $C_{c1}$  &  $x_2y$      &  $y(1-x_1)$        &$y(1-x_1-x_3')$  &$y(x_2'-x_1)$
 &$Eq_{c1}$ &  $x_3x_3'$   &  $x_2'(y-1)$        &$x_3'(1-x_1)$  &$(y-1)(x_2'+1)-x_2'$ \\

 $C_{c2}$  &  $x_2y$      &  $y(x_2-x_3')$        &$y(1-x_1-x_3')$  &$y(x_2'-x_1)$
 &$Eq_{c2}$ &  $x_3x_3'$   &  $x_2'(y-1)$        &$x_3(1-x_2')$  &$(y-1)(x_2'+1)-x_2'$ \\

 $C_{c3}$  &  $x_2y$      &  $x_3x_3'$        &$x_3+y$  &$x_3(1-x_2')$
 &$Eq_{c3}$ &  $x_3x_3'$   &  $x_2'(y-1)$        &$x_2+x_3'(1-x_2)$  &$x_1'(x_2-y)+1$ \\

 $C_{c4}$  &  $x_2y$      &  $x_3x_3'$        &$y(x_2-x_2')$  &$x_3(1-x_2')$
 &$Eq_{c4}$ &  $x_3x_3'$   &  $x_2'(y-1)$        &$x_1'(x_2-y)+1$  &$(1-x_2')(1-x_1-y)+1$ \\

 $C_{c5}$  &  $x_2y$      &  $x_3x_3'$        &$(x_3-y)(1-x_2-x_2')$  &$x_3(1-x_2')$
 &$Eq_{c5}$ &  $x_3x_3'$   &  $x_2'(y-1)$        &$x_3'(x_3-y)$  &$(1-x_2')(1-x_1-y)+1$ \\

 $C_{c6}$  &  $x_2y$      &  $x_3x_3'$        &$y(x_2-x_1')$  &$(y-x_3)(x_2+x_2'-1)$
 &$Eq_{c6}$ &  $x_3x_3'$   &  $(1-x_1')(x_3+y-1)$        &$x_3'(x_3+y+1)$  &$x_1'(x_2-y)+1$ \\

 $C_{c7}$  &  $x_2y$      &  $x_3x_3'$        &$x_2$  &$x_3(1-x_2')$
 &$Eq_{c7}$ &  $x_3x_3'$   &  $(1-x_1')(x_3+y-1)$        &$x_3(1-x_1')$  &$x_1'(x_2-y)+1$ \\

 $C_{d1}$  &  $x_2y$      &  $y(1-x_1-x_3')$        &$y(1-x_1)$  &$1-x_1-x_3'$
 &$Eq_{d1}$ &  $x_3x_3'$   &  $x_2'(y-1)$        &$x_3'(1-x_1)$  &$-x_2'$ \\

 $C_{d2}$  &  $x_2y$      &  $y(1-x_1-x_3')$        &$y(x_2-x_3')$  &$1-x_1-x_3'$
 &$Eq_{d2}$ &  $x_3x_3'$   &  $x_2'(y-1)$        &$x_3(1-x_2')$  &$-x_2'$ \\

 $C_{d3}$  &  $x_2y$      &  $x_3x_3'$        &$x_3+y$  &$x_3'(x_3+y-1)$
 &$Eq_{d3}$ &  $x_3x_3'$   &  $x_2'(y-1)$        &$x_2+x_3'(1-x_2)$  &$-x_2'$ \\

 $C_{d4}$  &  $x_2y$      &  $x_3x_3'$        &$y(x_2-x_2')$  &$x_3'(x_3+y-1)$
 &$Eq_{d4}$ &  $x_3x_3'$   &  $x_2'(y-1)$        &$(x_3-1)(1-x_1')$  &$-x_2'$ \\

 $C_{d5}$  &  $x_2y$      &  $x_3x_3'$        &$$  &$x_3'(x_3+y-1)$
 &$Eq_{d5}$ &  $x_3x_3'$   &  $x_2'(y-1)$        &$x_3'(x_3-y)$  &$(1-x_1')(x_3-1)$ \\

 $C_{d6}$  &  $x_2y$      &  $x_3x_3'$        &$y(x_2-x_1')$  &$x_3'(x_3+y-1)$
 &$Eq_{d6}$ &  $x_3x_3'$   &  $(1-x_1')(x_3+y-1)$        &$x_3'(x_3+y+1)$  &$(1-x_1')(x_3-1)$ \\

 $C_{d7}$  &  $x_2y$      &  $x_3x_3'$        &$x_2$  &$(1-x_1')(x_3-1)$
 &$Eq_{d7}$ &  $x_3x_3'$   &  $(1-x_1')(x_3+y-1)$        &$x_3(1-x_1')$  &$(x_2-x_3')(1-x_3)$ \\

 $C_{e1}$  &  $x_3x_3'$      &  $x_3'(1-x_1-y)$        &$x_3'(1-x_1)$  &$y+x_3'(1-y)$
 &$Eq_{e1}$ &  $x_2'(y-1)$   &  $(1-x_1')(x_3+y-1)$        &$(y-1)(1-x_1')$  &$(1-x_1')(y-x_1)$ \\

 $C_{e2}$  &  $x_3x_3'$      &  $x_3'(1-x_1-y)$        &$x_3'(1-x_1)$  &$(1-x_1')(1-x_1-y)$
 &$Eq_{e2}$ &  $x_2'(y-1)$   &  $(1-x_1')(x_3+y-1)$        &$(y-1)(1-x_1')$  &$x_3+y-1$ \\

 $C_{e3}$  &  $x_3x_3'$      &  $x_3'(1-x_1-y)$        &$x_3'(1-x_1)$  &$(1-x_2')(1-x_1-y)$
 &$Eq_{e3}$ &  $x_2'(y-1)$   &  $(1-x_1')(x_3+y-1)$        &$(y-1)(1-x_1')$  &$x_1'(x_2-y)+1$ \\

 $C_{e4}$  &  $x_3x_3'$      &  $x_3'(1-x_1-y)$        &$x_3'(1-x_1)$  &$-x_1x_3'$
 &$Eq_{e4}$ &  $x_2'(y-1)$   &  $(1-x_1')(x_3+y-1)$        &$(y-1)(1-x_1')$  &$(x_3-1)(1-x_1')$ \\

 $C_{f1}$  &  $x_3x_3'$      &  $x_3'(1-x_1-y)$        &$x_3'(x_3-y)$  &$y+x_3'(1-y)$
 &$Eq_{f1}$ &  $x_2'(y-1)$   &  $(1-x_1')(x_3+y-1)$        &$x_2'(x_3+y-1)$  &$(1-x_1')(y-x_1)$ \\

 $C_{f2}$  &  $x_3x_3'$      &  $x_3'(1-x_1-y)$        &$x_3'(x_3-y)$  &$(1-x_1')(1-x_1-y)$
 &$Eq_{f2}$ &  $x_2'(y-1)$   &  $(1-x_1')(x_3+y-1)$        &$x_2'(x_3+y-1)$  &$x_3+y-1$ \\

 $C_{f3}$  &  $x_3x_3'$      &  $x_3'(1-x_1-y)$        &$x_3'(x_3-y)$  &$(1-x_2')(1-x_1-y)$
 &$Eq_{f3}$ &  $x_2'(y-1)$   &  $(1-x_1')(x_3+y-1)$        &$x_2'(x_3+y-1)$  &$x_1'(x_2-y)+1$ \\

 $C_{f4}$  &  $x_3x_3'$      &  $x_3'(1-x_1-y)$        &$x_3'(x_3-y)$  &$-x_1x_3'$
 &$Eq_{f4}$ &  $x_2'(y-1)$   &  $(1-x_1')(x_3+y-1)$        &$x_2'(x_3+y-1)$  &$(1-x_1')(x_3-1)$ \\

 $C_{g2}$  &  $x_2$      &  $x_3x_3'$        &$(x_3-y)(x_3'-x_2)$  &$x_2-x_2'-x_3'(y-x_3)$

 &$Eq_{g1}$ &  $x_2'(y-1)$   &  $x_3x_3'$        &$(1-x_1')(x_3+y-1)$  &$(1-x_1')(y-x_1)$ \\

 $C_{g3}$  &  $x_2$      &  $x_3x_3'$        &$(x_3-y)(x_3'-x_2)$  &$(x_2-x_3')(1-x_3)$
 &$Eq_{g3}$ &  $x_2'(y-1)$   &  $x_3x_3'$        &$(1-x_1')(x_3+y-1)$  &$x_1'(x_2-y)+1$ \\

 $C_{g4}$  &  $x_2$      &  $x_3x_3'$        &$(x_3-y)(x_3'-x_2)$  &$(y-x_3)(x_2+x_2'-1)$
 &$Eq_{g4}$ &  $x_2'(y-1)$   &  $x_3x_3'$        &$(1-x_1')(x_3+y-1)$  &$(x_3-1)(1-x_1')$ \\

 $B_{a1}$  &  $x_3y$      &  $x_3'(y-1)$        &$x_2+y(1-x_2)$  &$(1-x_2')(y-1)$
 &$B_{a2}$ &  $x_3y$   &  $x_3'(y-1)$        &$y(x_3-x_2')$  &$(1-x_2')(y-1)$ \\

 $B_{a3}$  &  $x_3y$      &  $x_3'(y-1)$        &$y(1-x_1)$  &$(1-x_2')(y-1)$
 &$B_{a4}$ &  $x_3y$   &  $x_3'(y-1)$        &$x_3+x_2'-1$  &$(1-x_2')(y-1)$ \\

 $B_{a5}$  &  $x_3y$      &  $x_3'(y-1)$        &$y(x_3-x_1')$  &$x_3+x_2'-1$
 &$B_{a6}$ &  $x_3y$   &  $x_3'(y-1)$        &$y(x_3-x_3')$  &$x_3+x_2'-1$ \\

 $B_{a7}$  &  $x_3y$      &  $x_3'(y-1)$        &$x_3$  &$x_3+x_2'-1$
 &$B_{b1}$ &  $x_3y$   &  $x_3'(y-1)$        &$x_2+y(1-x_2)$  &$(1-y)(x_2-x_3')$ \\

 $B_{b2}$  &  $x_3y$      &  $x_3'(y-1)$        &$y(x_3-x_2')$  &$(1-y)(x_2-x_3')$
 &$B_{b3}$ &  $x_3y$   &  $x_3'(y-1)$        &$y(1-x_1)$  &$1-x_1-x_3'$ \\

 $B_{b4}$  &  $x_3y$      &  $x_3'(y-1)$        &$x_2+x_3-x_3'$  &$(1-y)(x_2-x_3')$
 &$B_{b5}$ &  $x_3y$   &  $x_3'(y-1)$        &$y(x_3-x_1')$  &$(1-y)(x_2-x_3')$ \\

 $B_{b6}$  &  $x_3y$      &  $x_3'(y-1)$        &$y(x_3-x_3')$  &$1-x_1-x_3'$
 &$B_{b7}$ &  $x_3y$   &  $x_3'(y-1)$        &$x_3$  &$1-x_1-x_3'$ \\

 $B_{c1}$  &  $x_3y$      &  $x_3'(y-1)$        &$x_2+y(1-x_2)$  &$1$
 &$B_{c2}$ &  $x_3y$   &  $x_3'(y-1)$        &$1$  &$1-y(1-x_2')$ \\

 $B_{c3}$  &  $x_3y$      &  $x_3'(y-1)$        &$y(x_3-x_2')$  &$1-y(1-x_2')$
 &$B_{c4}$ &  $x_3y$   &  $x_3'(y-1)$        &$y(1-x_1)$  &$1-y(1-x_2')$ \\

 $B_{c5}$  &  $x_3y$      &  $x_3'(y-1)$        &$y(x_3-x_1')$  &$1-y(1-x_2')$
 &$B_{c6}$ &  $x_3y$   &  $x_3-x_3'$        &$y(x_3-x_3')$  &$1$ \\

 $B_{c7}$  &  $x_3y$      &  $x_3-x_3'$        &$x_3$  &$1$
 &$B_{d1}$ &  $x_3y$   &  $x_3'(y-1)$        &$x_2+y(1-x_2)$  &$(y-1)(1-x_1')$ \\

 $B_{d2}$  &  $x_3y$      &  $x_3'(y-1)$        &$x_3-x_2'-x_3'$  &$(y-1)(1-x_1')$
 &$B_{d3}$ &  $x_3y$   &  $x_3'(y-1)$        &$y(x_3-x_2')$  &$x_3+x_1'-1$ \\

 $B_{d4}$  &  $x_3y$      &  $x_3'(y-1)$        &$y(1-x_1)$  &$(y-1)(1-x_1')$
 &$B_{d5}$ &  $x_3y$   &  $x_3'(y-1)$        &$y(x_3-x_1')$  &$(y-1)(1-x_1')$ \\

 $B_{d6}$  &  $x_3y$      &  $x_3-x_3'$        &$y(x_3-x_3')$  &$x_3+x_1'-1$
 &$B_{d7}$ &  $x_3y$   &  $x_3-x_3'$        &$x_3$  &$x_3+x_1'-1$ \\

 $B_{e1}$  &  $x_3'(y-1)$      &  $x_3-x_3'$        &$-x_3'$  &$1$
 &$B_{e2}$ &  $x_3'(y-1)$   &  $x_3-x_3'$        &$-x_3'$  &$x_3+x_1'-1$ \\

 $B_{e3}$  &  $x_3'(y-1)$      &  $x_3-x_3'$        &$-x_3'$  &$1-x_1-x_3'$
 &$B_{e4}$ &  $x_3'(y-1)$   &  $x_3-x_3'$        &$-x_3'$  &$x_3+x_2'-1$ \\

 $B_{f1}$  &  $x_3'(y-1)$      &  $x_3-x_3'$        &$(1-y)(x_3-x_3')$  &$1$
 &$B_{f2}$ &  $x_3'(y-1)$   &  $x_3-x_3'$        &$(1-y)(x_3-x_3')$  &$x_3+x_1'-1$ \\

 $B_{f3}$  &  $x_3'(y-1)$      &  $x_3-x_3'$        &$(1-y)(x_3-x_3')$  &$1-x_1-x_3'$
 &$B_{f4}$ &  $x_3'(y-1)$   &  $x_3-x_3'$        &$(1-y)(x_3-x_3')$  &$x_3+x_2'-1$ \\

 $B_{g1}$  &  $x_3y$      &  $x_3'(y-1)$        &$x_3-x_3'$  &$1-x_1-x_3'$
 &$B_{g2}$ &  $x_3y$   &  $x_3'(y-1)$        &$x_3-x_3'$  &$x_3+x_2'-1$ \\

 $B_{g3}$  &  $x_3y$      &  $x_3'(y-1)$        &$x_3-x_3'$  &$1$
 &$B_{g4}$ &  $x_3y$   &  $x_3'(y-1)$        &$x_3-x_3'$  &$x_3+x_1'-1$ \\

	\hline\hline
	\end{tabular}
\end{table}

\begin{table}[H]
	\footnotesize
	\centering
		\caption{The expressions of $H^{\sigma}_{R_{ij}}$  for the six representative diagrams in Fig~\ref{fig:FeynmanT}.
The ratios $r_0^{q,s}$ are defined by $r_0^{q,s}=m_0^{q,s}/M$.}
	\newcommand{\tabincell}[2]{\begin{tabular}{@{}#1@{}}#2\end{tabular}}
	\label{tab:amppc}
	\begin{tabular}[t]{lcc}
		\hline\hline
		$$  &$\frac{M_S}{16M^4}$&$\frac{M_P}{16M^4}$\\ \hline
		$H_{T_{c7}}^{LL}$
		&\tabincell{c}{$2\phi_q^A(x_3'(\Psi_3^{+-}(r_{\Lambda}-1)\Phi^T(x_2-y)-\Psi_4x_2'r_{\Lambda}(\Phi^A-\Phi^V))-2\Psi_2$\\$\Phi^A(r_{\Lambda}-1)(y-x_2)(x_3+y-1)-\Psi_3^{-+}x_2'r_{\Lambda}\Phi^T(x_3+y-1))$}
		&\tabincell{c}{$2\phi_q^A((r_{\Lambda}-1)(x_3+y-1)(2\Psi_2\Phi^A(y-x_2)+\Psi_3^{-+}x_2'\Phi^T)$\\$+x_3'r_{\Lambda}(\Psi_4x_2'(\Phi^A+\Phi^V)+\Psi_3^{+-}\Phi^T(y-x_2)))$}
		\\\\
		$H_{T_{c7}}^{SP}$
		&\tabincell{c}{$2\phi_q^A(x_2'r_{\Lambda}(\Psi_4x_3'(\Phi^A-\Phi^V)-\Psi_3^{-+}\Phi^T(x_3+y-1))$\\$+(r_{\Lambda}+1)(y-x_2)(2\Psi_2\Phi^A(x_3+y-1)-x_3'\Psi_3^{+-}\Phi^T))$}
		&\tabincell{c}{$2\phi_q^A(x_3'r_{\Lambda}(\Psi_3^{+-}\Phi^T(y-x_2)-\Psi_4x_2'(\Phi^A+\Phi^V))$\\$-(r_{\Lambda}+1)(x_3+y-1)(2\Psi_2\Phi^A(y-x_2)-\Psi_3^{-+}x_2'\Phi^T))$}
		\\\\

		$H_{C_{d7}}^{LL}$
		&\tabincell{c}{$r_{\Lambda}(3(x_3-1)\Phi^T(x_2\Psi_3^{+-}\phi_q^A+4r_0^q(\Psi_2-x_2\Psi_4)\phi_q^P)$\\$+2(x_3'-x_2)\Psi_3^{+-}r_0^q(\Phi^A-\Phi^V)\phi_q^P)$}
		&\tabincell{c}{$r_{\Lambda}(2r_0^q\phi_q^P((x_2-x_3')\Psi_3^{+-}(\Phi^A-\Phi^V)$\\$+6(x_3-1)\Phi^T(\Psi_2-x_2\Psi_4))+3x_2(x_3-1)\Psi_3^{+-}\Phi^T\phi_q^A)$}
		\\\\
		$H_{C_{d7}}^{SP}$
		&\tabincell{c}{$2(x_2-x_3')\Psi_3^{+-}r_0^qr_{\Lambda}(\Phi^A+\Phi^V)\phi_q^P$\\$-(x_3-1)(r_{\Lambda}-2)\Phi^T(x_2\Psi_3^{+-}\phi_q^A+4r_0^q(\Psi_2-x_2\Psi_4)\phi_q^P)$}
		&\tabincell{c}{$-x_2(x_3-1)\Psi_3^{+-}(r_{\Lambda}+2)\Phi^T\phi_q^A-2r_0^q\phi_q^P((x_2-x_3')$\\$\Psi_3^{+-}r_{\Lambda}(\Phi^A+\Phi^V)+2(x_3-1)(r_{\Lambda}+2)\Phi^T(\Psi_2-x_2\Psi_4))$}
		\\\\
		
		$H_{Eq_{a2}}^{LL}$
		&\tabincell{c}{$r_0^q((y-1)\Phi^A(\Psi_3^{-+}(r_{\Lambda}-1)(x_3+y-1)(\phi_q^P+\phi_q^T)+\Psi_3^{+-}r_{\Lambda}$\\$(\phi_q^P-\phi_q^T))+(y-1)\Phi^V(\Psi_3^{-+}(r_{\Lambda}-1)(x_3+y-1)(\phi_q^P+\phi_q^T)$\\$+\Psi_3^{+-}r_{\Lambda}(\phi_q^P-\phi_q^T))+2\Phi^T(x_3+y-1)(\phi_q^P+\phi_q^T)$\\$(2\Psi_2(y-1)(r_{\Lambda}-1)-\Psi_4(x_3'-1)r_{\Lambda}))$}
		&\tabincell{c}{$r_0^q(-(y-1)\Phi^A(\Psi_3^{-+}(r_{\Lambda}+1)(x_3+y-1)(\phi_q^P+\phi_q^T)$\\$+\Psi_3^{+-}r_{\Lambda}(\phi_q^P-\phi_q^T))-(y-1)\Phi^V(\Psi_3^{-+}(r_{\Lambda}+1)(x_3+y-1)$\\$(\phi_q^P+\phi_q^T)+\Psi_3^{+-}r_{\Lambda}(\phi_q^P-\phi_q^T))+2\Phi^T(x_3+y-1)$\\$(\phi_q^P+\phi_q^T)(2\Psi_2(y-1)(r_{\Lambda}+1)-\Psi_4(x_3'-1)r_{\Lambda}))$}
		\\\\
		$H_{Eq_{a2}}^{SP}$
		&\tabincell{c}{$r_{\Lambda}\Phi^T(\phi_q^A((y-1)\Psi_3^{-+}+(x_3'-1)\Psi_3^{+-}(x_3+y-1))+6(y-1)$\\$r_0^q(\Psi_2(x_3+y-1)-\Psi_4)(\phi_q^P-\phi_q^T))+\Phi^V((x_3'-1)\Psi_3^{+-}r_0^qr_{\Lambda}$\\$(x_3+y-1)(\phi_q^P+\phi_q^T)-2(y-1)(r_{\Lambda}-1)\phi_q^A(\Psi_2(x_3+y-1)$\\$-\Psi_4))+\Phi^A(2(1-y)(r_{\Lambda}-1)\phi_q^A(\Psi_2(x_3+y-1)-\Psi_4)$\\$-(x_3'-1)\Psi_3^{+-}r_0^qr_{\Lambda}(x_3+y-1)(\phi_q^P+\phi_q^T))$}
		&\tabincell{c}{$\phi_q^A(2\Psi_2(y-1)(r_{\Lambda}+1)(x_3+y-1)(\Phi^A+\Phi^V)-2\Psi_4(y-1)$\\$(r_{\Lambda}+1)(\Phi^A+\Phi^V)+r_{\Lambda}\Phi^T((y-1)\Psi_3^{-+}+(x_3'-1)$\\$\Psi_3^{+-}(x_3+y-1)))+r_0^qr_{\Lambda}(\Phi^A(x_3'-1)\Psi_3^{+-}(x_3+y-1)$\\$(\phi_q^P+\phi_q^T)-(x_3'-1)\Psi_3^{+-}\Phi^V(x_3+y-1)(\phi_q^P+\phi_q^T)$\\$+6(y-1)\Phi^T(\Psi_2(x_3+y-1)-\Psi_4)(\phi_q^P-\phi_q^T))$}
		\\\\

		$H_{Es_{d2}}^{LL}$
		&\tabincell{c}{$r_0^s\phi_s^P(2(x_3-1)(\Phi^A+\Phi^V)(\Psi_2(r_{\Lambda}-1)-\Psi_4x_1'r_{\Lambda})$\\$+\Phi^T(2(x_3-1)\Psi_3^{-+}(r_{\Lambda}-1)-(x_2'-1)\Psi_3^{+-}r_{\Lambda}))$\\$-\frac{1}{2}x_1'(1-x_3)\phi_s^A(2\Psi_4(r_{\Lambda}-1)(\Phi^A-\Phi^V)-\Psi_3^{-+}r_{\Lambda}\Phi^T)$}
		&\tabincell{c}{$r_0^s\phi_s^P(2(x_3-1)(\Phi^A+\Phi^V)(\Psi_2(r_{\Lambda}+1)-\Psi_4x_1'r_{\Lambda})$\\$+\Phi^T((x_2'-1)\Psi_3^{+-}r_{\Lambda}-2(x_3-1)\Psi_3^{-+}(r_{\Lambda}+1)))$\\$+\frac{1}{2}x_1'(1-x_3)\phi_s^A(2\Psi_4(r_{\Lambda}+1)(\Phi^A-\Phi^V)+\Psi_3^{-+}r_{\Lambda}\Phi^T)$}
		\\\\
		$H_{Es_{d2}}^{SP}$
		&\tabincell{c}{$r_0^s\phi_s^P(2(r_{\Lambda}-1)(\Phi^A-\Phi^V)(\Psi_4(x_2'-1)+(x_3-1)\Psi_2)$\\$-r_{\Lambda}\Phi^T(2x_1'(x_3-1)\Psi_3^{+-}-\Psi_3^{-+}(x_2'-1)))$\\$+\frac{1}{2}x_1'(x_3-1)\Psi_3^{+-}(r_{\Lambda}-2)\Phi^T\phi_s^A$}
		&\tabincell{c}{$2(r_{\Lambda}+1)r_0^s(\Phi^A-\Phi^V)(\Psi_4(x_2'-1)+(x_3-1)\Psi_2)\phi_s^P$\\$+\frac{1}{2}x_1'(x_3-1)\Psi_3^{+-}(r_{\Lambda}+2)\Phi^T\phi_s^A$\\$-r_{\Lambda}r_0^s\Phi^T\phi_s^P(2x_1'(1-x_3)\Psi_3^{+-}-\Psi_3^{-+}(x_2'-1))$}   \\\\		
		$H_{B_{b3}}^{LL}$
		&\tabincell{c}{$r_{\Lambda}(y(x_1+x_3'-1)\Phi^V(2\Psi_2\phi_q^A+r_0^q(\Psi_3^{-+}+\Psi_3^{+-})(\phi_q^P+\phi_q^T))$\\$-3\Phi^T((1-x_1)\phi_q^A(\Psi_3^{-+}+\Psi_3^{+-})+2r_0^q((1-x_1)\Psi_4(\phi_q^P+\phi_q^T)$\\$+\Psi_2y(\phi_q^T-\phi_q^P)))+y\Phi^A(1-x_1-x_3')(2\Psi_2\phi_q^A$\\$+r_0^q(\Psi_3^{-+}+\Psi_3^{+-})(\phi_q^P+\phi_q^T)))$}
		&\tabincell{c}{$r_{\Lambda}(y(x_3'-x_3-x_2)(\Phi^A-\Phi^V)(2\Psi_2\phi_q^A+r_0^q(\Psi_3^{-+}$\\$+\Psi_3^{+-})(\phi_q^P+\phi_q^T))-3\Phi^T((1-x_1)\phi_q^A(\Psi_3^{-+}+\Psi_3^{+-})$\\$+2r_0^q((1-x_1)\Psi_4(\phi_q^P+\phi_q^T)+\Psi_2y(\phi_q^T-\phi_q^P))))$}
		\\\\
		$H_{B_{b3}}^{SP}$
		&\tabincell{c}{$y(x_1+x_3'-1)r_{\Lambda}\Phi^V(2\Psi_2\phi_q^A+Mr_0^q(\Psi_3^{-+}+\Psi_3^{+-})(\phi_q^P+\phi_q^T))$\\$-y\Phi^A(1-x_1-x_3')r_{\Lambda}(2\Psi_2\phi_q^A+Mr_0^q(\Psi_3^{-+}+\Psi_3^{+-})$\\$(\phi_q^P+\phi_q^T))+(r_{\Lambda}-2)\Phi^T((1-x_1)\phi_q^A(\Psi_3^{-+}+\Psi_3^{+-})$\\$+2r_0^q((1-x_1)\Psi_4(\phi_q^P+\phi_q^T)+\Psi_2y(\phi_q^T-\phi_q^P)))$}
		&\tabincell{c}{$y(1-x_1-x_3')r_{\Lambda}(\Phi^A+\Phi^V)(2\Psi_2\phi_q^A+r_0^q(\Psi_3^{-+}+\Psi_3^{+-})$\\$(\phi_q^P+\phi_q^T))+(r_{\Lambda}+2)\Phi^T((1-x_1)\phi_q^A(\Psi_3^{-+}+\Psi_3^{+-})$\\$+2r_0^q((1-x_1)\Psi_4(\phi_q^P+\phi_q^T)+\Psi_2y(\phi_q^T-\phi_q^P)))$}
		\\\\
		
		$H_{B'_{a3}}^{LL}$
		&$x_3'r_{\Lambda}\phi_s^A(2\Psi_4x_3'(\Phi^A-\Phi^V)+3x_1\Psi_3^{+-}\Phi^T)$
		&$-r_{\Lambda}x_3'\phi_s^A(2\Psi_4x_3'(\Phi^A-\Phi^V)-3x_1\Psi_3^{+-}\Phi^T)$
		\\\\
		$H_{B'_{a3}}^{SP}$
		&$-x_3'\phi_s^A(x_1\Psi_3^{+-}(r_{\Lambda}-2)\Phi^T+2\Psi_4x_3'r_{\Lambda}(\Phi^A+\Phi^V))$
		&$x_3'\phi_s^A(2\Psi_4x_3'r_{\Lambda}(\Phi^A+\Phi^V)-x_1\Psi_3^{+-}(r_{\Lambda}+2)\Phi^T)$    \\
		\hline\hline
	\end{tabular}
\end{table}

\end{appendix}


\begin{thebibliography}{99}
\bibitem{Leutwyler:1997yr}
H.~Leutwyler,
On the 1/$N$-expansion in chiral perturbation theory,
Nucl. Phys. B, Proc. Suppl. \textbf{64}, 223 (1998).


\bibitem{Kaiser:2000gs}
R.~Kaiser and H.~Leutwyler,
Large $N_c$ in chiral perturbation theory,
Eur. Phys. J. C \textbf{17}, \textbf{623} 649 (2000).


\bibitem{Feldmann:1998vh}
T.~Feldmann, P.~Kroll, and B.~Stech,
Mixing and decay constants of pseudoscalar mesons,
Phys. Rev. D \textbf{58}, 114006 (1998).

\bibitem{Feldmann:1998sh}
T.~Feldmann, P.~Kroll, and B.~Stech,
Mixing and decay constants of pseudoscalar mesons: The sequel,
Phys. Lett. B \textbf{449}, 339 (1999).

\bibitem{Harari:1975ie}
H.~Harari,
Experimental consequences of a small mixture of $c\bar c$ in the $\eta'$  and $\eta$ mesons,
Phys. Lett. \textbf{60B}, 172  (1976).

\bibitem{Tsai:2011dp}
Y.~D.~Tsai, H.~n.~Li, and Q.~Zhao,
$\eta_c$ mixing effects on charmonium and $B$ meson decays,
Phys. Rev. D \textbf{85}, 034002 (2012).

\bibitem{Escribano:2007cd}
R.~Escribano and J.~Nadal,
On the gluon content of the $\eta$ and $eta'$ mesons,
J. High Energy Phys. 05 (2007) 006.

\bibitem{Mathieu:2009sg}
V.~Mathieu and V.~Vento,
Pseudoscalar glueball and $\eta-\eta'$ mixing,
Phys. Rev. D \textbf{81}, 034004 (2010).

\bibitem{Ahmady:1997fa}
M.~R.~Ahmady, E.~Kou, and A.~Sugamoto,
Non-spectator contribution: A mechanism for inclusive $B \to X_s \eta'$ and exclusive $B\to K^{(*)} \eta'$ decays,
Phys. Rev. D \textbf{58}, 014015 (1998).

\bibitem{Beneke:2002jn}
M.~Beneke and M.~Neubert,
Flavor singlet B-decay amplitudes in QCD factorization,
Nucl. Phys. \textbf{B651}, 225 (2003).

\bibitem{KLOE:2006guu}
F.~Ambrosino \textit{et al.} (KLOE Collaboration),
Measurement of the pseudoscalar mixing angle and $\eta'$ gluonium content with KLOE detector,
Phys. Lett. B \textbf{648}, 267 (2007).

\bibitem{Ke:2010htz}
H.~W.~Ke, X.~Q.~Li, and Z.~T.~Wei,
Determining the $\eta-\eta'$ mixing by the newly measured $BR(D(D_s)\to\eta(\eta')+\bar l+\nu_l$,
Eur. Phys. J. C \textbf{69}, 133 (2010).


\bibitem{Williamson:2006hb}
A.~R.~Williamson and J.~Zupan,
Two body $B$ decays with isosinglet final states in SCET,
Phys. Rev. D \textbf{74}, 014003 (2006).

\bibitem{Fleischer:2011ib}
R.~Fleischer, R.~Knegjens, and G.~Ricciardi,
Exploring CP violation and $\eta$-$\eta'$ mixing with the $B^0_{s,d} \to J/\psi \eta^{(\prime)}$ systems,
Eur. Phys. J. C \textbf{71}, 1798 (2011).

\bibitem{Feldmann:1999uf}
T.~Feldmann,
Quark structure of pseudoscalar mesons,
Int. J. Mod. Phys. A \textbf{15}, 159  (2000).

\bibitem{Zhu:2018bwp}
R.~Zhu, Y.~Ma, X.~L.~Han, and Z.~J.~Xiao,
Form factors for semileptonic $B_c$ decays into $\eta^{(')}$ and glueballs,
Phys. Rev. D \textbf{98}, 114035 (2018).

\bibitem{DiDonato:2011kr}
C.~Di Donato, G.~Ricciardi, and I.~Bigi,
$\eta-\eta'$ mixing\textemdash From electromagnetic transitions to weak decays of charm and beauty hadrons,
Phys. Rev. D \textbf{85}, 013016 (2012).

\bibitem{pdg2022}
R.L. Workman \textit{et al.} (Particle Data Group), Prog. Theor. Exp. Phys. \textbf{2022}, 083C01 (2022).

\bibitem{Ahmady:1998ws}
M.~R.~Ahmady and E.~Kou,
Possible large direct CP asymmetry in hadronic $B^{\pm} \to \pi^{\pm} \eta'$ decays,
Phys. Rev. D \textbf{59}, 054014 (1999).

\bibitem{Du:1997hs}
D.~s.~Du, C.~S.~Kim, and Y.~d.~Yang,
A new mechanism for $B^{\pm} \to \eta' K^{\pm}$ in perturbative QCD,
Phys. Lett. B \textbf{426}, 133  (1998).

\bibitem{Halperin:1997as}
I.~E.~Halperin and A.~Zhitnitsky,
$B \to K\eta' $ decay as unique probe of $\eta'$ meson,
Phys. Rev. D \textbf{56}, 7247  (1997).

\bibitem{Petrov:1997yf}
A.~A.~Petrov,
Intrinsic charm of light mesons and CP violation in heavy quark decay,
Phys. Rev. D \textbf{58}, 054004 (1998).

\bibitem{Khalil:2003bi}
S.~Khalil and E.~Kou,
A Possible Supersymmetric Solution to the Discrepancy between $B \to \phi K_s$ and $B \to \eta' K_s$ CP Asymmetries,
Phys. Rev. Lett. \textbf{91}, 241602 (2003).

\bibitem{Yang:2000ce}
M.~Z.~Yang and Y.~D.~Yang,
Revisiting charmless two-body B decays involving $\eta'$ and $\eta$,
Nucl. Phys. \textbf{B609}, 469 (2001).

\bibitem{Xiao:2008sw}
Z.~J.~Xiao, Z.~Q.~Zhang, X.~Liu, and L.~B.~Guo,
Branching ratios and CP asymmetries of $B \to K \eta^{(')}$ decays in the pQCD approach,
Phys. Rev. D \textbf{78}, 114001 (2008).

\bibitem{prd87094003}
Y.~Y.~Fan, W.~F.~Wang, S.~Cheng, and Z.~J.~Xiao,
Anatomy of $B \to K \eta^{(')}$ decays in different mixing schemes and effects of next-to-leading order contributions in the perturbative QCD approach,
Phys. Rev. D \textbf{87},   094003 (2013).

\bibitem{Charng:2006zj}
Y.~Y.~Charng, T.~Kurimoto, and H.~n.~Li,
Gluonic contribution to $B \to \eta'$ form factors,
Phys. Rev. D \textbf{74}, 074024 (2006);\textbf{78}, 059901(E) (2008).

\bibitem{Kou:2001pm}
E.~Kou and A.~I.~Sanda,
$B \to K \eta'$ decay in perturbative QCD,
Phys. Lett. B \textbf{525}, 240  (2002).

\bibitem{Akeroyd:2007fy}
A.~G.~Akeroyd, C.~H.~Chen, and C.~Q.~Geng,
$B \to \eta^{(')}(l^-\bar v_l,l^+l^-,K,K^*)$ decays in the quark-flavor mixing scheme,
Phys. Rev. D \textbf{75}, 054003 (2007).

\bibitem{Cheng:2008ss}
H.~Y.~Cheng, H.~n.~Li, and K.~F.~Liu,
Pseudoscalar glueball mass from $\eta-\eta'-G$ mixing,
Phys. Rev. D \textbf{79}, 014024 (2009).


\bibitem{Liu:2012ib}
X.~Liu, H.~n.~Li, and Z.~J.~Xiao,
Implications on $\eta$-$\eta'$-glueball mixing from $B_{d/s} \to J/\Psi \eta^{(')}$ decays,
Phys. Rev. D \textbf{86}, 011501(R) (2012).

\bibitem{LHCb:2012cw}
R.~Aaij \textit{et al.} (LHCb Collaboration),
Evidence for the decay $B^0\to J/\psi \omega$ and measurement of the relative branching fractions of $B^0_s$ meson decays to $J/\psi\eta$ and $J/\psi\eta^{'}$,
Nucl. Phys.  \textbf{B867}, 547  (2013).

\bibitem{LHCb:2014oms}
R.~Aaij \textit{et al.} (LHCb Collaboration),
Study of $\eta-\eta^{\prime}$ mixing from measurement of $B^0_{(s)} \rightarrow J/\psi \eta^{(\prime)}$ decay rates,
J. High Energy Phys. 01 (2015) 024.

\bibitem{LHCb:2015kmm}
R.~Aaij \textit{et al.} (LHCb Collaboration),
Search for the $\Lambda^0_b \rightarrow \Lambda \eta^\prime$ and $\Lambda^0_b \rightarrow \Lambda \eta$ decays with the LHCb detector,
J. High Energy Phys. 09 (2015) 006.

\bibitem{Wei:2009np}
Z.~T.~Wei, H.~W.~Ke, and X.~Q.~Li,
Evaluating decay rates and asymmetries of $\Lambda_b$ into light baryons in LFQM,
Phys. Rev. D \textbf{80}, 094016 (2009).

\bibitem{prd99054020}
J.~Zhu, Z.~T.~Wei, and H.~W.~Ke,
Semileptonic and nonleptonic weak decays of $\Lambda_b^0$,
Phys. Rev. D \textbf{99},   054020 (2019).


\bibitem{epjc76399}
C.~Q.~Geng, Y.~K.~Hsiao, Y.~H.~Lin, and Y.~Yu,
Study of $\Lambda _b\to \Lambda (\phi ,\eta ^{(\prime )})$ and $\Lambda _b\to \Lambda K^+K^-$ decays,
Eur. Phys. J. C \textbf{76}, 399 (2016).

\bibitem{plb598203}
M.~R.~Ahmady, C.~S.~Kim, S.~Oh, and C.~Yu,
Heavy baryonic decays of $\Lambda_b \to \Lambda \eta'$ decays,
Phys. Lett. B \textbf{598}, 203 (2004).

\bibitem{Datta:2002pk}
A.~Datta, H.~J.~Lipkin, and P.~J.~O'Donnell,
Nonstandard $\eta-\eta'$ mixing and the nonleptonic $B$ and $\Lambda_b$ decays to $\eta$ and $\eta'$,
Phys. Lett. B \textbf{544}, 145  (2002).

\bibitem{Schechter:1992iz}
J.~Schechter, A.~Subbaraman, and H.~Weigel,
Effective hadron dynamics: From meson masses to the proton spin puzzle,
Phys. Rev. D \textbf{48}, 339 (1993).

\bibitem{Escribano:2005qq}
R.~Escribano and J.~M.~Frere,
Study of the $\eta-\eta'$ system in the two mixing angle scheme,
J. High Energy Phys. 06 (2005) 029.

\bibitem{Bramon:1997va}
A.~Bramon, R.~Escribano, and M.~D.~Scadron,
The $\eta-\eta'$ mixing angle revisited,
Eur. Phys. J. C \textbf{7}, 271 (1999).

\bibitem{Gross:1979ur}
D.~J.~Gross, S.~B.~Treiman, and F.~Wilczek,
Light-quark masses and isospin violation,
Phys. Rev. D \textbf{19}, 2188 (1979).


\bibitem{Gusbin:1980gd}
D.~Gusbin,
$\pi^0 \eta \eta^\prime$ mixing effects in $K^+ \to \pi^+ \pi^0$ decay,
Phys. Rev. D \textbf{24}, 797 (1981).

\bibitem{Qian:2009dc}
W.~Qian and B.~Q.~Ma,
Tri-meson-mixing of $\pi$-$\eta$-$\eta'$ and $\rho$-$\omega$-$\phi$ in the light-cone quark model,
Eur. Phys. J. C \textbf{65}, 457  (2010).



\bibitem{Peng:2011ue}
T.~Peng and B.~Q.~Ma,
Tetramixing of vector and pseudoscalar mesons: A source of intrinsic quarks,
Phys. Rev. D \textbf{84}, 034003 (2011).







\bibitem{plb665197}
P.~Ball, V.~M.~Braun, and E.~Gardi,
Distribution amplitudes of the $\Lambda_b$ baryon in QCD,
Phys. Lett. B \textbf{665}, 197 (2008).

\bibitem{Ali:2012zza}
A.~Ali, C.~Hambrock, and A.~Y.~Parkhomenko,
Light-cone wave functions of heavy baryons,
Theor. Math. Phys. \textbf{170}, 2 (2012).

\bibitem{J. High Engry Phys.112013191}
G.~Bell, T.~Feldmann, Y.~M.~Wang, and M.~W.~Y.~Yip,
Light-cone distribution amplitudes for heavy-quark hadrons,
J. High Energy Phys. 11 (2013) 191.

\bibitem{J. High Engry Phys.022016179}
Y.~M.~Wang and Y.~L.~Shen,
Perturbative corrections to $\Lambda_b \to \Lambda$ form factors from QCD light-cone sum rules,
J. High Energy Phys. 02 (2016) 179.

\bibitem{epjc732302}
A.~Ali, C.~Hambrock, A.~Y.~Parkhomenko, and W.~Wang,
Light-cone distribution amplitudes of the ground state bottom baryons in HQET,
Eur. Phys. J. C \textbf{73}, 2302 (2013).

\bibitem{plb738334}
V.~M.~Braun, S.~E.~Derkachov, and A.~N.~Manashov,
Integrability of the evolution equations for heavy-light baryon distribution amplitudes,
Phys. Lett. B \textbf{738}, 334 (2014).

\bibitem{zpc42569}
V.~L.~Chernyak, A.~A.~Ogloblin, and I.~R.~Zhitnitsky,
Wave functions of octet baryons,
Z. Phys. C \textbf{42}, 569 (1989).


\bibitem{Liu:2014uha}
Y.~L.~Liu, C.~Y.~Cui, and M.~Q.~Huang,
Higher order light-cone distribution amplitudes of the $\Lambda$ baryon,
Eur. Phys. J. C \textbf{74}, 3041 (2014).

\bibitem{Liu:2008yg}
Y.~L.~Liu and M.~Q.~Huang,
Distribution amplitudes of $\Sigma$ and $\Lambda$ and their electromagnetic form factors,
Nucl. Phys. \textbf{A821}, 80 (2009).

\bibitem{J. High Engry Phys.020702016}
G.~S.~Bali, V.~M.~Braun, M.~G\"ockeler, M.~Gruber, F.~Hutzler, A.~Sch\"afer, R.~W.~Schiel, J.~Simeth, W.~S\"oldner, A.~Sternbeck \textit{et al.},
Light-cone distribution amplitudes of the baryon octet,
J. High Energy Phys. 02 (2016) 070.

\bibitem{prd89094511}
V.~M.~Braun, S.~Collins, B.~Gl\"a\ss{}le, M.~G\"ockeler, A.~Sch\"afer, R.~W.~Schiel, W.~S\"oldner, A.~Sternbeck, and P.~Wein,
Light-cone distribution amplitudes of the nucleon and negative parity nucleon resonances from lattice QCD,
Phys. Rev. D \textbf{89}, 094511 (2014).

\bibitem{epja55116}
G.~S.~Bali \textit{et al.} (RQCD Collaboration),
Light-cone distribution amplitudes of octet baryons from lattice QCD,
Eur. Phys. J. A \textbf{55}, 116 (2019).


\bibitem{Rui:2022sdc}
Z.~Rui, C.~Q.~Zhang, J.~M.~Li, and M.~K.~Jia,
Investigating the color-suppressed decays $\Lambda_b\rightarrow \Lambda \psi$ in the perturbative QCD approach,
Phys. Rev. D \textbf{106}, 053005 (2022).

\bibitem{Rui:2022jwu}
Z.~Rui, J.~M.~Li, and C.~Q.~Zhang,
Estimates of exchange topological contributions and $CP$-violating observables in $\Lambda_b\rightarrow \Lambda \phi$ decay,
Phys. Rev. D \textbf{107}, 053009 (2023).

\bibitem{Han:2022srw}
J.~J.~Han, Y.~Li, H.~n.~Li, Y.~L.~Shen, Z.~J.~Xiao, and F.~S.~Yu,
$\Lambda _b\rightarrow p$ transition form factors in perturbative QCD,
Eur. Phys. J. C \textbf{82},   686 (2022).

\bibitem{Chen:2005ht}
C.~H.~Chen and H.~N.~Li,
Nonfactorizable contributions to B meson decays into charmonia,
Phys. Rev. D \textbf{71}, 114008 (2005).

\bibitem{Rui:2016opu}
Z.~Rui, H.~Li, G.~x.~Wang, and Y.~Xiao,
Semileptonic decays of $B_c$ meson to S-wave charmonium states in the perturbative QCD approach,
Eur. Phys. J. C \textbf{76}, 564 (2016).


\bibitem{Ball:2004ye}
P.~Ball and R.~Zwicky,
New results on $B \to \pi, K, \eta$ decay form factors from light-cone sum rules,
Phys. Rev. D \textbf{71}, 014015 (2005).

\bibitem{Sun:2008ew}
J.~F.~Sun, D.~S.~Du, and Y.~L.~Yang,
Study of $B_c \to J/\psi \pi$, $\eta_c \pi$ decays with perturbative QCD approach,
Eur. Phys. J. C \textbf{60}, 107  (2009).


\bibitem{Ball:2007hb}
P.~Ball and G.~W.~Jones,
$B \to \eta^{(')}$ form factors in QCD,
J. High Energy Phys. 08 (2007) 025.

\bibitem{prd65074030}
C.~H.~Chou, H.~H.~Shih, S.~C.~Lee, and H.~n.~Li,
$\Lambda_b\rightarrow \Lambda J/\psi$ decay in perturbative QCD,
Phys. Rev. D \textbf{65}, 074030 (2002).

\bibitem{prd59094014}
H.~H.~Shih, S.~C.~Lee, and H.~n.~Li,
The $\Lambda_b \rightarrow p l \bar{\nu}$ decay in perturbative QCD,
Phys. Rev. D \textbf{59}, 094014 (1999).

\bibitem{prd61114002}
H.~H.~Shih, S.~C.~Lee, and H.~n.~Li,
Applicability of perturbative QCD to  $\Lambda_b \rightarrow \Lambda_c$  decays,
Phys. Rev. D \textbf{61}, 114002 (2000).

\bibitem{cjp39328}
H.~H.~Shih, S.~C.~Lee, and H.~N.~Li,
Asymmetry parameter in the polarized $\Lambda_b \rightarrow \Lambda_c l \bar{\nu}$ decay,
Chin. J. Phys. \textbf{39}, 328 (2001).

\bibitem{prd74034026}
X.~G.~He, T.~Li, X.~Q.~Li, and Y.~M.~Wang,
PQCD calculation for $\Lambda_b \rightarrow \Lambda \gamma$ in the standard model,
Phys. Rev. D \textbf{74}, 034026 (2006).

\bibitem{prd80034011}
C.~D.~Lu, Y.~M.~Wang, H.~Zou, A.~Ali, and G.~Kramer,
Anatomy of the pQCD approach to the baryonic decays $\Lambda_b\rightarrow p\pi, pK$,
Phys. Rev. D \textbf{80}, 034011 (2009).


\bibitem{Zhang:2022iun}
C.~Q.~Zhang, J.~M.~Li, M.~K.~Jia, and Z.~Rui,
Nonleptonic two-body decays of $\Lambda_b \to \Lambda_c \pi, \Lambda_c K$  in the perturbative QCD approach,
Phys. Rev. D \textbf{105},  073005 (2022).

\bibitem{Buchalla:1995vs}
G.~Buchalla, A.~J.~Buras, and M.~E.~Lautenbacher,
Weak decays beyond leading logarithms,
Rev. Mod. Phys. \textbf{68}, 1125 (1996).

\bibitem{Cheng:1996cs}
H.~Y.~Cheng,
Nonleptonic weak decays of bottom baryons,
Phys. Rev. D \textbf{56}, 2799  (1997); \textbf{99},  079901(E) (2019).

\bibitem{Ball:2006wn}
P.~Ball, V.~M.~Braun, and A.~Lenz,
Higher-twist distribution amplitudes of the K meson in QCD,
J. High Energy Phys. 05 (2006) 004.

\bibitem{Hsu:2007qc}
J.~F.~Hsu, Y.~Y.~Charng, and H.~n.~Li,
Okubo-Zweig-Iizuka-rule violation and $B \to \eta^{(')}K$ branching ratios,
Phys. Rev. D \textbf{78}, 014020 (2008).

\bibitem{prd95093001}
Y.~K.~Hsiao, Y.~Yao, and C.~Q.~Geng,
Charmless two-body anti-triplet $b$-baryon decays,
Phys. Rev. D \textbf{95},  093001 (2017).

\bibitem{Hsiao:2015cda}
Y.~K.~Hsiao, P.~Y.~Lin, C.~C.~Lih, and C.~Q.~Geng,
Charmful two-body anti-triplet $b$-baryon decays,
Phys. Rev. D \textbf{92}, 114013 (2015).


\end{thebibliography}
\end{document}